\documentstyle[11pt,aaspp4,epsf]{article}

\begin{document}

\title{BINDING ENERGY AND THE FUNDAMENTAL PLANE OF GLOBULAR CLUSTERS}
\author{Dean E. McLaughlin\altaffilmark{1}}
\affil{Department of Astronomy, 601 Campbell Hall\\
University of California, Berkeley, CA 94720-3411\\
dean@crabneb.berkeley.edu}

\altaffiltext{1}{Hubble Fellow}

\lefthead{McLaughlin}
\righthead{Fundamental Plane of Globular Clusters}

\begin{abstract}

A physical description of the fundamental plane of Galactic globular clusters
is developed which explains all empirical trends and correlations in a large
number of cluster observables and provides a small but complete set of truly
independent constraints on theories of cluster formation and evolution in 
the Milky Way.

In a very good first approximation, the internal structures and dynamics of
Galactic globulars are described by single-mass, isotropic
\markcite{kin66}King (1966) models and are thus characterized by four
nominally independent physical parameters. Within this theoretical framework,
it is shown that (1) 39 regular (non--core-collapsed) globulars with measured
central velocity dispersions all share a common core mass-to-light ratio,
$\Upsilon_{V,0}=1.45\,M_\odot\,L_\odot^{-1}$, and (2) 109 regular globulars
both with and without direct observations of $\Upsilon_{V,0}$ show a very
strong correlation between global binding energy and total luminosity,
regulated by Galactocentric position:
$E_b=7.2\times10^{39}\ {\rm erg}\ (L/L_\odot)^{2.05}(r_{\rm gc}/
8\,{\rm kpc})^{-0.4}$. The observational scatter about either of these two
constraints can be attributed fully to random measurement errors, making them
the {\it defining equations of a fundamental plane} (FP) to which
real clusters are confined in the larger, four-dimensional parameter space of
general King models. They are shown to underlie a pair of bivariate
correlations first found, and used to argue for the existence of a globular
cluster FP, by \markcite{djo95}Djorgovski (1995). A third, weaker correlation,
between clusters' total luminosities and King-model concentration parameters,
is related to the (non-random) distribution of globulars {\it on} the FP. With
$L$, $\Upsilon_{V,0}$, $E_b$, and the central concentration $c$ thus chosen as
the four physical quantities that define any single globular cluster, the
FP equations for $\Upsilon_{V,0}$ and $E_b(L,r_{\rm gc})$ are used to
{\it derive} expressions for {\it any other} observable in terms of only $L$,
$r_{\rm gc}$, and $c$. Results are obtained for generic King models and
applied specifically to the globular cluster system of the Milky Way.

\end{abstract}


\section{Introduction}

The properties of globular clusters in the Milky Way offer empirical
constraints not only on their own formation and evolution, but also
on the history of our Galaxy as a whole and, to a potentially large extent, on
the formation of generic star clusters---a key element of the star formation
process itself. The many observed correlations between various internal
structural and dynamical parameters of the clusters (core and half-light
radii, surface brightnesses, velocity dispersions, etc.) that have been found
by several authors (e.g., \markcite{bro84}Brosche \& Lentes 1984;
\markcite{kor85}Kormendy 1985; \markcite{vdb91}van den Bergh, Morbey, \&
Pazder 1991; \markcite{vdb94}van den Bergh 1994; \markcite{djo94}Djorgovski \&
Meylan 1994; \markcite{djo95}Djorgovski 1995; \markcite{bel96}Bellazzini et
al.~1996; \markcite{mur92}Murray \& Lin 1992) constitute an important set of
such constraints.

The apparently large number of these correlations (see especially
\markcite{djo94}Djorgovski \& Meylan 1994) can be somewhat misleading,
however, in that it is impossible for all of them to be independent. This
follows from the fact that Galactic globulars are structurally and dynamically
exceedingly simple, being very well described by single-mass, isotropic
\markcite{kin66}King (1966) models. Since these models are defined fully by
the specification of just four independent physical parameters, it is not
unreasonable to expect that the same should be true, at least in a
first-order approximation, of real clusters; but there can then be at most
three truly distinct correlations between their various intrinsic properties
(although these may be supplemented, in general, by dependences on
``external'' factors such as Galactocentric position or cluster metallicity).

It is clearly of interest, then, to systematically reduce the globular cluster
data in the Milky Way to the minimum number of independent physical
relationships required for a complete description of the observed properties
of the ensemble. The establishment of such a compact set of empirical
constraints---which should serve better as a target for theories of cluster
formation and evolution---is the main goal of this paper. To reach it, two
important results will be used as starting points.

First, the two best globular cluster correlations known to date (i.e., the
only two with scatter that can be fully accounted for by observational
uncertainties) are the bivariate dependences presented by
\markcite{djo}Djorgovski (1995) for a subset (roughly one-third) of the
Galactic globulars. One of these correlations involves only cluster core
parameters, while the other includes properties measured at their half-light
radii. \markcite{djo95}Djorgovski found these correlations by a statistical
principal-components analysis, and he offered no physical interpretation of
them; but he pointed out that their very existence suggests that globular
clusters may in practice be confined---much like elliptical galaxies and
other large stellar systems are---to a ``fundamental plane'' in the
higher-dimensional space of physical parameters available to them in
principle.

Second, an almost completely unexamined (but, as it happens, very closely
related) issue has to do with the dependence of binding energy, $E_b$, on mass
(or luminosity) among globular clusters. This was studied some time ago by
\markcite{sai79}Saito (1979), who also investigated the question for dwarf
and giant elliptical galaxies. He found that $E_b\propto M^{1.5}$ or
so---apparently consistent with very simple virial-theorem arguments---for
globulars and giant ellipticals alike. However, his observational sample
included only ten relatively bright globular clusters. The data required for a
calculation of binding energy are now available for many more clusters than
this and they are obviously of higher quality than those
available to \markcite{sai79}Saito. A total energy $E_b$ is the only obvious
physical property that is {\it not} routinely estimated from observations of
Galactic globulars; but it is arguably important---if for no other reason than
to complete a systemization of the data---to clarify the empirical
behavior of this quantity.

In what follows, catalogued data are used to estimate binding energies within
the context of single-mass, isotropic \markcite{kin66}King (1966) models for
109 ``regular'' (non--core-collapsed) globular clusters in the Milky Way. The
data and necessary aspects of the models are summarized in the next Section
and in a pair of Appendices. Section 3 presents the binding energies
themselves, and also derives core mass-to-light ratios, $\Upsilon_{V,0}$, for
39 of these clusters with measured central velocity dispersions. It is shown
that (1) $\Upsilon_{V,0}$ is a constant for all clusters, and (2) $E_b$
correlates tightly with total cluster luminosity (or mass), scaling as
$E_b\propto L^{2.05\pm0.1}$. This is significantly (if not surprisingly)
different from \markcite{sai}Saito's (1979) result, and perhaps also from the
relation obeyed by elliptical galaxies (although this also appears not
to have been reviewed in any detail since \markcite{sai79}Saito's work).

Section 3 also establishes the choice of $L$, $E_b$, $\Upsilon_{V,0}$,
and the King-model concentration parameter $c\equiv\log\,(r_t/r_0)$, as the
four ``independent'' physical properties used to define any globular cluster.
The tight constraints found on $\Upsilon_{V,0}$ and $E_b(L)$ then imply that
real clusters have (at most) only two free parameters, and they do indeed
occupy a fundamental plane in the theoretical space of King models. The
influence of environment on this plane, as reflected in cluster Galactocentric
radii and metallicities, is also examined in \S3; it is shown to be confined to
a dependence on $r_{\rm gc}$ in the {\it normalization} of the $E_b$--$L$
scaling.

In \S4, an orthogonal coordinate system called $\epsilon$-space (analogous to
the so-called $\kappa$-space of elliptical galaxies; \markcite{ben92}Bender,
Burstein, \& Faber 1992) is constructed to obtain a face-on view of the
fundamental plane. This shows that globulars are not distributed uniformly
over it; essentially, the parameter $c$, which
describes the internal density profile of a cluster, is also correlated with
total luminosity or mass (see also \markcite{vdb94}van den Bergh 1994;
\markcite{bel96}Bellazzini et al.~1996). Alternate representations of the
fundamental plane, in the form of a number of bivariate correlations between
cluster core and half-light parameters, are also developed in \S4. It is
shown that the two correlations of \markcite{djo95}Djorgovski (1995) are
equivalent to the two constraints $\Upsilon_{V,0}=constant$ and $E_b\propto
L^{2.05}$ and, thus, that \markcite{djo95}Djorgovski identified the same
fundamental plane that is developed here.

Section 5 and Appendix A then use the defining equations of the fundamental
plane, along with manipulations based on generic properties of King
model, to derive a set of simple relations that ultimately describe the
variation of {\it any} globular cluster observable as a function of $L$, $c$,
and $r_{\rm gc}$ {\it only}. These may themselves be manipulated to explain
any observed correlation between any other combination of cluster variables.
Thus, the fundamental plane indeed reduces a large array of globular cluster
data to a very small set of independent constraints for models of their
formation and evolution. These are set out in \S6, which summarizes the paper.

It should be noted that the choice of ``basic'' cluster properties adopted
here---total luminosity (mass), binding energy, mass-to-light ratio, and
concentration parameter---is not a unique possibility, nor is it necessary in
any rigorous sense. It will be noted, for example, that the scaling of binding
energy with luminosity is essentially equivalent to other, previously known
correlations between half-light radii and luminosities ($R_h\sim L^0$;
\markcite{vdb91}van den Bergh et al.~1991) and between core velocity
dispersions and luminosities ($\sigma_0\sim L^{0.52}$; e.g.,
\markcite{djo94}Djorgovski \& Meylan 1994). However, this does not make a
description in terms of $E_b$ any less correct or valid. In fact, this
framework seems particularly amenable to theoretical studies of cluster
evolution and formation both: Many simulations of the former follow the
evolution of $c$ and $E_b$ over time, and a discussion of the $E_b(L,r_{\rm
gc})$ dependence in the latter context will be taken up in a subsequent paper
(McLaughlin, in preparation).

\section{Data and Model Calculations}

It is well known that \markcite{kin66}King's (1966) theoretical sequence of
isotropic, lowered isothermal spheres of single-mass stars (see also
\markcite{bin87}Binney \& Tremaine 1987) provides an excellent match to the
luminosity profiles of most Milky Way globulars and a good description
of their internal kinematics. This specific model framework is therefore
adopted for analyses throughout this paper. For
self-consistency, and in order to make use of as large an empirical database
as possible, the complications of velocity anisotropy and multi-mass stellar
populations are not considered (although theoretical models incorporating
these features have, of course,
been constructed [\markcite{mic63}Michie 1963; \markcite{gun79}Gunn \& Griffin
1979] and applied to a number of Galactic globulars [see \markcite{pry93}Pryor
\& Meylan 1993]).

\markcite{har96}Harris (1996) maintains an on-line catalogue\footnotemark
\footnotetext{{\tt http://physun.physics.mcmaster.ca/Globular.html}. The
version used in this paper is the revision of 22 June 1999.}
of King-model parameters and other properties of 147 Galactic globular
clusters; it is in part a synthesis and updating of earlier compilations by
\markcite{tra93}Trager, Djorgovski, \& King (1993) and
\markcite{dja93}Djorgovski (1993a) (see also \markcite{tra95}Trager, King,
\& Djorgovski 1995). Among other quantities, the catalogue lists the
clusters' heliocentric distances and their Galactocentric positions,
$r_{\rm gc}$; their foreground reddenings, $E(B-V)=A_V/3.1$, and absolute
magnitudes, $M_V$; metallicities, [Fe/H]; projected
half-light (or effective) radii, $R_h$; central $V$-band surface brightnesses,
$\mu_{V,0}$ (not corrected for extinction in the raw catalogue); scale
radii,\footnotemark\ $r_0$,
\footnotetext{The radii $r_0$ are referred to by \markcite{har96}Harris
(1996), and by \markcite{tra93}Trager et al.~(1993) and
\markcite{dja93}Djorgovski (1993), as core radii, and they are denoted
$r_c$ in these catalogues. However, the tabulated values in fact correspond to
the King-model scale radii $r_0$ defined by equation (\ref{eq:23}), and they
are quantitatively different from the projected half-power radii of
low-concentration clusters.}
and concentration parameters, $c\equiv\log\,\left(r_t/r_0\right)$ (with $r_t$
the tidal radius), derived from fits of
\markcite{kin66}King (1966) models to cluster surface-brightness
profiles; and central luminosity densities, $j_0$ (in $V$-band $L_\odot$
pc$^{-3}$), obtained from the measured $c$, $r_0$, and extinction-corrected
$\mu_{V,0}$ (see, e.g., \markcite{dja93}Djorgovski 1993a, and Appendix A
below).

The \markcite{kin66}King-model structural parameters are not available for six
of the globulars in \markcite{har96}Harris' (1996) tables; another cluster
(Djorgovski 1) has no measured metallicity; and no central surface brightness
is given for one more (the Pyxis cluster). These objects were therefore
excluded a priori from the data set ultimately used here. Of the 139 clusters
remaining, 30 are identified (following \markcite{tra93}Trager et al.~1993)
as having obvious or suspected post--core-collapse (PCC) morphologies, i.e.,
central density cusps. As \markcite{tra93}Trager et al.~discuss,
such clusters are not particularly well described by King
models---a concentration parameter of $c=2.5$ has generally been assigned to
them {\it arbitrarily}---and their tabulated scale radii $r_0$ and core
surface brightnesses $\mu_{V,0}$ are most likely over- and underestimates,
respectively. Thus, although most of the calculations described below have
been performed for the PCC clusters using the data as given, the results
should {\it not} be taken at face value; and while these objects are shown
on many of the plots in this paper, they are never included in any more
detailed analyses (e.g., least-squares fits).

This leaves 109 regular, King-model clusters as the main focus of
attention. Of these, 103 have Galactocentric radii $r_{\rm gc} < 40$ kpc;
4 have $65\la r_{\rm gc}\la 100$ kpc; and two (the faint Palomar 4 and AM1)
lie at $r_{\rm gc}\simeq110$--120 kpc. Estimates of the central line-of-sight
velocity dispersions, $\sigma_{p,0}$, are given for roughly one-third (39/109)
of this restricted sample, as well as for 17 of the 30 PCC clusters, by
\markcite{pry93}Pryor \& Meylan (1993). These additional data are also
employed here.

The recent study of \markcite{ros98}Rosenberg et al.~(1998) has been used
to replace some of \markcite{har96}Harris' (1996) numbers for the
low-luminosity cluster Palomar 12. Specifically, from a King-model fit to
star counts, \markcite{ros98}Rosenberg et al.~derive $c=1.08$ and
$r_0=0\farcm63$ (or 3.5 pc, for a heliocentric distance of 19.1 kpc), as
compared to the $c=1.94$ and $r_0=0\farcm20$ given by \markcite{har96}Harris.
With $\mu_{V,0}=20.6$, according to \markcite{har96}Harris, this then
implies (following the procedure outlined in Appendix A below) a central
luminosity density $j_0\simeq32\,L_\odot\,{\rm pc}^{-3}$, lower than the
previous value by a factor of roughly three. After incorporating these few
changes for this one cluster, the final database is similar to that
adopted by \markcite{djo94}Djorgovski \& Meylan (1994), and the subsets
thereof in \markcite{djo95}Djorgovski (1995) and \markcite{bel98}Bellazzini
(1998), for their recent investigations of the Milky Way globular cluster
system.

Expositions of observational uncertainties can be found in
\markcite{tra93}Trager et al.~(1993) and \markcite{pry93}Pryor \& Meylan
(1993), and the situation may be summarized roughly as follows: scale and
(projected) half-light radii are known to typical precisions of
$\Delta\,\log\,r_0\simeq\pm0.1$ dex and $\Delta\,\log\,R_h\simeq\pm0.1$ dex;
central velocity
dispersions, to within $\Delta\,\log\,\sigma_{p,0}\simeq\pm0.07$ dex; central
surface brightnesses, to $\Delta\,\mu_{V,0}\simeq\pm0.3$ mag arcsec$^{-2}$;
and concentrations, to $\Delta\,c\simeq\pm0.2$ dex.
Uncertainties in any derived quantities follow immediately from these
estimates. For example, the central luminosity density is obtained from
$\mu_{V,0}=const.-2.5\,\log\,\left(j_0r_0\right)$ (eqs.~[\ref{eq:a2}],
[\ref{eq:a3}]), and thus $\Delta\,\log\,j_0\simeq\pm0.2$ dex. Similarly, the
average surface brightness within a half-light radius is defined as
$\langle\mu_V\rangle_h=const.-2.5\,\log\,\left(L/R_h^2\right)$
(see eq.~[\ref{eq:a5}]), so that $\Delta\,\langle\mu_V\rangle_h
\simeq\pm0.5$ mag arcsec$^{-2}$.
There are also uncertainties in cluster distance moduli---$\pm0.15$ mag
may be representative; see the discussion at \markcite{har96}Harris' (1996)
website---and thus in Galactocentric distances
$r_{\rm gc}$ and total luminosities $L$ (note that integrated apparent $V$
magnitudes are generally good to a few hundredths of a magnitude). These
are generally not dominant in the error budgets discussed here, however,
and for the most part they can be ignored.

In order to refer cluster data to isotropic King models, it is necessary
first to relate the observed velocity dispersion $\sigma_{p,0}$ to the model
{\it scale velocity}, $\sigma_0$. This latter quantity is defined through the
distribution function assumed for the cluster stars, $f(E)\propto
\sigma_0^{-3}\,\left[\exp\left(E/\sigma_0^2\right)-1\right]$, and through the
dimensionless cluster potential, $W(r)\equiv\left[\phi(r_t)-\phi(r)\right]/
\sigma_0^2$ (\markcite{kin66}King 1966; \markcite{bin87}Binney \& Tremaine
1987). $W(r)$ and a scaled density profile, $\rho/\rho_0$ vs.~$r/r_0$, follow
from integrating Poisson's equation with this distribution function and a
specified value of $W_0\equiv W(r=0)$. With $r_t$ the radius at which $W=\rho
=0$, there is then a one-to-one relation between $W_0$ and the concentration
parameter $c\equiv\log\,\left(r_t/r_0\right)$. In the limit of high
concentration, the King model is a normal isothermal sphere, $\sigma_0$ is the
true one-dimensional velocity dispersion inside the cluster, and $\sigma_{p,0}=
\sigma_0$. For low concentrations $c\la 1$, however, the outer, non-isothermal
parts of a cluster influence projected quantities even at its center, and
$\sigma_{p,0}<\sigma_0$. This effect is illustrated in the upper left-hand
panel of Fig.~\ref{fig1}. The numerical integrations summarized there have
been used (given the observed concentrations $c$) to convert the central
dispersions $\sigma_{p,0}$ to model $\sigma_0$-values for the
39 regular and 17 PCC clusters in \markcite{pry93}Pryor \& Meylan (1993). In
most cases, this is not a large correction: the \markcite{pry93}Pryor \&
Meylan clusters all have $c>0.75$, and $0.8<\sigma_{p,0}/\sigma_0<1$. It does
have the effect of adding slightly to the observational uncertainty, however:
with $\Delta\,c=\pm0.2$ and $\Delta\,\log\,\sigma_{p,0}=\pm0.07$, the typical
errorbar on $\log\,\sigma_0$ is roughly $\pm0.09$ dex.

\placefigure{fig1}

The binding energies of Milky Way globulars are calculated in \S3 from the
basic definition (cf.~\markcite{kin66}King 1966; \markcite{sai79}Saito 1979)
\begin{equation}
E_b \equiv -{1\over{2}}\,\int_0^{r_t}4\pi r^2\rho\phi\,dr =
{1\over{2}}\,\int_0^{r_t}4\pi r^2\rho
\left[{{GM}\over{r_t}}+\sigma_0^2\,W(r)\right]\ dr\ ,
\label{eq:21}
\end{equation}
where $M$ is the total mass and the sign has been chosen to make $E_b>0$ for 
bound clusters. A dimensionless binding energy follows directly from this
and is a function only of $c$ (through $W_0$) for an isotropic and
single-mass King model:
\begin{equation}
{\cal E}(c)\equiv {{GE_b}\over{\sigma_0^4\,r_0}} =
{{81}\over{2}}\,{{r_0}\over{r_t}}\,
\left[\int_0^{r_t} {{\rho}\over{\rho_0}}\,\left({r\over{r_0}}\right)^2\
{{dr}\over{r_0}}\right]^2 +
{9\over{2}}\,\int_0^{r_t} {{\rho}\over{\rho_0}}\,\left({r\over{r_0}}\right)^2
W(r)\ {{dr}\over{r_0}}\ .
\label{eq:22}
\end{equation}
Here the central mass density $\rho_0$ is related to $r_0$ and $\sigma_0$ by
the {\it model definition}
\begin{equation}
r_0^2\equiv{{9\sigma_0^2}\over{4\pi G \rho_0}}\ .
\label{eq:23}
\end{equation}
The function ${\cal E}$ is shown in the upper right panel of Fig.~\ref{fig1}.
Over the range $0.5\le c\le 2.5$ appropriate to Galactic globulars, an
observational uncertainty of $\Delta\,c=\pm0.2$ translates to an
r.m.s.~variation of $\pm0.2$ dex in ${\cal E}$ as well.

If the mass density profile $\rho(r)/\rho_0$ is replaced by its luminosity
equivalent, $j(r)/j_0$, the light intensity $I$ at a given projected radius
$R$ follows from the standard integral ${\cal I}(R)\equiv I(R)/\left(j_0r_0
\right)=2\,\int_R^{r_t}\left(j/j_0\right)\left(r/r_0\right)\left(r^2/r_0^2-
R^2/r_0^2\right)^{-1/2}\,dr/r_0$. The central surface density ${\cal I}_0$ is
again a function of $c$ only, and it is shown in the bottom left panel of
Fig.~\ref{fig1}. The full profile ${\cal I}(R)$ also uniquely predicts the
radius ${\cal R}(c)\equiv R_h/r_0$ which contains half of a cluster's total
luminosity in projection. This is plotted against the concentration parameter
at the bottom right of Fig.~\ref{fig1}. Both ${\cal I}_0(c)$ and ${\cal R}(c)$
figure prominently in the analysis of \S\S4 and 5 below.

It will also be necessary to refer to another feature of single-mass King-model
clusters, namely, the dimensionless total luminosity ${\cal L}(c)\equiv
L/j_0r_0^3=\int_0^{r_t}\,4\pi\left(j/j_0\right)\left(r/r_0\right)^2\,dr/r_0$.
This quantity is shown for a range of central concentrations in the top panel
of Fig.~\ref{fig2}. The lower half then compares the
calculated luminosity---$L_{\rm mod}={\cal L}(c)j_0r_0^3$---to that directly
observed for the 139 globular clusters of
\markcite{har96}Harris (1996). The agreement for the 109 regular
clusters is excellent (as was also noted by \markcite{dja93}Djorgovski 1993a):
$\langle\log\,L_{\rm mod}-\log\,L\rangle=-0.017$ with an r.m.s.~scatter,
$\Delta=0.25$ dex, that is less than the observational errorbar of
$\pm0.3$.\footnotemark
\footnotetext{This errorbar, drawn in the bottom panel of Fig.~\ref{fig2},
follows from (conservatively) combining in quadrature the uncertainties
of $\pm0.1$ in $\log\,r_0$; $\pm0.12$ in $\log\,\left(j_0r_0\right)$; and
$\pm0.16$ dex in $\log\,{\cal L}(c)$ (which follows from the shape of
${\cal L}$ in the interval $0.5\le c\le 2.5$, given $\Delta\,c=\pm0.2$ dex).}
Figure \ref{fig2} can therefore be viewed as confirming that regular globular
clusters are indeed well described by King models---and as showing the
difficulties with PCC clusters, whose luminosities are often underestimated
on the basis of their crude King-model parametrizations.

\placefigure{fig2}

As has already been suggested, the concentration parameters measured for
Galactic globular clusters span a reasonably wide range of values, from
$c\simeq0.5$ to $c\simeq2.5$ (\markcite{har96}Harris 1996); moreover, there
is a strong {\it systematic} quality to the variation of $c$ from cluster
to cluster (see, e.g., Fig.~\ref{fig6} below). Thus, Figs.~\ref{fig1} and
\ref{fig2} also serve to illustrate the {\it dynamical and structural
non-homology} that characterizes the Milky Way globular cluster system: The
relationships between global and local properties of a cluster (i.e., any of
the ratios plotted in these Figures) obviously depend quantitatively on its
internal density and velocity profiles. But the details of these profiles
are quite sensitive to the King-model parameter $c$ (or, equivalently,
$W_0$), so that even {\it dimensionless} ratios such as ${\cal E}$,
${\cal R}$, and ${\cal L}$ can differ significantly between clusters; one
globular cannot, in general, be turned into another simply by applying a
single scaling factor to its basic physical parameters. 

This is a natural consequence of the fact that globular clusters are well
described by King models (it would not be the case if, for example,
these objects were singular isothermal spheres), and it is clearly key
to a full understanding of cluster structures and a complete explanation of
the various correlations between their properties. Although this point has
always been appreciated in principle (e.g., \markcite{djo95}Djorgovski 1995;
\markcite{bel98}Bellazzini 1998), its consequences are fully explored for the
first time in this paper. Throughout, the functions of $c$ just
discussed are used repeatedly to derive and manipulate physical parameters
of the Milky Way globulars, thus always taking into account---or, in some
sense, correcting for---any non-homology. Whenever an evaluation of
${\cal E}$, ${\cal I}_0$, ${\cal R}$, or ${\cal L}$ is required, it is
obtained from an observed $c$-value by interpolating on the numerical model
curves plotted in Figs.~\ref{fig1} and \ref{fig2}. (For convenience, however,
Appendix B also gives simple approximations to these dimensionless quantities
as polynomial functions of the concentration $c$.)
It is worth noting that analogous corrections for non-homology in elliptical
galaxies may remove some significant part of the observed ``tilt'' of their
fundamental plane (see \markcite{gra97}Graham \& Colless 1997;
\markcite{bus97}Busarello et al.~1997).

\section{Two Basic Properties of Galactic Globular Clusters}

The full spatial structure and internal dynamics of a single-mass, isotropic
\markcite{kin66}King (1966) model cluster are set by specifying four
independent physical parameters. The four most convenient, for purposes of
comparison with observations of real clusters, are the central concentration
$c$ [or the central potential depth $W_0$, to fix the overall shape of a
surface brightness profile $\mu(R)\sim-2.5\,\log\,I(R)$]; the scale radius
$r_0$ [for horizontal normalization in a plot of $\mu(R)$ vs.~projected radius
$R$]; the central luminosity $j_0$ (the vertical normalization, related to
$\mu_0$ as described above and in Appendix A); and the core mass-to-light
ratio $\Upsilon_0$ (for normalization of the model to observed line-of-sight
velocity dispersions). More generally, however, {\it any} set of four
linearly independent combinations of these quantities will serve equally well
as a physical ``basis'' for the sequence of King models. (Such additional
factors as metallicity or galactocentric position are quite separate from
any model characterization of a cluster's internal structure, and they
are best viewed as ``external'' physical parameters.) As will be described
below, the basis ultimately chosen here comprises the concentration, $c$; the
(logarithmic) mass-to-light ratio, $\log\,\Upsilon_0$; total luminosity,
$\log\,L$ (which agrees well with the model quantity $\left[ \log\,j_0 +
3\,\log\,r_0 + \log\,{\cal L}(c)\right]$); and binding energy, $\log\,E_b=
const. + \left[ 2\,\log\,\Upsilon_0 + 2\,\log\,j_0 + 5\,\log\,r_0 +
\log\,{\cal E}(c)\right]$.

At any rate, the implication is that, insofar as Galactic globulars can be
approximated by the simplest King models---that is, to the extent that
internal velocity anisotropies and a range of constituent stellar masses can
be ignored as second-order corrections---they constitute a nominally
four-parameter family of stellar systems. As was mentioned in \S1, however,
there exist many
correlations between various properties of globular clusters in the Milky Way,
implying that these objects in reality inhabit only a small part of the
physical space potentially available to them in principle.
Indeed, as \markcite{djo95}Djorgovski (1995) and \markcite{bel98}Bellazzini
(1998) have suggested, and as will now be shown in a different manner, there
exist at least two (and perhaps three) independent constraints linking
the four basic parameters of globular clusters, so that they are confined
to a fundamental plane (perhaps a line) in the theoretical 4-space of
King models.

\subsection{A Constant Core Mass-to-Light Ratio}

As was discussed in \S2, the central line-of-sight velocity
dispersions $\sigma_{p,0}$ compiled for 39 regular and 17 PCC globular
clusters by \markcite{pry93}Pryor \& Meylan (1993) have been converted to
King-model scale velocities $\sigma_0$ using the concentration parameters
$c$ given by \markcite{har96}Harris (1996). With measurements of all three
of $\sigma_0$, $r_0$, and $j_0$
thus in hand, self-consistent mass-to-light ratios $\Upsilon_{V,0}$
may be calculated for fully one-third of the Milky Way globulars via the
core-fitting procedure of \markcite{ric86}Richstone \& Tremaine (1986): Given
$\sigma_0$ and $r_0$ for any one cluster, equation (\ref{eq:23}) yields a core
mass density $\rho_0$, from which follows $\Upsilon_{V,0}\equiv \rho_0/j_0$.

Figure \ref{fig3} plots $\rho_0$ against $j_0$ for the \markcite{pry93}Pryor
\& Meylan dataset, with the regular clusters represented by circles
(open symbols use the directly observed $\sigma_{p,0}$ in place of the
corrected $\sigma_0$ in eq.~[\ref{eq:23}]) and PCC objects shown as open
squares. The straight line traces the relation $\log\,\rho_0=0.16+\log\,j_0$,
obtained by a least-squares fit to the non--core-collapsed clusters (filled
circles) {\it only}. Clearly, the intercept in this equation is the mean
mass-to-light ratio $\langle\log\,\Upsilon_{V,0}\rangle$. Moreover, the
r.m.s.~scatter about this fit ($\Delta=0.2$ dex) can be attributed entirely
to random observational uncertainties;\footnotemark
\footnotetext{Given $\Delta\,\log\sigma_0\simeq\pm0.09$, $\Delta\,\log\,r_0=
\pm0.1$, and $\Delta\,\log\,(j_0r_0)=0.12$, and with $\Upsilon_{V,0}\propto
\sigma_0^2/j_0r_0^2$ the uncertainty $\Delta\log\,\Upsilon_{V,0}$ is typically
$\pm0.24$ dex even if the constituent errors are uncorrelated.}
there is no evidence for any significant variation in $\Upsilon_{V,0}$ from
one cluster to another.
(The apparent departure of the densest core-collapsed clusters from the mean
line in Fig.~\ref{fig3} is certainly not entirely real,
and it may be completely spurious: forcing a King-model fit to these objects
leads to underestimates of their core densities $j_0$, and thus to the
underestimates of their total luminosities in Fig.~\ref{fig2} and to
overestimates of $\Upsilon_{V,0}$ here. Once again, PCC clusters
are never included in any of the quantitative analyses in this paper.)

\placefigure{fig3}

This result has two implications. First, almost as a practical matter, it
can confidently be assumed, even in the absence of direct velocity-dispersion
measurements, that {\it all other} (non-PCC) Galactic globulars share the
same mass-to-light ratio, viz.
\begin{equation}
\log\,\Upsilon_{V,0}=\log\,\left(\rho_0/j_0\right)=0.16\pm0.03
\ \ \ \ \ [M_\odot\,L_\odot^{-1}]\ ,
\label{eq:24}
\end{equation}
where the uncertainty is just $0.2/\sqrt{39}$. This mean $\Upsilon_{V,0}=(1.45
\pm0.1)\ M_\odot\,L_\odot^{-1}$ is the same as the direct
average of the values given for individual clusters by \markcite{pry93}Pryor
\& Meylan (1993), even though these authors define the mass-to-light ratio
differently (as the ratio of central {\it surface} densities) and fit
different (multi-mass) King models to the raw velocity and surface-brightness
data. Their numbers also show no evidence for any significant dependence of
$\Upsilon_{V,0}$ on $j_0$, or on cluster luminosity. Figure \ref{fig3} is also
roughly consistent with the mean $\Upsilon_{V,0}\simeq1.2\ M_\odot\,
L_\odot^{-1}$ found by \markcite{man91}Mandushev, Spassova, \& Staneva (1991)
(who, however, claim marginal evidence for a weak trend,
$\Upsilon_{V,0}\propto L^{0.08\pm0.06}$ with large scatter).

The second consequence of Fig.~\ref{fig3} and equation (\ref{eq:24}) is an
effective reduction of the dimensionality of the parameter space available
to Galactic globular clusters, from four to three; in practice, {\it the core
mass-to-light ratio is not a free physical variable}. This point is taken up
again in \S4.2 below, where it is shown that the constraint
$\log\,\Upsilon_{V,0}\equiv0.16$ is equivalent to one of
\markcite{djo95}Djorgovski's (1995) bivariate correlations defining the
globular cluster fundamental plane.

\subsection{Binding Energy vs.~Luminosity}

Following the discussion at the beginning of this Section, the King-model
parameters remaining to characterize Galactic globulars are $c$, $r_0$,
and $j_0$---or any three linearly independent combinations of these quantities
plus a constant $\Upsilon_{V,0}$. The remainder of this Section, as well as
\S4, is devoted to establishing total luminosity $L$ (or mass
$\Upsilon_{V,0}L$) and binding energy $E_b$ as replacements for $r_0$ and
$j_0$ as two of the axes
in the parameter space of King models. There are several justifications
for this replacement. First, there exists a previously unrecognized,
but exceptionally strong, correlation between $E_b$ and $L$ in the Milky Way
cluster system. This correlation further reduces the dimensionality of the
space inhabited by real globulars, and it provides an immediate
physical explanation for \markcite{djo95}Djorgovski's (1995) second equation
for the fundamental plane (\S4.2). Second, an emphasis on binding energy in
particular facilitates comparison with the expected properties of gaseous
protoclusters (McLaughlin, in preparation), i.e., this quantity is a natural
focus for theoretical investigations of the origin and evolution of the
globular cluster fundamental plane. And third, a view in terms of binding
energy might profitably be extended to galaxies and clusters of galaxies,
perhaps to yield new insight into their fundamental planes as well.

Given equations (\ref{eq:22}) and (\ref{eq:23}), $E_b$ may be calculated for
a King-model globular cluster using measurements of $c$ and any two of
$\sigma_0$ (which, again, is inferred from the directly observed $\sigma_{p,0}$
and $c$), $r_0$, and $\rho_0\equiv \Upsilon_{V,0}j_0$ in one of the three
equivalent relations,
\begin{equation}
E_b(\sigma_0,r_0) = 4.639\times10^{45}\,{\rm erg}\,
\left({{\sigma_0}\over{{\rm km\,s}^{-1}}}\right)^4
\left({{r_0}\over{{\rm pc}}}\right)\,{\cal E}(c)
\label{eq:31}
\eqnum{\ref{eq:31}a}
\end{equation}
\begin{equation}
E_b(\sigma_0,\rho_0) =  5.995\times10^{46}\,{\rm erg}\,
\left({{\sigma_0}\over{{\rm km\,s}^{-1}}}\right)^5
\left({{\Upsilon_{V,0}j_0}\over{M_\odot\,{\rm pc}^{-3}}}\right)^{-1/2}\,
{\cal E}(c)
\eqnum{\ref{eq:31}b}
\end{equation}
\begin{equation}
E_b(r_0,\rho_0) = 1.663\times10^{41}\,{\rm erg}\,
\left({{r_0}\over{{\rm pc}}}\right)^5
\left({{\Upsilon_{V,0}j_0}\over{M_\odot\,{\rm pc}^{-3}}}\right)^2\,
{\cal E}(c)\ .
\eqnum{\ref{eq:31}c}
\addtocounter{equation}{+1}
\end{equation}
With $\Upsilon_{V,0}$ a constant, equation (\ref{eq:31}c) is of
particular interest in that it allows for an estimate of binding energy based
entirely on photometry, i.e., velocity-dispersion measurements are not
explicitly required. The function ${\cal E}(c)$ is shown in Fig.~\ref{fig1}
above, and it can be approximated by equation (\ref{eq:b2})
below. As was mentioned in \S2, an r.m.s.~variation of $\pm0.2$ dex in $c$
translates to $\Delta\,\log\,{\cal E}(c)=\pm0.2$; thus, the typical
measurement errors in $c$, $\log\,r_0$, and $\log\,j_0$ result in a
total observational uncertainty of $\pm0.45$--0.75 dex
(depending on whether or not the individual errors are correlated)
in the logarithm of $E_b$.

Figure \ref{fig4} plots the concentrations and the binding energies, calculated
using each of equations (\ref{eq:31}) (supplemented by eq.~[\ref{eq:24}]),
against luminosity and mass for the
39 King-model globulars (circles) and 17 PCC clusters (squares) with
$\sigma_0$ known from \markcite{pry93}Pryor \& Meylan (1993). Note that
the total masses used in this Figure are taken directly from
\markcite{pry93}Pryor \& Meylan, who derived them from fits of anisotropic,
multi-mass King models to the cluster data. These masses are therefore highly
model-dependent, and in fact it is not clear what relevance they have in the
analysis of this paper, which treats globulars as isotropic, single-mass
clusters (in particular, the equality $M=\Upsilon_{V,0} L$ is implicit here,
but it does not necessarily hold for the numbers of \markcite{pry93}Pryor \&
Meylan 1993). Nevertheless, the right column of Fig.~\ref{fig4} demonstrates
that correlations between cluster concentration and luminosity/mass, and
between $E_b$ and $L$ or $M$, exist independently of specific model
assumptions.

\placefigure{fig4}

The top row of Fig.~\ref{fig4} shows a rough trend of increasing central
concentration with increasing cluster luminosity. Such a correlation has also
been noted by, e.g., \markcite{vdb94}van den Bergh (1994),
\markcite{djo94}Djorgovski \& Meylan (1994), and \markcite{bel96}Bellazzini
et al.~(1996), and it was recognized in essence long ago, by
\markcite{sha27}Shapley \& Sawyer (1927). This dependence is visibly weaker
than that of $E_b$ on $L$, an impression that is quantified by the difference
in the Spearman rank correlation coefficients $s$ for the various
relationships. (These nonparametric measures of correlation strength,
specified in each panel of Fig.~\ref{fig4}, are computed for
the regular, King-model clusters [filled circles] only.) A linear
dependence can be fit to these data---$c\simeq const. +
0.4\,\log\,(L/L_\odot)$---but the r.m.s.~scatter about it is $\Delta=0.3$,
significantly larger than the observational errorbar $\Delta\,c=\pm$0.2 dex.
Clearly, $c$ is not completeley independent of $\log\,L$, but the relation
between the two does not appear to be one-to-one (or, if such a strong link
does exist, it is nonlinear).

On the other hand, $\log\,E_b$ and $\log\,L$ are very strongly correlated.
Indeed, if the correlation coefficients $s$ in the lower six panels of
Fig.~\ref{fig4} are compared to those given by \markcite{djo94}Djorgovski \&
Meylan (1994), \markcite{djo95}Djorgovski (1995), and
\markcite{bel98}Bellazzini (1998) for a large number of empirical relations
between other observables of Milky Way clusters, it can quickly be seen that
{\it the correlation between binding energy and luminosity (or mass) is as
strong as or stronger than any other correlation between properties of
Galactic globulars}. In addition to this, it is clear that a simple power-law
dependence of $E_b$ on $L$ or $M$ completely describes the data. Least-squares
fits (for the regular clusters only) of $\log\,E_b$ against either $\log\,L$
or the $\log\,M$ from \markcite{pry93}Pryor \& Meylan (1993) are drawn in
Fig.~\ref{fig4}. The r.m.s.~scatter of $\log\,E_b$ about the straight lines
in these six fits always lies within the observational errorbar ($\pm0.5$ dex)
shown in the second row of the Figure.

The three fits of $\log\,E_b$ against $\log\,L$ are statistically
indistinguishable. This is only to be expected; it reconfirms the main
underlying assumption here---that globular clusters are well described by
isotropic, single-mass King models---and the finding of \S3.1, that the
core mass-to-light ratio is an essentially constant $\log\,\Upsilon_{V,0}=
0.16$. Figure \ref{fig5} specifically compares the binding energies computed
using measurements of $r_0$ and $\rho_0$ (eq.~[\ref{eq:31}c]) to those
calculated from $\sigma_0$ and $r_0$ (eq.~[\ref{eq:31}a]). Since the ratio
of these is just $\left(4\pi G \Upsilon_{V,0} j_0 r_0^2/9\sigma_0^2\right)^2$,
it should be---and is---equal to unity in the mean, with an r.m.s.~scatter
($\Delta=0.4$ dex) just twice that found in Fig.~\ref{fig3} for the basic
identity $\Upsilon_{V,0}=\rho_0/j_0$.

\placefigure{fig5}

If $\log\,E_b$ is then specified, for each of the 39 regular clusters in
\markcite{pry93}Pryor \& Meylan's (1993) sample, as the direct mean of the
three values derived from equations (\ref{eq:31}), a least-squares regression
against cluster luminosity results in
\begin{equation}
\log\,(E_b/{\rm erg}) =
(39.82\pm0.75) + (2.05\pm0.15)\,\log\,\left(L/L_\odot\right)\ ,
\label{eq:32}
\end{equation}
with a Spearman rank correlation coefficient $s=0.92$ and an r.m.s.~scatter
in $\log\,E_b$ of $\pm0.43$ dex. This relation is significantly different (the
uncertainties in eq.~[\ref{eq:32}] are $\pm1\sigma$) from the $E_b\propto
L^{1.5}$ proposed by \markcite{sai79}Saito (1979).

It is perhaps worth noting that the three fits of $\log\,E_b$ vs.~$\log\,M$
in Fig.~\ref{fig4} are also statistically identical to each other, but that
they differ formally
from the $E_b(L)$ fit: $E_b\propto M^{1.8\pm0.1}$ is indicated, as opposed to
$E_b\propto L^{2.05\pm0.15}$. At first glance, the difference in these
exponents might seem to imply a varying {\it global}
mass-to-light ratio, $(M/L)_V\propto L^{0.14\pm0.11}$ or so. Such a result
is obviously of very low statistical significance, however, and of dubious
origin besides. It could well be completely an artifact of {\it a comparison
between different King models}, namely, the single-mass and isotropic version
used here to compute $E_b$, vs.~the multi-mass and anisotropic variant used by
\markcite{pry93}Pryor \& Meylan (1993) to obtain $M$. In addition, a more
direct inspection of the \markcite{pry93}Pryor \& Meylan masses shows no clear
evidence for any systematic variation of global $(M/L)_V$ with cluster
luminosity (consistent with the lack of any such trend in $\Upsilon_{V,0}$
from Fig.~\ref{fig3}).
As a whole, then, the current data and models still show no conclusive signs
of any variation in global mass-to-light ratios among Milky Way globulars.
The rest of this paper will therefore discuss cluster binding energies
explicitly as a function of luminosity only, and it will be taken as given
that $E_b$ scales with total mass in the same way.

Figure \ref{fig6} shows the concentration $c$ and the binding energy
$E_b$ (calculated from eq.~[\ref{eq:31}c] with $\log\,\Upsilon_{V,0}
\equiv0.16$) as functions of luminosity for all of the 139 globulars with
data taken from \markcite{har96}Harris (1996). As before, the 109 regular
clusters are plotted as filled circles, and the 30 PCC objects as open squares.
Evidently, the correlations seen in Fig.~\ref{fig4} extend to the whole of the
Milky Way cluster system. In particular, the power-law nature and high
statistical significance of the $E_b$ vs.~$L$ scaling are unchanged from the
smaller sample of \markcite{pry93}Pryor \& Meylan. It is now found that
\begin{equation}
\log\,\left(E_b/{\rm erg}\right) =
(39.89\pm0.38) + (2.05\pm0.08)\,\log\,\left(L/L_\odot\right)\ ,
\label{eq:33}
\end{equation}
where the uncertainties are again $\pm1\sigma$ estimates. The r.m.s.~scatter
about this fit (to the regular clusters only) is $\Delta=0.53$ dex in
$\log\,E_b$, consistent with the combination of purely random measurement
errors in $c$, $\log\,r_0$, and $\log\,j_0$. With a Spearman rank statistic of
$s=0.93$, the correlation between $E_b$ and $L$ is better by far than those
between any other cluster properties in this expanded (and now essentially
complete) dataset.

\placefigure{fig6}

The rough correspondence between $c$ and $\log\,L$ seen in Fig.~\ref{fig4}
also persists for regular, King-model clusters---$c\approx const.+0.4\,\log\,L$
is again indicated---but this remains less significant (smaller $s$) and more
scattered (the r.m.s.~deviation deviation from the straight line in the top
of Fig.~\ref{fig6} is $\Delta\simeq0.35$ dex, again exceeding the typical
observational uncertainty in $c$) than the $E_b$--$L$ relation.

The cluster luminosities plotted in Figs.~\ref{fig4} and \ref{fig6} are
the integrated absolute magnitudes tabulated by \markcite{har96}Harris (1996),
and they are {\it independent} of any King-model fits; they have
{\it not} been computed from the formula $L={\cal L}(c) j_0 r_0^3$.
Nevertheless, Fig.~\ref{fig2} above demonstrated that the model luminosities
do correspond very closely to the directly observed values. Thus, a point
of potential concern here is that a plot of $E_b$ against $L$ is effectively
one of $(j_0^2 r_0^5)$ against $(j_0 r_0^3)$, raising the possibility
that any correlation might be either trivial or spurious. Figure
\ref{fig7} illustrates that the situation is rather more subtle than this in
reality, and that the $E_b(L)$ dependence found here is indeed nontrivial.
The top two panels of this Figure show $\log\,r_0$ and $\log\,(j_0r_0)$ (which
again is closely related to the central surface brightness $\mu_{V,0}$)
vs.~total luminosity for all of \markcite{har96}Harris' (1996) clusters.
It is immediately obvious that $r_0$ and $j_0r_0$ each correlate with
$L$---the dashed lines drawn describe $r_0\propto L^{-0.3}$ and $j_0r_0\propto
L^{1.25}$---but the r.m.s.~scatter $\Delta$ in these plots is much larger
than the observational errorbars (drawn in the upper left corners) in either
case. The point of physical interest is that the {\it scatter} in $\log\,r_0$
is (anti)correlated with that in $\log\,(j_0r_0)$. One example of this has
already been shown in Fig.~\ref{fig2}, with its plot of ${\cal L}(c)j_0r_0^3$
against $L$; another is now seen in the bottom panel of Fig.~\ref{fig7}, which
shows explicitly that the specific combination $(j_0^2r_0^5)\propto E_b/
{\cal E}(c)$ is much more tightly correlated with cluster luminosity [the
errorbar in the upper left, $\pm0.4$ dex, represents the quadrature sum of
the uncertainties in $\log\,(j_0^2r_0^2)$ and $\log\,r_0^3$]. This already
suggests that the relation between $E_b$ and $L$ is the more fundamental one,
and that the (degraded) $r_0$--$L$ and $j_0r_0$--$L$ correlations are properly
viewed as deriving from it.

\placefigure{fig7}

The details of this derivation are discussed in \S5 and Appendix A, along with
other examples of similarly weak or scattered monovariate structural
correlations in the Galactic globular cluster system. A key consideration is
the effect of the clusters' structural non-homology (see \S2). Thus---for
example---the bottom of Fig.~\ref{fig7} is still not equivalent to the lower
half of Fig.~\ref{fig6}, because $j_0^2r_0^5$ differs from $E_b$ by a
{\it non-constant} factor of ${\cal E}(c)$ (eq.~[\ref{eq:31}c]). Indeed, the
straight line in Fig.~\ref{fig7} represents the scaling $j_0^2r_0^5\propto
L^{1.7}$; it is only because $\log\,{\cal E}\sim 0.85\,c$ ({\it very}
roughly, from Fig.~\ref{fig1} above) and $c\sim 0.4\,\log\,L$ that
$E_b\propto L^{2.05}$ obtains in the end. Note also that the r.m.s.~scatter
$\Delta$ in $\log\,E_b$ vs.~$\log\,L$ is actually slightly {\it smaller} than
that in $\log\,(j_0^2r_0^5)$ vs.~$\log\,L$, despite the large scatter in the
$c$--$\log\,L$ correlation; this is further evidence that the $E_b(L)$
relation is nontrivial and physically significant.

There are other ways to see this dependence, however, and it might have been
anticipated on the basis of some previous work. In particular, Fig.~\ref{fig8}
shows the projected half-light radii of the clusters in
\markcite{har96}Harris (1996) as a function of their total luminosities and
(3D) Galactocentric radii $r_{\rm gc}$. This comparison, which is not new
(cf.~\markcite{vdb91}van den Bergh et al.~1991; \markcite{vdb95}van den Bergh
1995), shows that $R_h$ depends only weakly on $L$ but tends to increase
systematically with $r_{\rm gc}$ [the dashed line in the upper right panel of
this Figure has the equation $\log\,(R_h/{\rm pc})=0.23+0.4\,\log\,
(r_{\rm gc}/{\rm kpc})$]. Moreover, it turns out that the latter variation
is responsible for what little dependence $R_h$ does appear to have on $L$:
as Fig.~\ref{fig8} also shows (see also \markcite{vdb91}van den Bergh et
al.~1991), the normalized quantity $R_h^*\equiv R_h(r_{\rm gc}/8\,{\rm kpc})
^{-0.4}$ is essentially independent of cluster luminosity.\footnotemark
\footnotetext{The relation claimed by \markcite{ost97}Ostriker \& Gnedin
(1997), $R_h\propto L^{-0.63}$ for clusters with $5\le r_{\rm gc}\le 40$ kpc,
is not consistent with the earlier analysis of \markcite{vdb91}van den Bergh
et al.~(1991), and it is {\it not confirmed} here. Again, the plot of
$R_h^*$ against $L$ shows that {\it at a fixed Galactocentric radius}, cluster
sizes are quite insensitive to their total masses. This point is important
for discussions of cluster destruction timescales, especially in the low-mass
regime.}
In fact, with a rank correlation coefficient of $s\simeq-0.1$, $R_h^*$
and $L$ are about as close to being perfectly {\it un}correlated ($s=0$) as
$E_b$ and $L$ are to being perfectly correlated ($s=1$).

\placefigure{fig8}

The point of this is that $E_b\propto j_0^2r_0^5{\cal E}(c)$, $L=j_0
r_0^3{\cal L}(c)$, and $R_h=r_0{\cal R}(c)$ together give $E_b\propto ({\cal E}
{\cal R}/{\cal L}^2)\,L^2/R_h$. But it happens that the non-homology factor
$({\cal E} {\cal R}/{\cal L}^2)$ varies very little over the range of
concentrations $0.5\le c\le 2.5$ appropriate for regular Galactic
globulars (see Figs.~\ref{fig1} and \ref{fig2} above, or Fig.~\ref{fig14}
below). The simple scaling $E_b\propto L^2/R_h$, which is self-evident if the
question of non-homology is ignored altogether, is then quite
accurate; and the weak dependence of $R_h$ on $L$---even if the influence of
Galactocentric radius is also ignored---leads inevitably to the rough
expectation, $E_b\propto L^2$. That is, Figs.~\ref{fig8} and \ref{fig6} are
equivalent characterizations of the Milky Way globular clusters; $E_b\sim
L^2$ if and only if $R_h\sim constant$ for King-model clusters, and
conceptually one constraint is no more correct or fundamental
than the other. As a practical matter, however, note that the r.m.s.~scatter
about the mean $\log\,R_h^*$ in Fig.~\ref{fig8} is about twice the
measurement uncertainty of $\pm0.1$ dex, whereas the scatter about the
line $E_b\propto L^{2.05}$ in Fig.~\ref{fig6} is basically comparable to the
observational errorbar. The source of the scatter in Fig.~\ref{fig8} is
discussed in \S5 (Fig.~\ref{fig17}), which also considers another known
correlation ($\sigma_0\propto L^{0.52}$; see Fig.~\ref{fig18} and
\markcite{djo94}Djorgovski \& Meylan 1994) that is equivalent to the binding
energy--luminosity scaling found here.

The rest of this paper keeps its focus on binding energy, rather than
half-light radius or velocity dispersion explicitly, as one of the four
defining parameters of globular clusters; and the $E_b(L)$ relation, rather
than the constancy of $R_h$, will be used (along with the constancy of
$\Upsilon_{V,0}$) to define the fundamental plane. Figure \ref{fig8} makes it
clear that this course is not strictly necessary, but it has been adopted for
the convenience and insight it offers (as has already been suggested, a
discussion in terms of $E_b$ can greatly facilitate theoretical studies of
cluster formation and evolution).

Before the globular cluster fundamental plane can be constructed explicitly,
however, Fig.~\ref{fig8} raises another issue to be addressed: Given that
$E_b\propto L^2/R_h$ and $R_h\propto r_{\rm gc}^{0.4}$, it is clear that 
``secondary'' quantities such as Galactocentric position or cluster
metallicity---which are extraneous to the formal King-model fitting
process---are not completely disconnected from the basic cluster attributes
$\Upsilon_{V,0}$, $c$, $L$, and $E_b$. What, then, is the total effect
of such environmental factors on these fundamental parameters and the
relationships between them?

\subsection{The Influences of Galactocentric Position and Metallicity}

Figure \ref{fig9} shows the relative insensitivity of $\Upsilon_{V,0}$,
$L$, and $c$ to $r_{\rm gc}$ and [Fe/H] for globular clusters in the Milky Way.
The top row incorporates only the sample of \markcite{pry93}Pryor \& Meylan
(1993), with $\Upsilon_{V,0}\equiv 9\,\sigma_0^2/(4\pi G j_0r_0^2)$ as in
\S3.1. Clearly, a cluster's core mass-to-light ratio is quite detached from
its metallicity. There is perhaps a hint of a slight decrease in
$\Upsilon_{V,0}$ towards larger Galactocentric radii, but this is not
statistically significant. Thus, although it would be interesting to examine
the question more closely by obtaining velocity data for the remaining
two-thirds of the Galactic globulars, it will be assumed here that
$\Upsilon_{V,0}$ is completely independent of $r_{\rm gc}$. (Note that the
values for the PCC clusters [open squares] are overestimates, due to the
underestimation of their central densities $j_0$ [cf.~Fig.~\ref{fig3}]. Also,
the point at $r_{\rm gc}\simeq90$ kpc corresponds to NGC 2419, which appears
as an outlier in a plot of $\sigma_0$ against $L$ [see Fig.~\ref{fig18} below],
and for which $\Upsilon_{V,0}$ may have been underestimated as a result.)

\placefigure{fig9}

The middle panels of Fig.~\ref{fig9} illustrate the well known facts
(e.g., \markcite{djo94}Djorgovski \& Meylan 1994) that globular cluster
luminosities are completely uncorrelated with [Fe/H]---there is no
mass-metallicity relation---and very weakly anticorrelated with Galactocentric
position. However, it can be seen that the latter result does not reflect any
clear, systematic dependence of $L$ on $r_{\rm gc}$, but appears instead to be
driven by a handful of ``excess'' faint objects at large radii ($r_{\rm gc}\ga
10$ kpc) which are not equally represented in the inner parts of the Galaxy.
This slight imbalance may be due to dynamical evolution from an initial
distribution of cluster luminosities which was much more insensitive to
$r_{\rm gc}$ (see, e.g., \markcite{mcl96}McLaughlin \& Pudritz 1996), with a
larger fraction of the low-mass clusters inside the Solar circle having been
disrupted by disk shocking and evaporation over a Hubble time. Or,
alternatively, since some of the outer-halo globulars are somewhat younger
than average, it could be that clusters formed with lower average masses at
lower redshifts (S.~van den Bergh, private communication). But whatever its
cause, this effect does {\it not} change the basic scalings inferred in
Fig.~\ref{fig6} for concentration and binding energy as functions of
luminosity.

Although a plot of $c$ alone against $r_{\rm gc}$ shows a rough
anticorrelation (\markcite{djo94}Djorgovski \& Meylan 1994), this is simply
a result of the tendencies for $c$ to decrease towards lower $L$ ($c\approx
const.+0.4\,\log\,L$, from Figs.~\ref{fig4} and \ref{fig6}) and for the
lowest-luminosity clusters to lie preferentially at large Galactocentric
radii. The bottom panels of Fig.~\ref{fig9} correct for this, showing that
{\it at a given cluster luminosity}, $c$ is essentially independent of
$r_{\rm gc}$. Equivalently, the mean relation between concentration and
luminosity does not change with Galactocentric position (see also
Figs.~\ref{fig12} and \ref{fig13} below). It is obvious, too, that $c$
is---like $\Upsilon_{V,0}$ and $L$---impervious to [Fe/H].

Figure \ref{fig10} shows that similar remarks apply to the {\it scaling},
though not to the normalization, of the $E_b(L)$ relation. The top plot does
show a tendency for the lowest-energy King-model clusters to be found at the
largest radii; but this is due in part to the preponderance of faint clusters
in the outer parts of the Galaxy. If the overall dependence $E_b\propto
L^{2.05}$ is removed, the influence of Galactocentric position alone can
be isolated: the straight line in the bottom of Fig.~\ref{fig10} has the
equation $\log\,(E_b/{\rm erg})-2.05\,\log(L/L_\odot)=40.22-0.4\,
\log(r_{\rm gc}/{\rm kpc})$. The slope of this line is uncertain by perhaps
$\pm0.1$; it is, of course, just that expected from the proportionality
$E_b\propto L^2/R_h$ and the systematic increase of cluster radii with
$r_{\rm gc}$ (Fig.~\ref{fig8}). Put another way---and as can be confirmed
either by plotting $E_b\,r_{\rm gc}^{0.4}$ against $L$ or by fitting
$E_b$ vs.~$L$ directly in a series of narrow radial bins---the {\it slope} of
the $E_b(L)$ relation in the Milky Way does not vary with Galactocentric
position, but the normalization decreases systematically towards larger
$r_{\rm gc}$ (see also Fig.~\ref{fig12}). All the data are described
completely by
\begin{equation}
\log\,\left(E_b/{\rm erg}\right) =
\left[(39.86\pm0.40) - 0.4\,\log\,\left(r_{\rm gc}/8\,{\rm kpc}\right)\right]
+ (2.05\pm0.08)\,\log\,\left(L/L_\odot\right)\ ,
\label{eq:34}
\end{equation}
in which the errobars are $\pm1\sigma$ and all clusters are normalized to a
common $r_{\rm gc}=8$ kpc for convenience. Note that the r.m.s.~scatter
about this relation, $\Delta=0.49$ dex, is $\sim$10\% lower than that in
equation (\ref{eq:33}) and Fig.~\ref{fig6}, which ignored the $r_{\rm gc}$
dependence. It can be attributed entirely to the effects of random measurement
errors in $r_0$, $j_0$, and $c$, even under the most conservative assumption
that these uncertainties are uncorrelated. Also, the intercepts in equations
(\ref{eq:32}) and (\ref{eq:33}) are consistent with equation (\ref{eq:34})
given the median Galactocentric radii of the \markcite{pry93}Pryor \& Meylan
(1993) sample of regular clusters ($\overline{r_{\rm gc}}=8.7$ kpc) and
the full \markcite{har96}Harris (1996) dataset
($\overline{r_{\rm gc}}=7.0$ kpc).

\placefigure{fig10}

Finally, Fig.~\ref{fig11} shows that metallicity plays no role in setting a
cluster's total energy, a result which is completely in keeping with the
irrelevance of [Fe/H] to any other aspect of the internal structure and
dynamics of globular clusters (see \markcite{djo94}Djorgovski \& Meylan 1994
for further examples).

\placefigure{fig11}

\subsection{Summary}

To a very good first approximation, Galactic globular clusters can be treated
as realizations of single-mass, isotropic \markcite{kin66}King (1966) models.
A complete basis for the physical description of the ensemble is therefore
provided by the four quantities $\log\,\Upsilon_{V,0}$, $c$, $\log\,L$, and
$\log\,E_b$. While this particular choice of variables is not necessary---any
set of four linearly independent King-model parameters is permissible, in
principle---it is certainly sufficient. Moreover, it has the advantage of
immediately revealing two strong empirical constraints on the properties of
Galactic globulars (each of which will, of course, be recovered in any
alternative parametrization of King-model space): (1) the core mass-to-light
ratio is a constant, $\log\,(\Upsilon_{V,0}/M_\odot\,L_\odot^{-1})=0.16\pm
0.03$, and (2) clusters' binding energies are set by their total luminosities
and Galactocentric positions, through $\log\,E_b=39.86-0.4\,(\log\,r_{\rm gc}/
8\,{\rm kpc})+2.05\,\log\,L$. As far as the current data can discriminate, the
Milky Way cluster system appears almost {\it perfectly} to obey these two
relations---deviations from them can be attributed entirely to random
measurement errors---and it therefore populates a two-dimensional subspace in
the four-dimensional volume of King models. The effects of metallicity and
environment on individual cluster properties are summarized
{\it completely} (again, within the limits of current data) by the
$r_{\rm gc}$-dependence in the normalization of the $E_b(L)$ relation. 

Equations (\ref{eq:24}) and (\ref{eq:34}) thus {\it define} a fundamental
plane for Galactic globulars. Various representations of this are developed in
the next Section. Essentially, central concentration and total luminosity or
mass are then left to determine the distribution of clusters {\it on} the
plane; but even this is not random, as there also exists a rough correlation
between $c$ and $\log\,L$ (Fig.~\ref{fig6}). This dependence is rather weak
by comparison with the constraints on $\log\,\Upsilon_{V,0}$ and $\log\,E_b$,
and a one-to-one $c(\log\,L)$ relation has {\it not} yet been found in the
data; but, as will also be discussed in \S4, the situation does suggest the
intriguing possibility that Galactic globulars may actually fall along, or
at least have evolved from, something closer to a ``fundamental straight line''
(cf.~\markcite{bel98}Bellazzini 1998, who suggested something similar but
with a much different physical interpretation).

\section{The Fundamental Plane}

For much of this Section, it will be taken as given that $\Upsilon_{V,0}$
is a constant for Galactic globulars, as in equation (\ref{eq:24}). This
immediately removes the core mass-to-light ratio as a physical variable, and 
the parameter space remaining available to the clusters is only
three-dimensional.

The preceding discussion and Fig.~\ref{fig6} suggest a physically
transparent view of the globular cluster fundamental plane (FP) as a
thin, tilted slice of $(\log\,L,\log\,E_b,c)$ space---the volume left by the
constraint $\log\,\Upsilon_{V,0}=constant$---which is seen edge-on along the
$c$-axis (in the $\log\,L$--$\log\,E_b$ plane) and closer to face-on along the
$E_b$ axis (in the $c$--$\log\,L$ plane). As was discussed in \S3.3, however,
the normalization of the $E_b(L)$ relation decreases systematically with
increasing Galactocentric radius. Thus, a plot of $\log\,E_b$ vs.~$\log\,L$
for the entire Milky Way cluster system, with its wide range of $r_{\rm gc}$,
actually provides an edge-on view of a {\it collection} of separate and
distinct FPs which intersect the $\log\,E_b$ axis at different points. Figure
\ref{fig6} can therefore be improved by introducing the normalized quantity
(cf.~eq.~[\ref{eq:34}])
\begin{equation}
\log\,E_b^* \equiv \log\,E_b + 0.4\,\log\left(r_{\rm gc}/8\,{\rm kpc}\right)\ ,
\label{eq:40}
\end{equation}
which simply removes all environmental influences on the FP.

Figure \ref{fig12} plots $c$ and $\log\,E_b^*$ against $\log\,L$ for the 109
regular and 30 PCC clusters catalogued by \markcite{har96}Harris (1996),
divided according to whether they lie within the Solar circle ($R_0=8$ kpc)
or outside it. Two points are apparent from this. First, it is confirmed that
the $r_{\rm gc}$ dependence in $E_b$ is adequately described by a scaling
close to $E_b\propto r_{\rm gc}^{-0.4}$, and that the mean increase of $c$
with $L$ is, although rough, essentially independent of Galactocentric position
(cf.~\S3.3). The solid straight lines in both bottom panels follow the
relation given by equation (\ref{eq:34}) above; the dashed lines show the
$3\sigma$ limits on the fitted slope and intercept. Clearly, clusters at
$r_{\rm gc}<8$ kpc and $r_{\rm gc}>8$ kpc obey this equation equally well.
Similarly, the dashed lines in the top right panel of Fig.~\ref{fig12} are
linear least-squares fits of $c$ against $\log\,L$ (shallower slope) and of
$\log\,L$ against $c$ (steeper slope) for the regular clusters at $r_{\rm gc}>
8$ kpc (excluding Palomar 1, the most obvious outlier). These same lines are
then re-drawn in the top left panel of the Figure, where they are seen to be
equally acceptable as crude descriptions of $c$ vs.~$\log\,L$ for the
King-model clusters with $r_{\rm gc}<8$ kpc.

\placefigure{fig12}

Second, it can be seen from Fig.~\ref{fig12} that {\it both $c$ and $\log\,
E_b^*$ are more strongly correlated with $\log\,L$ for clusters beyond the
Solar circle} than for those within it: the Spearman rank correlation
coefficients are higher, and in the case of $c$ vs.~$\log\,L$, the
r.m.s.~scatter about the linear fits is smaller. This point has already been
discussed in relation to the central concentrations
by \markcite{bel96}Bellazzini et al.~(1996) and \markcite{ves97}Vesperini
(1997), who note that globulars at $r_{\rm gc}<8$ kpc have been subjected
to much stronger dynamical evolution than have those at larger radii
(evaporation is faster and disk shocks are more severe in denser regions of
the Galaxy). Thus, the very existence of a $c$--$\log\,L$ correlation among
the more distant clusters suggests that it has a largely primordial origin.
These authors' quantitative numerical simulations confirm in more detail that
``evolutionary processes are unlikely to play a dominant role in establishing
the $c$--$\log\,M$ correlation'' (\markcite{ves97}Vesperini 1997). The same
calculations show explicitly that the main effect of the evolution within
$r_{\rm gc}=8$ kpc is to increase the {\it scatter} in the final version of
an initial $c(\log\,L)$ dependence. This is consistent with the appearance of
the data in the top half of Fig.~\ref{fig12}. (It must be noted again, however,
that the r.m.s.~scatter of measured $c$ values about the linear regressions at
$r_{\rm gc}>8$ kpc is still substantially larger than the typical errorbar of
$\pm0.2$ dex. It remains unclear whether there might exist a {\it nonlinear}
$c$--$\log\,L$ correlation that is significantly tighter, or whether there is
simply a real scatter in cluster concentrations at any given luminosity.)

Similarly, the very strong correlation of $\log\,E_b^*$ with $\log\,L$ for the
globulars at $r_{\rm gc}>8$ kpc implies that this dependence, too, may have
been set largely at the time of cluster formation. If this is correct, then it
seems that the main effect of the stronger evolution within the Solar circle
may have been a greater fractional depopulation of the low-mass tail of the
cluster distribution, rather than any significant change in a power-law
scaling between $E_b$ and $L$. (It is emphasized again that the r.m.s.~scatter
in the data of both lower panels of Fig.~\ref{fig12}
is essentially the same as, or even smaller than, the typical observational
errorbar on $\log\,E_b$.) Indeed, an initial scaling close to $E_b\propto
L^2$ {\it can} be expected to be roughly preserved during the course of
dynamical evolution of an entire cluster system. Direct calculations (see the
discussion and references in \markcite{mur92}Murray \& Lin 1992;
\markcite{har94}Harris \& Pudritz 1994) show that the half-mass radii of
individual clusters are fairly well preserved during dynamical evolution,
even over the course of a Hubble time, i.e., $dR_h/dt\approx0$; but this and
$E_b\propto L^2/R_h$ together imply $d\,\log\,E_b/dt\approx2\,d\,\log\,L/dt$.
Thus, as a cluster loses mass to evaporation and tidal shocking it should move
down and to the left in the $(\log\,L,\log\,E_b^*)$ plane, along a line roughly
parallel to the initial correlation (if this was indeed roughly $\sim L^2$,
and assuming that any changes in $0.4\,\log\,r_{\rm gc}$ can be neglected).

To put this another way, it is plausible that {\it clusters remain more or
less in the fundamental plane as they evolve} and, thus, that globular
clusters may have appeared on the FP as a result of the formation process.
In addition, the relation between $c$ and $\log\,L$ implies that the
distribution of globulars over the FP is non-random and also reflects something
of initial conditions. To see this properly, however, requires a further
manipulation of the basic cluster variables.

\subsection{$\mathbf{\epsilon}$-Space}

By construction, the $\log\,L$, $\log\,E_b^*$, and $c$ axes are mutually
orthogonal in the three-dimensional cluster parameter space selected by
$\log\,\Upsilon_{V,0}=constant$. And in the Milky Way, the view of the FP in
a plot of $\log\,E_b^*$ vs.~$\log\,L$ is nearly perfectly edge-on  (again, this
follows from the fact that random measurement errors can account fully for the
scatter of observed clusters about the relation $\log\,E_b^*=39.86+2.05
\log\,L$). The FP is therefore canted at quite a steep angle [$\arctan\,
(2.05)=64^\circ$] to the $(\log\,L, c)$ plane, and the top panels of
Figs.~\ref{fig7} and \ref{fig12} do {\it not} offer a truly face-on view of
it; plots of $c$ against $\log\,L$ suffer from strong projection
effects and may not, in general, accurately reflect the true dispersion of
globulars over the actual surface of the FP. (Plots of $c$ vs.~$\log\,E_b^*$
would be better, but that plane still makes an angle of $26^\circ$ with the FP
and projection effects would still be present.)
In order to be as rigorous as possible---and hopefully to allow eventually for
some connection with the fundamental plane(s) of other stellar
systems---it desirable to correct for this. This is easily done in a
procedure analogous to that used by \markcite{ben92}Bender et al.~(1992; see
also \markcite{bur97}Burstein et al.~1997) to define a three-dimensional
``$\kappa$-space'' for elliptical galaxies. The present construct for
globulars is based specifically on cluster binding energies and will therefore
be referred to as $\epsilon$-space.

The easiest way to a face-on view of the FP is to define a new coordinate
system by rotating the $\log\,L$--$\log\,E_b^*$ plane through $64^\circ$
(counterclockwise) about the $c$ axis. The new (and still mutually
orthogonal) axes thus obtained are proportional to
\begin{eqnarray}
\epsilon_1 & \equiv & \log\,E_b^* -  2.05\,\log\,L \nonumber \\
\epsilon_2 & \equiv & 2.05\,\log\,E_b^* +  \log\,L \\
\epsilon_3 & \equiv & c  \nonumber
\label{eq:eps}
\end{eqnarray}
Alternatively, of course, Galactic globulars are expected, on the basis of
equation (\ref{eq:34}), to lie on a line $\epsilon_1=const.$; and the line
$\epsilon_2=const.$ is perpendicular to this in the original
$\log\,L$--$\log\,E_b^*$ plane. In this $\epsilon$-space, then, edge-on views
of the globular cluster fundamental plane are obtained by plotting
$\epsilon_1$ against either $\epsilon_2$ or $\epsilon_3$, while the
$(\epsilon_2,\epsilon_3)$ plane provides the face-on view.

This is shown in Fig.~\ref{fig13} for the full sample of 139 globular clusters
from \markcite{har96}Harris (1996). The two edge-on views of the fundamental
plane are drawn as the bold line $\epsilon_1=39.86$. The scatter of the
King-model clusters about this line in both the $(\epsilon_1, \epsilon_2)$ and
the $(\epsilon_1,\epsilon_3)$ planes are just the residuals due to measurement
errors. The apparent tendency for the 30 PCC clusters to fall below the FP is,
again, probably spurious: the underestimation of $j_0$ leads to low values of
$E_b^*$ and $L$, and thus of $\epsilon_1$, for these objects; and
$\epsilon_3=c$ has been {\it arbitrarily} set to 2.5 for many of them.

\placefigure{fig13}

The plot of $\epsilon_3$ against $\epsilon_2$ confirms that Galactic globulars
are not distributed uniformly throughout their fundamental plane. The
correlation of $c$ with $\log\,L$ clearly carries over into one between
$\epsilon_3$ and $\epsilon_2$. Moreover, because the clusters do occupy a
fairly narrow swath on the FP, the potential problems with projection in
the $c$--$\log\,L$ correlation are actually not too severe: the dashed line
drawn in Fig.~\ref{fig13} has the equation $\epsilon_3=-(12.5\pm3.2)+(0.13
\pm0.03)\,\epsilon_2$, which corresponds (given the definition of $\epsilon_2$
and the scaling $E_b^*\propto L^{2.05}$) to $c\simeq0.68\,\log\,L+const.$
(cf.~Fig.~\ref{fig12}). This particular relation follows from fitting only
the regular, King-model clusters at $r_{\rm gc}>8$ kpc, where, again similar
to the case for $c$ vs.~$\log\,L$, the correlation between $\epsilon_3$ and
$\epsilon_2$ is strongest: The Spearman statistic is $s=0.8$ for the globulars
outside the Solar circle, and the r.m.s.~scatter about the best linear fit
is $\Delta\approx0.3$ dex; but $s=0.4$ and $\Delta\simeq0.4$ dex inside 8 kpc.

Again taking the view that cluster properties in the outer halo better
reflect initial conditions, it seems quite clear that Galactic globulars
formed in something more like a ``fundamental band'' than a plane per se.
Moreover, it is
even plausible (but not proven) that they might have been born along a {\it
fundamental straight line}, a one-dimensional locus in $\epsilon$-space.
(This is a term coined by \markcite{bel98}Bellazzini 1998, who arrived at a
similar conclusion from a very different argument. See Figs.~\ref{fig16} and
\ref{fig19} below for further discussion of \markcite{bel98}Bellazzini's
interpretation of the globular cluster FP.) That is, it is conceivable that
globular clusters formed through a simple process, governed almost completely
by the mass and Galactocentric position of a gaseous protocluster, which
naturally resulted not only in $\Upsilon_{V,0}=const.$ and $E_b^*\propto
L^{2.05}$, but also in some (unknown) one-to-one relation between
central concentration and luminosity.

The viability of this interpretation depends to some extent---similar to what
was concluded above---on the possibility that there may be some nontrivial
function of $\epsilon_3=c$ which correlates very tightly with $\epsilon_2$,
i.e., which shows scatter within the observational uncertainties (at least for
the outer-halo clusters) and which would therefore serve better as the third
axis in $\epsilon$-space. On the other hand, while globulars at $r_{\rm gc}>8$
kpc should have been less influenced by dynamical evolution, it is unrealistic
to expect that they could have been completely untouched by it. Thus, another
possibility is that some or all of the ``excess'' scatter of
clusters in the fundamental $(\epsilon_2, \epsilon_3)$ plane is indeed
irreducible, but that it arose during evolution from an initial cluster
distribution that showed no intrinsic scatter. Theoretical studies of cluster
evolution will be key in deciding this issue and will be of considerable
importance, whatever the outcome, for theories of cluster formation.

It is obviously also of interest to understand the relation between the
present $\epsilon$-space and the corresponding $\kappa$-space defined for
elliptical galaxies (\markcite{ben92}Bender et al.~1992) and studied for
other hot stellar systems (\markcite{bur97}Burstein et al.~1997)---or,
equivalently, to explain the connection between the bivariate correlations
for globular clusters (\markcite{djo95}Djorgovski 1995, and \S4.2 just below)
and the analogous relations for ellipticals (\markcite{djo87}Djorgovski \&
Davis 1987; \markcite{dre87}Dressler et al.~1987; see \markcite{pah98}Pahre,
de Carvalho, \& Djorgovski 1998 for a comprehensive recent discussion). A
potentially important difference is that one full dimension---the $\epsilon_3
\equiv c$ axis---of the observational space employed here is given over to a
parameter describing the broken homology of globulars, whereas the
corresponding structural factor is incorporated into all three axes of the
space used for ellipticals (see \markcite{ben92}Bender et al.~1992). While
explicit corrections can certainly be made for the effects of non-homology in
galaxies (\markcite{gra97}Graham \& Colless 1997; \markcite{bus97}Busarello et
al.~1997), these are perhaps less easily pictured in $\kappa$-space than in
$\epsilon$-space. The construction of an $\epsilon$-space for galaxies would
require the use of a single family of structural and dynamical models that is
highly accurate in its description of any individual system (in order to
estimate total luminosities/masses and global binding energies by
extrapolation from direct observations of the galaxy cores).

That real differences exist between the fundamental planes of globular clusters
and elliptical galaxies is clear from contrasts in certain of their monovariate
correlations, or projections of their FPs (e.g., \markcite{kor85}Kormendy
1985; \markcite{djb93}Djorgovski 1993b). What remains to be seen is the extent
to which these disparities might result from fundamentally different $E_b(L)$
relations, as opposed to any number of other physical distinctions (examples
being a varying core mass-to-light ratio in ellipticals [\markcite{vdm91}van
der Marel 1991]; the presence of dark matter in the galaxies [see
\markcite{kri97}Kritsuk 1997]; and possibly different systematics in the
structural non-homology). It may be significant, in this regard, that
\markcite{bus97}Busarello et al.~1997 find evidence that the {\it kinetic}
energy per unit mass measured within an effective radius in ellipticals scales
as $\left(E_k/M\right)\propto\sigma_0^{1.6}$ or so, whereas the total
$\left(E_k/M\right)$ in globular clusters can be roughly---although not
perfectly---estimated as $\left(E_b/L\right)\propto L^{1.05}\sim \sigma_0^2$
(see Fig.~\ref{fig18} below for the relation between $\sigma_0$ and $L$).

\subsection{Bivariate Correlations}

\markcite{djo95}Djorgovski (1995) used a statistical, principal-components
analysis to infer the existence of two strong, bivariate correlations
involving core and half-light properties of Galactic globular clusters
with measured central velocity dispersions (i.e., those in the catalogue of
\markcite{pry93}Pryor \& Meylan 1993):
\begin{equation}
\log\,\sigma_{p,0}-0.45\,\log\,r_0=-(0.20\pm0.01)\,\mu_{V,0}+(4.17\pm0.2)
\label{eq:djo}
\eqnum{\ref{eq:djo}a}
\end{equation}
and
\begin{equation}
\log\,\sigma_{p,0}-0.7\,\log\,R_h=-(0.24\pm0.02)\,\langle\mu_V\rangle_h+
(4.83\pm0.26)\ ,
\eqnum{\ref{eq:djo}b}
\addtocounter{equation}{+1}
\end{equation}
where $\langle\mu_V\rangle_h\equiv26.362-2.5\,\log\,\left(L/2\pi R_h^2\right)$
is the average surface brightness within a projected half-light radius;
$\sigma_{p,0}$ is in units of km s$^{-1}$; and $r_0$, $R_h$ are measured in pc.
It was the existence of these correlations that originally led
\markcite{djo95}Djorgovski to conclude that the Galactic globulars are
confined to a fundamental plane. But the statistical analysis does not,
by itself, offer any physical insight into the FP. (\markcite{djo95}Djorgovski
suggested, as did \markcite{bel98}Bellazzini [1998] after him, that
eq.~[\ref{eq:djo}a] is a reflection of the ``pure virial theorem'' applied to
cluster {\it cores}. However, as will be discussed further below, this is not
a fully satisfactory interpretation.) Indeed, note that equations
(\ref{eq:djo}) appear superficially to constitute two constraints on five
different observables, and thus it is not clear a priori that they need define
a plane of any kind. Now, however, with a specific physical basis
($\log\,\Upsilon_{V,0}$, $c$, $\log\,E_b$, and $\log\,L$) having been chosen
to describe globular clusters in a King-model framework, it is possible to
reduce the number of distinct variables in these equations to just four,
and to show explicitly---by comparing with the bivariate $(\sigma_0, r_0,
\mu_{V,0})$ and $(\sigma_0, R_h, \langle\mu_V\rangle_h)$ correlations
{\it expected} to arise from each of the two empirical relations defining the
fundamental plane---that equation (\ref{eq:djo}a) derives from equation
(\ref{eq:24}) above while (\ref{eq:djo}b) is a reflection of equation
(\ref{eq:32}). Thus, \markcite{djo95}Djorgovski's (1995) correlations {\it
together} identify the same globular cluster FP that has been developed here.

The basic definition of the King radius (eq.~[\ref{eq:23}]), together with
a constant $\log\,\Upsilon_{V,0}=0.16$ and the definition of surface
brightness, yields a relation between $\sigma_0$ ({\it not} exactly the
$\sigma_{p,0}$ used by \markcite{djo95}Djorgovski in establishing his
correlations), $r_0$, and $\mu_{V,0}$. The details of this derivation are
given in Appendix A; from equation (\ref{eq:a9a}),
\begin{equation}
\log\,\sigma_0 - 0.5\,\log\,r_0=-0.2\,\mu_{V,0}-0.5\,\log\,{\cal I}_0(c)
+4.241\ .
\label{eq:41}
\eqnum{\ref{eq:41}a}
\end{equation}
Here and throughout, $\sigma_0$ is in units of km s$^{-1}$, $r_0$ is in pc,
and $\mu_{V,0}$ is in mag arcsec$^{-2}$. The function ${\cal I}_0$ is the
dimensionless central surface density for a King model of concentration $c$;
it is given in Fig.~\ref{fig1} above and in equation (\ref{eq:b3}) below.
An equivalent expression, in terms of cluster half-light radii and surface
brightnesses, follows from defining $\langle\mu_V\rangle_h$ and from accounting
for non-homology in the ratio ${\cal R}(c)\equiv R_h/r_0$ (see Fig.~\ref{fig1}
and eq.~[\ref{eq:b4}]); according to equation (\ref{eq:a9b}),
\begin{equation}
\log\,\sigma_0 - 0.5\,\log\,R_h=-0.2\,\langle\mu_V\rangle_h
-0.5\,\log\,\left[{\cal L}(c)/{\cal R}(c)\right]+4.640\ ,
\eqnum{\ref{eq:41}b}
\addtocounter{equation}{+1}
\end{equation}
where ${\cal L}$ is the dimensionless luminosity given in Fig.~\ref{fig2} and
equation (\ref{eq:b5}).

Similarly, the analysis in Appendix A (see eqs.~[\ref{eq:a10a}] and
[\ref{eq:a10b}]) shows that the finding $\log\,(E_b/{\rm erg})=39.82+2.05\,
\log\,(L/L_\odot)$ (eq.~[\ref{eq:32}]; for clusters in the catalogue of
\markcite{pry93}Pryor \& Meylan 1993 specifically) is equivalent to
\begin{equation}
\log\,\sigma_0-0.775\,\log\,r_0=-0.205\,\mu_{V,0}
+0.25\,\log\,\left[{\cal L}(c)^{2.05}/{\cal E}(c)\,{\cal I}_0(c)^{2.05}\right]
+3.943
\label{eq:42}
\eqnum{\ref{eq:42}a}
\end{equation}
(with the dimensionless energy ${\cal E}$ again discussed in \S2 and Appendix
B), and to
\begin{equation}
\log\,\sigma_0-0.775\,\log\,R_h=-0.205\,\langle\mu_V\rangle_h 
- 0.25\,\log\,\left[{\cal E}(c)/{\cal R}(c)\right]
+4.352\ .
\eqnum{\ref{eq:42}b}
\addtocounter{equation}{+1}
\end{equation}
Note that, in general, the $r_{\rm gc}$ dependence in the normalization of
the $E_b(L)$ relation (eq.~[\ref{eq:34}]) should appear as an additional
term ($-0.1\,\log\,r_{\rm gc}+constant$) on the right-hand side of equations
(\ref{eq:42}). In order to make a more direct comparison with the analysis of
\markcite{djo95}Djorgovski (1995), however, the effect (which is evidently
quite small anyway) has been acknowledged only implicitly here, by using a
normalization of the $E_b$--$L$ scaling that is appropriate for the median
Galactocentric position of the globulars in the \markcite{pry93}Pryor \&
Meylan database.

Figure \ref{fig14} confirms the validity of equations (\ref{eq:41}b) and
(\ref{eq:42}b) as a representation of the globular cluster FP in terms
of half-light quantities. The left panels here show the non-homology
functions $0.5\,\log\,\left({\cal L}/{\cal R}\right)$ and $0.25\,\log\,
\left({\cal E}/{\cal R}\right)$, as obtained in general from numerical
integrations of King models (curves) and evaluated specifically for the
central concentrations of the 39 regular clusters (points) in the sample of
\markcite{pry93}Pryor \& Meylan (1993). The functions vary only slightly
among these clusters, and the effects of non-homology happen to be effectively
suppressed in this projection of the FP. The mean values indicated for each
function may therefore be applied to derive nearly constant intercepts in
equation (\ref{eq:41}b) for $\Upsilon_{V,0}=const.$ and equation
(\ref{eq:42}b) for $E_b\propto L^{2.05}$. The right panels of Fig.~\ref{fig14}
then make a direct comparison between the resulting FP ``predictions''
(solid lines) and the \markcite{pry}Pryor \& Meylan data. The agreement is
clearly quite good and, as expected, the r.m.s.~scatter of the data about the
expected correlations is within the range of observational uncertainty.

\placefigure{fig14}

The top panels of Fig.~\ref{fig14}, and equation (\ref{eq:41}b), also represent
the basic scaling $L\propto (\sigma_0^2R_h)({\cal L}/{\cal R})$, which
ultimately---given that $\Upsilon_{V,0}$ is a constant independent of $L$---is
just an expression for the total {\it mass} of a King model. Since the ratio
${\cal L}/{\cal R}$ does not vary particularly strongly or systematically
with $c$ (or, therefore, with $L$), this result appears to differ significantly
from the $L\propto\left(\sigma_0^2R_h\right)^{0.7-0.8}$ that
\markcite{sch93}Schaeffer et al.~(1993) find for elliptical galaxies and Abell
clusters (although note that these authors claim that their scaling holds for
Galactic globulars as well). This is another indication that the globular
cluster fundamental plane may indeed differ from those of these other
systems. Whether this particular contrast is due to fundamentally
different trends of $\Upsilon_{V,0}$ with $L$, or to different structural
systematics (i.e., to the non-homology term ${\cal L}/{\cal R}$ perhaps being
more sensitive to $L$ in the larger systems), is unclear at this point.

The two left panels of Fig.~\ref{fig14} further show the product
${\cal E}{\cal R}/{\cal L}^2$ to be very nearly constant as a function
of $c$, and thus they confirm that the expectation $E_b\propto L^2/R_h$
for homologous clusters actually is a good approximation for King-model
globulars as well (as was discussed in \S3.2 above; see Fig.~\ref{fig8}).

Figure \ref{fig15} illustrates the accuracy of equations (\ref{eq:41}a) and
(\ref{eq:42}a) for the globular cluster FP. Again, the panels on the left show
the appropriate non-homology functions for generic King models and for the
regular clusters of \markcite{pry93}Pryor \& Meylan (1993). The top plot shows
(as does Fig.~\ref{fig1}) that ${\cal I}_0$ is a very weak function of $c$.
Non-homology is therefore a rather minor concern for this FP equation as well,
and the mean value of ${\cal I}_0$ can be used to estimate a roughly constant
intercept for equation (\ref{eq:41}a); this then compares very favorably to
the (non-PCC) data in the top right panel of Fig.~\ref{fig15}. The bottom
panels, however, show that things are not always so simple. The logarithm of
${\cal L}^{2.05}/{\cal E}{\cal I}_0^{2.05}$ in the left panel increases
systematically (and close to linearly, as the dashed line indicates) with
concentration parameter $c$, and hence with cluster luminosity (and, thus,
with central surface brightness as well; see eqs.~[\ref{eq:51}] below). This
function is therefore {\it not} well approximated by a constant, and if it is
treated as such---if it is simply assigned its mean value in equation
(\ref{eq:42}a), as was done in the other examples here---the neglected
non-homology manifests itself as a slope in the observed correlation between
$(\log\,\sigma_0-0.775\,\log\,r_0)$ and $\mu_{V,0}$ which is steeper than the
naive expectation.

\placefigure{fig15}

Equation (\ref{eq:42}b) for $(\log\,\sigma_0-0.775\,\log\,R_h)$ vs.~$\langle
\mu_V\rangle_h$---a result of $E_b\propto L^{2.05}$---comes closest to the
basic form of \markcite{djo95}Djorgovski's (1995) equation (\ref{eq:djo}b).
The fact that his analysis zeroed in on a half-light correlation with
the slightly different combination $(\log\,\sigma_{p,0}-0.7\,\log\,R_h)$ may
itself be understood as a consequence of the fact that $E_b\sim L^{2.05}$ for
the Galactic globulars: since $R_h$ is then so nearly constant as a function of
luminosity, a correlation between $0.7\,\log\,R_h$ and $\langle\mu_V\rangle_h$
is not significantly worse, statistically, than one involving $0.775\,\log\,
R_h$. Meanwhile, equation (\ref{eq:41}a) for $(\log\,\sigma_0-0.5\,\log\,r_0)$
vs.~$\mu_{V,0}$ is clearly equivalent to equation (\ref{eq:djo}a) from
\markcite{djo95}Djorgovski (1995), which therefore is another statement of the
fact that the King-model clusters have an essentially constant
$\Upsilon_{V,0}=(1.45\pm0.10)\,M_\odot\, L_\odot^{-1}$.

Finally, as was noted above, it has been claimed (e.g.,
\markcite{djo95}Djorgovski 1995; \markcite{bel98}Bellazzini 1998; see also
\markcite{djo94}Djorgovski \& Meylan 1994) that equation (\ref{eq:djo}a)
(or eq.~[\ref{eq:41}a]) might follow from globular cluster {\it cores},
viewed as dynamically distinct entities in their own right, being
``virialized'' with a constant mass-to-light ratio. However,
this is not entirely accurate. If cores are defined as those parts
of clusters within the volume $r\le r_0$, direct integrations of King models
show that they do not satisfy the simplest version of the virial theorem; that
is, if kinetic energy is denoted $E_k$, then $2E_k(r_0)\ne E_b(r_0)$ in
general. This is shown in Fig.~\ref{fig16}, which plots the ratio $2E_k/E_b$
as a function of $r/r_0$ for three representative King models. [Binding energy
is defined, as in eq.~(\ref{eq:21}), by $E_b(r)=-(1/2)\int_0^r 4\pi r^2\rho
\phi\,dr$; the kinetic energy is computed from $E_k(r)=(3/2)\int_0^r 4\pi
r^2\rho\sigma^2\,dr$.] This may explain why \markcite{bel98}Bellazzini (1998),
who assumed $2E_k/E_b\equiv1$ at $r=r_0$, inferred a mass-to-light ratio of
only $\Upsilon_{V,0}\simeq0.7\,M_\odot\,L_\odot^{-1}$---a factor of two too
small---from his analysis of \markcite{djo}Djorgovski's (1995) bivariate
correlation for cluster cores. This Section has shown that, while
$\Upsilon_{V,0}$ is indeed a constant in Galactic globulars, the precise
form of the correlation between $(\log\,\sigma_0-0.5\,\log\,r_0)$ and
$\mu_{V,0}$ depends on the definition of $r_0$ in equation (\ref{eq:23});
but although the basic scaling there does arise generically from a dimensional
analysis of Poisson's equation (or, equivalently, from the virial theorem {\it
including surface terms} which vanish only at the {\it tidal} radius), the
normalization (which helps set the intercept in eq.~[\ref{eq:djo}a] or
eq.~[\ref{eq:41}a]) is a {\it convenience} specific to \markcite{kin66}King
(1966) models.

\placefigure{fig16}

\section{Other Correlations}

Sections 3 and 4 have presented the main results of this paper: The binding
energies of 109 regular globular clusters in the Milky Way, calculated
within the theoretical context of single-mass, isotropic \markcite{kin66}King
(1966) models, correlate very tightly with total luminosities, and decrease
systematically with increasing Galactocentric radius (eq.~[\ref{eq:34}]). This
result and the fact (eq.~[\ref{eq:24}]) that the core mass-to-light ratio
is a constant (at least for the 39 regular clusters where it has been
directly measured) then imply the existence of a fundamental plane for globular
clusters, one which has an immediate and clear interpretation even
while accounting for the bivariate cluster correlations discovered by
\markcite{djo95}Djorgovski (1995). This new physical view of the FP is
expected to aid in developing theories of cluster formation and evolution.

Meanwhile, on perhaps a more pragmatic note, it is a mathematical
necessity, requiring no further proof once a King-model framework has been
adopted, that {\it any correlation} between {\it any set of cluster
observables} can be obtained simply by treating only $r_{\rm gc}$, $L$ and
(to rather a lesser degree) $c$ as independent variables and then manipulating
the constraints on $E_b(L, r_{\rm gc})$ and $\Upsilon_{V,0}$ according to
generic properties of King models. (As was concluded in \S3.3, all cluster
attributes are independent of metallicity, a fact which---although interesting
in its own right---justifies the neglect of [Fe/H] in this discussion.) No
other empirical trend contains any physical information beyond the intrinsic
properties of these models and the defining equations (\ref{eq:24}) and
(\ref{eq:34}) of the globular cluster FP. This point is developed quite
generally in Appendix A, and it is now illustrated in brief for the Milky Way
cluster system specifically.

Equations (\ref{eq:a11})--(\ref{eq:a16}) give expressions for various physical
quantities in terms of the basis chosen in this paper for King-model parameter
space, i.e., as functions of $L$, $c$, $\Upsilon_{V,0}$, and $E_b$, allowing
for arbitrary $\Upsilon_{V,0}$ and any power-law scaling $E_b=
A\,(L/L_\odot)^{\gamma}$. Applied to the Milky Way cluster system, with
$\log\,\Upsilon_{V,0}\equiv0.16$, $\gamma=2.05$, and $\log\,(A/{\rm erg})=
39.86-0.4\,\log(r_{\rm gc}/8\,{\rm kpc})$, the analysis of Appendix A
therefore results in the following observable dependences on $L$,
$r_{\rm gc}$, and $c$:
\begin{equation}
\log\,r_0=-0.05\,\log\,L + 0.4\,\log\,(r_{\rm gc}/8\,{\rm kpc})
- \log\,\left({\cal L}^2/{\cal E}\right) + 1.681
\label{eq:51}
\eqnum{\ref{eq:51}a}
\end{equation}
\begin{equation}
\log\,R_h=-0.05\,\log\,L + 0.4\,\log\,(r_{\rm gc}/8\,{\rm kpc})
- \log\,\left({\cal L}^2/{\cal E}{\cal R}\right) + 1.681
\eqnum{\ref{eq:51}b}
\end{equation}
\begin{equation}
\log\,j_0=1.15\,\log\,L - 1.2\,\log(r_{\rm gc}/8\,{\rm kpc})
+ \log\,\left({\cal L}^5/{\cal E}^3\right) - 5.042
\eqnum{\ref{eq:51}c}
\end{equation}
\begin{equation}
\mu_{V,0}=-2.75\,\log\,L + 2\,\log\,(r_{\rm gc}/8\,{\rm kpc})
-2.5\,\log\left({\cal L}^3{\cal I}_0/{\cal E}^2\right) + 34.766
\eqnum{\ref{eq:51}d}
\end{equation}
\begin{equation}
\langle\mu_V\rangle_h=-2.75\,\log\,L + 2\,\log\,(r_{\rm gc}/8\,{\rm kpc})
- 5\,\log\,\left({\cal L}^2/{\cal E}{\cal R}\right) + 36.761
\eqnum{\ref{eq:51}e}
\end{equation}
\begin{equation}
\log\,\sigma_0=0.525\,\log\,L - 0.2\,\log\,(r_{\rm gc}/8\,{\rm kpc})
-0.5\,\log\,\left({\cal E}/{\cal L}\right) - 1.872
\eqnum{\ref{eq:51}f}
\addtocounter{equation}{+1}
\end{equation}
for cluster radii in pc, luminosities in $L_\odot$, surface brightnesses in
mag arcsec$^{-2}$, luminosity densities in $L_\odot\,{\rm pc}^{-3}$, and
velocities in km s$^{-1}$. 

\markcite{djo94}Djorgovski \& Meylan (1994) present many ``monovariate''
correlations for Galactic globulars, of the type $\log\,r_0$ vs.~$M_V
=4.83-2.5\,\log\,L$; $\log\,\sigma_0$ vs.~$\mu_{V,0}$; $\log\,j_0$
vs.~$\log\,r_{\rm gc}$; and
so on. All of these can be traced back to some combination of equations
(\ref{eq:51}), and thus to King-model definitions and the two
fundamental-plane relations. (Additional quantities considered by
\markcite{djo94}Djorgovski \& Meylan include dynamical relaxation
times, which are derived from measurements of other cluster parameters and
thus could easily be included in the analysis here.) Clearly, however, the
separation of $L$- and $r_{\rm gc}$-dependences in many cases might give a
skewed or degraded impression of the true, underlying physical relationship
responsible for a correlation; and so can the neglect of structural and
dynamical non-homology, i.e., of the increase of (and scatter in) $c$ as a
function of $L$, and the concomitant variations in ${\cal L}(c)$,
${\cal E}(c)$, ${\cal R}(c)$, and ${\cal I}_0(c)$ from one cluster to
another.

One example of this is shown in Fig.~\ref{fig17}, which follows up the FP
prediction of equation (\ref{eq:51}a) to improve significantly on the rather
poor anticorrelation between scale radius and luminosity that was seen in
Fig.~\ref{fig7} above for the full set of clusters taken from
\markcite{har96}Harris (1996). It is now clear
that the only reason {\it any} such correlation appears is because the
non-homology term $\log\,\left[{\cal L}(c)^2/{\cal E}(c)\right]$ increases
(very roughly) as $const.+0.7\,c\approx const.+0.3\,\log\,L$ (top panel of
Fig.~\ref{fig17}), far outweighing the explicit dependence on $\log\,L$ in
equation (\ref{eq:51}a). The correlation plotted in the bottom panel of
this Figure takes this and the expected dependence on Galactocentric position
fully into account. The solid line drawn through the data here is just
equation (\ref{eq:51}a). The dashed line is a formal least-squares fit to
the 109 regular-cluster points. Although it has the expected slope of $-0.05$,
its intercept is offset slightly downwards, a discrepancy which
traces back to the small---and purely observational---bias introduced by
assuming that $L={\cal L}(c)j_0r_0^3$ exactly for the Galactic globulars
(see the discussion around Fig.~\ref{fig2}). The r.m.s.~scatter of the data
points about either of these lines is a factor of almost 2 smaller than in
a plot of $\log\,r_0$ against $\log\,L$ alone and appears to be fully
attributable to random measurement errors.\footnotemark
\footnotetext{The observational errorbar on $\log\,r_0$ is $\pm0.1$ dex, and
an uncertainty of $\pm0.2$ in $c$ translates to about $\pm0.15$ in
$\log({\cal L}^2/{\cal E})$. In addition, the derivation of equation
(\ref{eq:51}a) in Appendix A involves squaring the equality $L={\cal L}j_0
r_0^3$, thereby introducing additional scatter of $2\times0.25/\sqrt{109}
\simeq0.05$ dex about the mean line (see Fig.~\ref{fig2}). Even added
conservatively, in quadrature, these three terms lead to an expected
r.m.s.~scatter of $\Delta\simeq0.2$ dex, essentially that found
in the bottom of Fig.~\ref{fig17}.}

\placefigure{fig17}

It is worth noting that this approach also explains the ``large'' scatter of
$\log\,R_h^*$ values in Fig.~\ref{fig8} above ($\Delta=0.21$ dex, as opposed
to the observational errorbar of $\pm0.1$ on $\log\,R_h$ alone): the
scatter about the prediction of equation (\ref{eq:51}b) is guaranteed to be
the same as that about equation (\ref{eq:51}a), since the former comes from
adding $\log\,{\cal R}(c)$ to each side of the latter.

All of the other equations above can be similarly dissected, but it is
particularly worth looking at the correlation of velocity scale with total
luminosity (eq.~[\ref{eq:51}f]). This is compared to cluster data from the
smaller sample of \markcite{pry93}Pryor \& Meylan (1993) in Fig.~\ref{fig18}.
In the top panel, non-homology and the expected variation with Galactocentric
radius are ignored. A fairly strong correlation persists in this case,
however, because the ratio ${\cal E}/{\cal L}$ depends only weakly
on $c$ (as can be inferred from the similar shapes of the individual curves
in Figs.~\ref{fig1} and \ref{fig2} above) and because all
but one of the regular (non-PCC) \markcite{pry93}Pryor \& Meylan clusters
lie within $2.6\la r_{\rm gc}\la 29$ kpc [so that the term $0.2\,\log\,
(r_{\rm gc}/8\,{\rm kpc})$ ranges only from $-0.1$ to $+0.1$ or so]. The
exception to this is the bright cluster NGC 2419, marked on Fig.~\ref{fig18},
which is located at $r_{\rm gc}=91.5$ kpc according to \markcite{har96}Harris
(1996). This partly explains its appearance as an outlier relative to the
straight line drawn through the top panel of the Figure:
$$\log\,\sigma_0=0.525\,\log\,L-1.928\ ,$$
which follows from applying an average $0.5\,\log\,({\cal E}/{\cal L})\simeq
0.049$ and a median $r_{\rm gc}=8.7$ kpc (both evaluated using the 39 regular
clusters in the sample) to equation (\ref{eq:51}f). Clearly, this simple
version of the full relation already provides a reasonable description of the
data (aside from NGC 2419), and the monovariate correlation is a rather
faithful representation of the FP scaling $E_b^*\propto L^{2.05}$ and the
constancy of $\Upsilon_{V,0}$.

\placefigure{fig18}

The situation nevertheless improves somewhat if ${\cal E}/{\cal L}$ and
$r_{\rm gc}$ are treated properly as variables. This is done in the
bottom panel of Fig.~\ref{fig18}, which shows a stronger and more significant
$\sigma_0$--$L$ correlation (i.e., the r.m.s.~scatter $\Delta$ is lower and
the Spearman rank correlation coefficient $s$ is higher), and in which NGC
2419 appears as a less extreme outlier (though an outlier all the same).
The solid line plotted here traces the full equation (\ref{eq:51}f), and it is
statistically indistinguishable from a formal least-squares fit to the regular
clusters (dashed line). The scatter of these data about either line is again
that expected given the typical observational uncertainties.\footnotemark
\footnotetext{In this case, the errorbar on $\log\,\sigma_0$ is $\pm0.09$ dex
(\S2). However, the intercept in equation (\ref{eq:51}f) includes a
contribution from $0.5\,\log\,\Upsilon_{V,0}$ (see the general
eq.~[\ref{eq:a16}]), which is further subject to errors in $\log\,\sigma_0$
(because $\Upsilon_{V,0}\propto \sigma_0^2$; see \S3.1). The total
expected scatter in Fig.~\ref{fig18} is therefore $2\times\Delta\,
(\log\,\sigma_0)=0.18$.}

As was mentioned above, equations (\ref{eq:51}) can be combined in various
ways to produce a large number of other correlations. For instance, the
sum of (\ref{eq:51}a) and (\ref{eq:51}c) gives the full dependence
underlying the rough correlation between $\log\,(j_0r_0)$ and $\log\,L$ seen
in Fig.~\ref{fig7} above. Another example---and the last to be considered
here---is found by eliminating $\log\,r_{\rm gc}$ between equations
(\ref{eq:51}a) and (\ref{eq:51}d) above:
\begin{equation}
\log\,r_0 - 0.2\,\mu_{V,0} =
0.5\,\log\,L - 0.5\,\log\,\left({\cal L}/{\cal I}_0\right) - 5.272\ .
\label{eq:52}
\end{equation}
Alternatively, the manipulations leading to equation (\ref{eq:a13}) below
show that this relation is really just an expression for the total luminosity
of a King model, combined with the definition of surface brightness; as such,
it is completely independent of any fundmamental plane specifications.

Equation (\ref{eq:52}) relates to the correlation between scale
radius and core surface brightness, shown in the top left panel of
Fig.~\ref{fig19}, for the 109 regular and 30 PCC clusters from the catalogue of
\markcite{har96}Harris (1996). This is of interest because
\markcite{bel98}Bellazzini (1998) has claimed that the correlation (along with
one between $\sigma_0$ and $\mu_{V,0}$, which can actually be obtained from
eqs.~[\ref{eq:51}d] and [\ref{eq:51}f]) results from Galactic globulars having
a ``constant core mass.'' However, this conclusion was based on the assumption
that cores satisfy the virial theorem in the form $2 E_k(r_0)=E_b(r_0)$, and
this was shown above (Fig.~\ref{fig16}) {\it not} to be true in general.
Moreover, the basic hypothesis of a constant $M_0\equiv\Upsilon_{V,0}j_0r_0^3$
is not borne out by the data. The top right panel of Fig.~\ref{fig19}, for
instance, shows clearly that measured values of $j_0r_0^3$ range over more
than two orders of magnitude---far in excess of the observational
errorbar---in the Milky Way cluster system. \markcite{bel98}Bellazzini (1998)
also recognizes this but implies, in essence, that the scatter, although
unarguably real, is consistent with more or less random excursions from a
line of constant core mass. However, the bottom left panel of Fig.~\ref{fig19}
shows that this is not the case either: If it were true that $j_0r_0^3\sim
const.$, such that $r_0\sim (j_0r_0)^{-1/2}$ with some intrinsic scatter, then
the quantity $(\log\,r_0 - 0.2\,\mu_{V,0})$ would have to be {\it uncorrelated}
with any other cluster property; but instead it shows a significant dependence
on the total cluster luminosity.

\placefigure{fig19}

The point is that {\it both} the monovariate $r_0$--$\mu_{V,0}$ correlation,
{\it and} a correlation between $\log\,r_0-0.2\mu_{V,0}$ and $\log\,L$, are
expected on the basis of equation (\ref{eq:52}). But this says nothing about a
constant cluster core mass; rather, it implies that King-model clusters at a
given {\it total} luminosity (and with a single central concentration) will
always show a  correlation of the form $\log\,r_0\sim0.2\mu_{V,0}$, {\it by
definition}. The rougher trends and the large scatter in the top half of
Fig.~\ref{fig19} reflect the superposition of many such correlations for
clusters with a wide range of total $L$ in the Milky Way, modified somewhat by
systematics in the non-homology terms ${\cal L}$ and ${\cal I}_0$ that result
from the range in $c$-values as well.

The solid line in the bottom left of Fig.~\ref{fig19} is just equation
(\ref{eq:52}) with an intercept evaluated assuming a constant
$\log\,({\cal L}/{\cal I}_0)=1.007$, the average for \markcite{har96}Harris'
(1996) King-model clusters; the dashed line is a least-squares fit to the
data. The two differ---and the scatter about either exceeds the observational
errorbvars on the data---because the ratio ${\cal L}/{\cal I}_0$ has a
significant dependence on $c$, and hence on $\log\,L$ (see Figs.~\ref{fig1} and
\ref{fig2}); the cluster non-homology cannot be ignored here. The bottom right
corner of Fig.~\ref{fig19} therefore shows the full correlation expected on
the basis of equation (\ref{eq:52}). The agreement with the data is now
excellent; the solid, ``model'' line is indistinguishable from a least-squares
fit to the regular clusters, and the r.m.s.~scatter is fully within the realm
of random measurement errors. [This last plot is equivalent to one of
$({\cal L}j_0r_0^3)^{1/2}$ against $(L/L_\odot)^{1/2}$, so the observed and
expected scatter both are just half those in the bottom panel of
Fig.~\ref{fig2} above.]

Thus, the observed correlation between $r_0$ and $\mu_{V,0}$ in globular
clusters is controlled entirely by the generic behavior of total luminosity
in King models, and by the systematic (but scattered) increase of $c$ with
$L$ found in the Milky Way. There is no need to postulate a constant core
mass. Indeed, much more generally, in the scheme developed here---where
globulars are defined completely by $L$, $c$, $\Upsilon_{V,0}$, and
$E_b^*$---there is no {\it room} for such new constraints not already provided
by the equations of the fundamental plane.

\section{Summary}

If they are described by single-mass, isotropic \markcite{kin66}King (1966)
models, globular clusters are fully defined, in general, by specifying just
four independent physical parameters. These were chosen in this
paper to be a mass-to-light ratio ($\log\,\Upsilon_{V,0}$), total
binding energy ($\log\,E_b$), central concentration [$c=\log\,(r_t/r_0)$],
and total luminosity ($\log\,L$). It has been shown that (1) all 39
regular (non--core-collapsed) Galactic globular clusters with measured core
velocity dispersions have a common mass-to-light ratio (\S3.1),
$$\log\,(\Upsilon_{V,0}/M_\odot\,L_\odot^{-1})=0.16\pm0.03\ ;$$
and (2) if it is assumed that this also holds for the other regular
clusters in the Milky Way with no velocity data, but for which measured
King-model structural parameters are available, the full cluster system (109
objects in all, excluding core-collapsed members) shows a very well-defined
dependence of binding energy on total luminosity, modulated by Galacocentric
radius (\S3.2):
$$\log\,\left(E_b/{\rm erg}\right)=
\left[\left(39.86\pm0.40\right)-0.4\,\log\,\left(r_{\rm gc}/8\,{\rm kpc}
\right)\right]+\left(2.05\pm0.08\right)\,\log\,\left(L/L_\odot\right)\ .$$
The scatter about each of these relations is fully accounted for by the typical
observational errorbars on $\log\,\Upsilon_{V,0}$ ($\simeq0.2$ dex) and
$\log\,E_b$ ($\simeq0.5$ dex).

With $\Upsilon_{V,0}$ essentially a fixed constant and $E_b$ known
precisely (at a given Galactocentric position) as a function of $L$, only
two of the four basic cluster properties are truly independent; globular
clusters in the Milky Way are confined to a narrow, two-dimensional
subregion---a fundamental plane (FP)---in the larger, four-dimensional space
of King models. The distribution of clusters on the plane is then determined
essentially by their $c$- and $L$-values. They are not scattered randomly,
however, as there also exists a correlation between $c$ and $\log\,L$. Still,
this does {\it not} necessarily mean that the locus of globular clusters is
only one-dimensional: a linear $c$--$\log\,L$ relation is of poorer quality
than either of the empirical constraints on $\log\,\Upsilon_{V,0}$ and
$\log\,E_b$, and although there {\it might} exist some more complicated
function which gives a comparably one-to-one dependence of concentration on
luminosity, one has yet to be found (\S\S3 and 4).

Regardless, the mean trend of $c$ vs.~$\log\,L$ is independent of
Galactocentric position and cluster metallicity, as is the value of
$\log\,\Upsilon_{V,0}$ and the slope of the correlation between $\log\,E_b$
and $\log\,L$ (\S3.3). The normalization of the $E_b(L)$ scaling is also
independent of [Fe/H], but it decreases towards larger Galactocentric
radii. The equation above accounts for this, and the $r_{\rm gc}$ dependence
there describes the {\it full extent of environmental influences} on
the globular cluster FP.

It was shown in \S4 that cluster concentration parameters (and, to a
lesser extent, binding energies) correlate more tightly with luminosity for
globulars at $r_{\rm gc}>8$ kpc than for those at smaller
radii. Thus, since dynamical evolution has likely been less effective at
erasing initial conditions outside the Solar circle---and since simple
arguments indicate that clusters could remain on or close to the FP as they
evolve---it may well be that most properties of the fundamental plane were set
more or less at the time of globular cluster formation. Quantitative
calculations of cluster evolution over a Hubble time in the Galactic potential
should be applied to check this.

A three-dimensional ``$\epsilon$-space'' was constructed in \S4.1 from a simple
transformation of the cluster parameter space left after specifying
$\log\,\Upsilon_{V,0}=constant$ (and after removing the effects of
Galactocentric radius) in order to obtain a directly face-one view of the FP.
It could be interesting to attempt a similar construction for elliptical
galaxies and clusters of galaxies, i.e., to interpret their fundamental planes
in terms of binding energy as well, in order to make a more direct comparison
with the globular cluster FP.

The equations for $\Upsilon_{V,0}$ and for $E_b$ as a function of $L$, which
define the FP, were shown in \S4.2 to be equivalent to the two strong bivariate
cluster correlations used by \markcite{djo95}Djorgovski (1995) to argue for
the existence of a fundamental plane for globular clusters.
\markcite{djo95}Djorgovski (and \markcite{bel98}Bellazzini 1998) offered a
simple virial-theorem argument (which was shown here to be incomplete) as an
explanation for one of these correlations, but did not interpret the other.
The results of this paper have put his results on a firm physical footing.

Finally, since $\Upsilon_{V,0}$, $c$, $E_b$, and $L$ completely define
globular clusters, any correlations connecting any of their other properties,
or any other trends with Galactocentric radius, necessarily derive from
generic properties of \markcite{kin66}King (1966) models combined with the
empirical FP relations and their environmental dependence. Appendix A outlined
such derivations quite generally, and specialized results for the Milky Way
cluster ensemble were presented in \S5.

The picture that has been developed here is a simplification in that it
works specifically in the context of single-mass and isotropic models for
globular clusters. This approximation is evidently an excellent one, but
because of it the results of this paper can say nothing about the stellar mass
functions or possible velocity anisotropy in observed clusters. Nor have the
kinematics of clusters within the Milky Way (e.g., orbital
eccentricities, or bulk rotation of the metal-rich subsystem) been considered
in any way. These issues aside, however, the preceding discussion may be
distilled into a small set of four main facts to be explained by theories of
cluster formation and evolution:

\begin{itemize}

\item All cluster properties are independent of metallicity.

\item The core mass-to-light ratio, $\Upsilon_{V,0}=(1.45\pm0.1)\,M_\odot\,
L_\odot$, does not vary significantly with cluster luminosity or
Galactocentric position.

\item The concentration parameter, which controls the shape of a cluster's
internal density profile, correlates with luminosity. The dependence of $c$
on $\log\,L$ is much more significant outside the Solar circle than inside,
but even there the relationship is not obviously one-to-one. Its basic form
is, however, independent of $r_{\rm gc}$.

\item Binding energy is intrinsically a function of luminosity and is
regulated by Galactocentric position: $E_b\propto L^{2.05}r_{\rm gc}^{-0.4}$.
Since there is no evidence for significant variations in the {\it global}
mass-to-light ratios of clusters in the Milky Way, $E_b$ is inferred to scale
with total mass in the same way as with luminosity.

\end{itemize}

Once again, the balance of current evidence suggests that these characteristics
of the fundamental plane---and, thus, most of the systematics of the Galactic
globular cluster system---were essentially fixed by the cluster formation
process. Moreover, it is not implausible that this process was controlled
largely by a single intrinsic protocluster parameter---the initial gas
mass---and adjusted by an external influence depending on Galactocentric
radius. Further discussion along these lines is left for future work
(McLaughlin, in preparation). Regardless of any interpretive details, however,
it clearly will be important to determine the extent to which the globular
cluster systems of other galaxies can be described in the same simple terms
that apply in the Milky Way.

\acknowledgments

This work was supported by NASA through grant number HF-1097.01-97A awarded by
the Space Telescope Science Institute, which is operated by the Association of
Universities for Research in Astronomy, Inc., for NASA under contract
NAS5-26555.

\appendix

\section{KING MODELS AND GLOBULAR CLUSTER CORRELATIONS}

The analysis in this paper refers to eleven physical parameters
which may be observed or derived for globular clusters: the central
line-of-sight velocity dispersion $\sigma_{p,0}$; the scale radius $r_0$,
which is closely related (but not always identical) to the core radius,
at which the luminosity surface density falls to half its central
value; the central surface brightness $\mu_{V,0}$, which is related to the
central $V$-band intensity, $I_0$, in units of $L_\odot\,{\rm pc}^{-2}$; the
central luminosity density $j_0$ ($L_\odot\,{\rm pc}^{-3}$); the central
mass-to-light ratio $\Upsilon_{V,0}$, which gives the central {\it mass}
density, $\rho_0=\Upsilon_{V,0}j_0$, and which must equal the global
mass-to-light ratio in (idealized) clusters of single-mass stars; the
projected half-light radius $R_h$ and the surface brightness $\langle\mu_V
\rangle_h$ averaged within that aperture; the central concentration $c$,
which describes the global shape of the cluster's surface brightness profile;
the total luminosity $L$; and the cluster binding energy, $E_b$.

There are, of course, other quantities of interest (e.g., core and half-mass
relaxation times; average velocity dispersion within the half-light radius),
but within the context of \markcite{kin66}King's (1966) model for
globular clusters, these other parameters can all be derived from some subset
of the eleven just listed. Indeed, if the isotropic, single-mass King models
are adopted a priori as complete descriptions of the internal structures of
Galactic globular clusters, then only {\it four} of these eleven variables are
truly independent. These are (1) the mass-to-light ratio $\Upsilon_{V,0}$; (2)
the concentration parameter $c=\log\,(r_t/r_0)$, which is directly related to
the depth of a cluster's potential well; and two other parameters---chosen
here to be (3) binding energy and (4) total luminosity---required to normalize
a dimensionless model to an observed object. The reduction to this
physical basis requires seven definitions and model relations. (See also
\markcite{dja93}Djorgovski 1993a for an outline of this procedure, and
\markcite{kin66}King 1966 or \markcite{bin87}Binney \& Tremaine 1987 for more
detailed descriptions of the models themselves.)

First, recall that, where they are available, the directly observed central
velocity dispersions of Milky Way globulars (\markcite{pry93}Pryor \& Meylan
1993) have {\it already} been converted to \markcite{kin66}King-model scale
velocities $\sigma_0$ (in \S2 above; see Fig.~1 and eq.~[\ref{eq:b1}]). Also,
the so-called core radii tabulated by \markcite{har96}Harris (1996; also
\markcite{tra93}Trager et al.~1993 and \markcite{dja93}Djorgovski 1993a) are
in fact the $r_0$ referred to here as scale radii ($r_0$ differs from the
true projected half-power radius in low-concentration clusters).

Given these things, the definition of $r_0$ (\markcite{kin66}King 1966) yields
a cluster's central {\it mass} density:
\begin{equation}
\Upsilon_{V,0}j_0=9\sigma_0^2/(4\pi G r_0^2)\ ,
\label{eq:a1}
\end{equation}
where the basic scaling $\rho_0\propto(\sigma_0^2/Gr_0^2)$ follows simply from
dimensional analysis, but the coefficient $(9/4\pi)$ is specific to the King
model. A central {\it luminosity} volume density alone is derived from an
observed $r_0$, $c$, and central surface brightness, by way of the trivial
(model-independent) definition
\begin{equation}
\mu_{V,0}=26.362-2.5\,\log\,(I_0/L_\odot\,{\rm pc}^{-2})
\label{eq:a2}
\end{equation}
and the model relation
\begin{equation}
I_0={\cal I}_0(c)\,j_0r_0\ ,
\label{eq:a3}
\end{equation}
where ${\cal I}_0$ is the dimensionless function of concentration shown in the
bottom left panel of Fig.~\ref{fig1} and approximated by equation
(\ref{eq:b3}). Equation (\ref{eq:a1}) is used in \S3.1 to infer the central
mass-to-light ratio $\Upsilon_{V,0}$ for the 39 non--core-collapsed clusters
catalogued by \markcite{pry93}Pryor \& Meylan (1993).

The projected half-light radius, King scale radius, and central concentration
are linked through the function
\begin{equation}
{\cal R}(c)=R_h/r_0\ ,
\label{eq:a4}
\end{equation}
which is drawn in the bottom right panel of Fig.~\ref{fig1} and approximated
in equation (\ref{eq:b4}) below. The average surface brightness within $R_h$
is then given (for $L$ in $L_\odot$ and $R_h$ in pc) by
\begin{equation}
\langle\mu_V\rangle_h=26.362-2.5\,\log\,(L/2\pi R_h^2)\ ,
\label{eq:a5}
\end{equation}
where the total cluster luminosity is given in terms of $c$, $j_0$ and $r_0$
by
\begin{equation}
L={\cal L}(c)\,j_0r_0^3\ .
\label{eq:a6}
\end{equation}
(See Fig.~\ref{fig2} and eq.~[\ref{eq:b5}] for the dimensionless function
${\cal L}$.) Finally, the cluster binding energy is expressed as
(cf.~eq.~[\ref{eq:22}])
\begin{equation}
E_b ={{\sigma_0^4r_0}\over{G}}\,{\cal E}(c) =
\left({9\over{4\pi}}\right)^{1/2}\,{{\sigma_0^5
\left(\Upsilon_{V,0}j_0\right)^{-1/2}}\over{G^{3/2}}}\,{\cal E}(c) =
\left({{4\pi}\over{9}}\right)^2\,G\left(\Upsilon_{V,0}j_0\right)^2r_0^5\,
{\cal E}(c)\ ,
\label{eq:a7}
\end{equation}
with ${\cal E}$ the function in equation (\ref{eq:b2}) and the top right
panel of Fig.~\ref{fig1}. (The last two equalities in eq.~[\ref{eq:a7}] follow
from the definition [\ref{eq:a1}].)

To repeat, then, a King-model globular cluster is defined fully by
its core mass-to-light ratio, central concentration, binding energy, and
total luminosity. Real clusters therefore can be expected to inhabit a
nominally four-dimensional parameter space. However, the dimensionality of
this space is reduced by one for every independent constraint on the four
fundamental quantities. Thus, given relations of the type inferred in \S3
for the Milky Way clusters, 
\begin{equation}
\Upsilon_{V,0}={\rm constant}\ \ \ \ \ \ \ {\rm and}\ \ \ \ \ \ \ 
E_b=A\,\left(L/L_\odot\right)^{\gamma}\ ,
\label{eq:a8}
\end{equation}
globulars constitute an essentially two-parameter family
of objects (although, in principle, $\Upsilon_{V,0}$ or the $E_b(L)$ relation
may further depend on external, environmental variables; in our Galaxy, $A
\propto r_{\rm gc}^{-0.4}$ according to \S3.3 above.) That is, having
specified $L$ and $c$ for an object (at some given $r_{\rm gc}$), its binding
energy $E_b$ and mass-to-light ratio $\Upsilon_{V,0}$ follow immediately from
equation (\ref{eq:a8}). The product $\left(j_0r_0^3\right)$ is then given
uniquely by equation (\ref{eq:a6}), and the combination
$\left(j_0^2r_0^5\right)$ by equation (\ref{eq:a7}). This sets
the values of $j_0$ and $r_0$ individually, and all other cluster parameters
follow from equations (\ref{eq:a1})--(\ref{eq:a5}). If there exists a third
basic relation, $c=c(\log\,L)$---a situation which is strongly suggested,
although not proven, by current data on the Galactic cluster system---then
luminosity or mass alone (again, at a given $r_{\rm gc}$ in our Galaxy)
controls the globular cluster sequence.

It is clear, as a result of this, that {\it any correlations between any
observed or derived cluster parameters must trace back to some
combination of the equations (\ref{eq:a1})--(\ref{eq:a7}) characterizing
King models, and/or to relations between the fundamental physical variables
$\Upsilon_{V,0}$, $c$, $E_b$, and $L$}. Several examples of such observable
correlations are now derived quite generally; their application to the Milky
Way globular cluster system specifically is discussed in \S\S4 and 5 above.

\subsection{Bivariate Correlations}

Equation (\ref{eq:a1}) can be viewed either as a definition of $r_0$ or as an
expression for $\Upsilon_{V,0}$ in terms of $\sigma_0$, $r_0$, and $j_0$.
Either way, it may be re-written as
$$\left(4\pi G/9\right)\,\Upsilon_{V,0}\,j_0r_0=\sigma_0^2r_0^{-1}\ ;$$
or, in appropriate units,
$$\left({{\sigma_0}\over{{\rm km}\,{\rm s}^{-1}}}\right)^2\,
\left({{r_0}\over{{\rm pc}}}\right)^{-1} =
5.9870\times10^{-3}\,
\left({{\Upsilon_{V,0}}\over{M_\odot\,L_\odot^{-1}}}\right)\,
\left({{j_0r_0}\over{L_\odot\,{\rm pc}^{-2}}}\right)\ .$$
Taking the logarithm of both sides and using equations (\ref{eq:a2}) and
(\ref{eq:a3}) to introduce the observable $\mu_{V,0}$ then yields
\begin{equation}
\log\,\sigma_0-0.5\,\log\,r_0=-0.2\,\mu_{V,0}-0.5\,\log\,{\cal I}_0(c)
+0.5\,\log\,\Upsilon_{V,0}+4.1610\ .
\label{eq:a9a}
\end{equation}
Given a constant $\log\,\Upsilon_{V,0}$ (eq.~[\ref{eq:24}]) and a very slowly
varying ${\cal I}_0(c)$, this becomes an essentially bivariate correlation
corresponding to one of \markcite{djo95}Djorgovski's (1995) two equations for
the fundamental plane of Galactic globular clusters (\S4.2). It may also be
put in terms of half-light cluster radii and average surface brightness by
noting a relation between $\mu_{V,0}$ and $\langle\mu_V\rangle_h$ that follows
from King-model definitions: Equations (\ref{eq:a2}), (\ref{eq:a3}),
(\ref{eq:a6}), (\ref{eq:a4}), and (\ref{eq:a5}) combined give
\begin{eqnarray}
\mu_{V,0} & = & 26.362-2.5\,\log\,
\left[{{{\cal I}_0(c)\,j_0r_0}\over{L_\odot\,{\rm pc}^{-2}}}\right]
 = 26.362-2.5\,\log\,
\left[2\pi\,{{{\cal I}_0(c){\cal R}(c)^2}\over{{\cal L}(c)}}\,
{{L/2\pi R_h^2}\over{L_\odot\,{\rm pc}^{-2}}}\right] \nonumber \\
 & = & \langle\mu_V\rangle_h-2.5\,\log\,\left(2\pi\right)-
2.5\,\log\left[{{{\cal I}_0(c){\cal R}(c)^2}\over{{\cal L}(c)}}\right]\ ,
\label{eq:mumu}
\end{eqnarray}
and substitution of this in equation (\ref{eq:a9a}) yields
\begin{equation}
\log\,\sigma_0-0.5\,\log\,R_h=-0.2\,\langle\mu_V\rangle_h-
0.5\,\log\,\left[{\cal L}(c)/{\cal R}(c)\right]
+0.5\,\log\,\Upsilon_{V,0}+4.5601\ .
\label{eq:a9b}
\end{equation}
as another expression of equation (\ref{eq:a1}).

A power-law scaling of binding energy with total luminosity leads to another
pair of equivalent bivariate correlations. With $E_b=A\,\left(
L/L_\odot\right)^{\gamma}$ as in equation (\ref{eq:a8}), the first
equality in equation (\ref{eq:a7}) reads
$$\sigma_0^4r_0=GA\,\left(L/L_\odot\right)^{\gamma} {\cal E}(c)^{-1}\ ,$$
where, again, the normalization $A$ in general may vary with Galactocentric
position (or with any other environmental factor appropriate to a given
dataset). Applying equation (\ref{eq:a6}) and inserting the usual units, this
becomes
$$\left({{\sigma_0}\over{{\rm km}\,{\rm s}^{-1}}}\right)^4\,
\left({{r_0}\over{{\rm pc}}}\right)^{1-2\gamma} =
2.1558\times10^{-46}\,\left({A\over{{\rm erg}}}\right)\,
\left[{{{\cal L}(c)^{\gamma}}\over{{\cal E}(c)}}\right]\,
\left({{j_0\,r_0}\over{L_\odot\,{\rm pc}^{-2}}}\right)^{\gamma}\ ,$$
so that taking the logarithm and using the definition of $\mu_{V,0}$
(eqs.~[\ref{eq:a2}] and [\ref{eq:a3}]) yields
\begin{eqnarray}
\log\,\sigma_0-0.25\,\left(2 \gamma-1\right)\,\log\,r_0 & = &
-(0.1\,\gamma)\,\mu_{V,0}
+0.25\,\log\,\left[{\cal L}(c)^{\gamma}/{\cal E}(c)\,
{\cal I}_0(c)^{\gamma}\right] \nonumber \\
 & & + 0.25\,\log\,A + 2.6362\,\gamma - 11.4166\ .
\label{eq:a10a}
\end{eqnarray}
And finally, a version that refers to half-light cluster quantities is
obtained by applying equations (\ref{eq:a4}) and (\ref{eq:mumu}):
\begin{eqnarray}
\log\,\sigma_0-0.25\,\left(2\gamma-1\right)\,\log\,R_h & = &
-(0.1\,\gamma)\,\langle\mu_V\rangle_h
-0.25\,\log\,\left[{\cal E}(c)/{\cal R}(c)\right] \nonumber \\
 & & + 0.25\,\log\,A + 2.8357\,\gamma - 11.4166\ ,
\label{eq:a10b}
\end{eqnarray}
which is shown in \S4.2 to correspond to \markcite{djo95}Djorgovski's (1995)
second equation for the globular cluster fundamental plane in the Milky Way.

\subsection{``Monovariate'' Correlations}

Given the suite of definitions listed above
(eqs.~[\ref{eq:a1}]--[\ref{eq:a7}]), plus the two fundamental-plane
constraints of equation (\ref{eq:a8}), any globular cluster observables can be
expressed as functions of central concentration and total luminosity (and
any environmental factors, such as $r_{\rm gc}$, which in general can be
subsumed in eq.~[\ref{eq:a8}]). Some of these relations can appear as
monovariate cluster correlations if non-homology and variations in $c$ are
ignored.

One such expression follows from the third equality in equation (\ref{eq:a7}).
Together with equations (\ref{eq:a8}) and (\ref{eq:a6}), this reads
$${{E_b}\over{{\rm erg}}}=1.6627\times10^{41}\,
\left({{\Upsilon_{V,0}}\over{M_\odot\,L_\odot^{-1}}}\right)^2\,
\left({{j_0}\over{L_\odot\,{\rm pc}^{-3}}}\right)^2\,
\left({{r_0}\over{{\rm pc}}}\right)^5\, {\cal E}(c)=
{A\over{{\rm erg}}}\,
\left({L\over{L_\odot}}\right)^{\gamma-2}
\left[{{\cal L}(c)\,{j_0\,r_0^3}\over{L_\odot}}\right]^2\ ,$$
or, after taking the logarithm of both sides,
\begin{equation}
\log\,r_0=(2-\gamma)\,\log\,L - \log\,\left[{\cal L}(c)^2/{\cal E}(c)\right]
-\log\,A + 2\,\log\,\Upsilon_{V,0} + 41.2208\ .
\label{eq:a11}
\end{equation}
As \S5 discusses, this does translate to a rough correlation between $r_0$
and $L$ in the Milky Way (see Figs.~\ref{fig7} and \ref{fig17}). It also can
be used to derive the dependence of $R_h={\cal R}(c)\,r_0$ on $L$ and $c$:
\begin{equation}
\log\,R_h=(2-\gamma)\,\log\,L - \log\,\left[{\cal L}(c)^2/{\cal E}(c)
{\cal R}(c)\right] - \log\,A + 2\,\log\,\Upsilon_{V,0} + 41.2208\ ,
\label{eq:a12}
\end{equation}
accounting for the behavior of observed cluster half-light radii in
Fig.~\ref{fig8} above. (Like $r_0$, $R_h$ is measured in pc.)

Correlations involving surface brightness follow from the basic definitions
(\ref{eq:a2}), (\ref{eq:a3}), and (\ref{eq:a6}), which state that
\begin{eqnarray}
\log\,r_0 & = & 
0.2\,\mu_{V,0} + 0.5\,\log\,(j_0r_0^3) + 0.5\,\log\,{\cal I}_0(c) - 5.2724
\nonumber \\
 & = & 0.2\,\mu_{V,0} + 0.5\,\log\,L -
0.5\,\log\,\left[{\cal L}(c)/{\cal I}_0(c)\right] - 5.2724
\nonumber
\end{eqnarray}
for $j_0$ in units of $L_\odot\,{\rm pc}^{-3}$ and $r_0$ in pc. This is itself
a bivariate correlation; it is discussed further in \S5, in connection with
a rough correlation between $r_0$ and $\mu_{V,0}$ observed in the Milky Way
cluster system (see Fig.~\ref{fig19}). But it and equation (\ref{eq:a11}) also
lead to
\begin{equation}
\log\,j_0=\left(3\gamma-5\right)\,\log\,L +
\log\,\left[{\cal L}(c)^5/{\cal E}(c)^3\right]+3\,\log\,A -
6\,\log\,\Upsilon_{V,0} - 123.662
\label{eq:a13}
\end{equation}
or to
\begin{equation}
\mu_{V,0}=2.5 \left(3-2\gamma\right)\,\log\,L-2.5\,\log\left[{\cal L}(c)^{3}
{\cal I}_0(c)/{\cal E}(c)^2\right]-5\,\log\,A+10\,\log\,\Upsilon_{V,0}
+232.466\ ,
\label{eq:a14}
\end{equation}
which corresponds to the correlation between $\log\,j_0r_0$ and $\log\,L$ in
Fig.~\ref{fig7} above. In addition, equations (\ref{eq:a14}) and
(\ref{eq:mumu}) together, or equations (\ref{eq:a12}) and (\ref{eq:a5})
together, imply
\begin{equation}
\langle\mu_V\rangle_h=2.5\left(3-2\gamma\right)\,\log\,L -
5\,\log\,\left[{\cal L}(c)^2/{\cal E}(c){\cal R}(c)\right]-5\,\log\,A +
10\,\log\,\Upsilon_{V,0} + 234.461\ . 
\label{eq:a15}
\end{equation}

Finally, a correlation between core velocity dispersion and total cluster
luminosity can be obtained directly from the definition $E_b=\sigma_0^4
r_0 {\cal E}(c)/G$, the empirical scaling $E_b/{\rm erg} = A
(L/L_\odot)^{\gamma}$, and equation (\ref{eq:a11}):
\begin{equation}
\log\,\sigma_0=
0.5(\gamma-1)\,\log\,L -
0.5\,\log\,\left[{\cal E}(c)/{\cal L}(c)\right] +
0.5\,\log\,A - 0.5\,\log\,\Upsilon_{V,0} - 21.7218
\label{eq:a16}
\end{equation}
for $\sigma_0$ in units of km s$^{-1}$. This version of the $E_b(L)$
relation in the Milky Way is compared to data in Fig.~\ref{fig18} above.

The six quantities on the left-hand sides of equations
(\ref{eq:a11})--(\ref{eq:a16}) are all expressed in terms of the
King-model basis $L$, $c$, $\Upsilon_{V,0}$, and $E_b=A(L/L_\odot)^{\gamma}$
(once more, any effects of, say, Galactocentric position and metallicity are
easily allowed to enter through these variables; see \S5). Moreover, any
correlation between any other set of globular cluster observables may be
derived from equations (\ref{eq:a11})--(\ref{eq:a16})---the results of
\S A.1, for example, could be thus obtained. As for {\it any} collection
of single-mass, isotropic King-model clusters, there is no {\it new} physical
content in empirical scalings beyond those existing just among $L$, $c$,
$\Upsilon_{V,0}$ and $E_b$. This follows necessarily from the fact that this
chosen basis is (by definition) complete.

\section{DIMENSIONLESS FUNCTIONS OF CENTRAL CONCENTRATION}

This Appendix lists ad hoc polynomial fits that describe the variation of
certain derived quantities in single-mass, isotropic \markcite{kin66}King
(1966) models, as functions of the central concentration $c\equiv\log\,
(r_t/r_0)$. These fits have been obtained by comparing with numerical
integrations of King models with $0.12\la c\la 3.6$, or $0.5\le W_0\le 16$
for $W_0$ the (normalized) central potential. The fits generally will
{\it not} be reliable if extrapolated beyond this range in $c$, but all
Galactic globular clusters are described by $0.5\le c\le 2.5$.

The projected, or line-of-sight, velocity dispersion at the center of a model
cluster is given in terms of the velocity {\it scale parameter}
$\sigma_0$---which differs from the true one-dimensional velocity dispersion
for low-concentration models---by
\begin{equation}
\log\,\left(\sigma_{p,0}/\sigma_0\right)=-14.203\,\log\,\left[
1+0.11313\times10^{-1.1307\,c}\right]\ ,
\label{eq:b1}
\end{equation}
which deviates from the curve in Fig.~\ref{fig1} above by less than 0.005 dex
for $c\ga0.5$.

The dimensionless binding energy (eq.~[\ref{eq:22}] above, and the top right
panel of Fig.~\ref{fig1}) may be fit by
\begin{eqnarray}
\log\,{\cal E}\equiv \log\,\left(GE_b/\sigma_0^4r_0\right) & = &
-1.64893+2.82056\,c+9.38926\,c^2-26.0275\,c^3+30.6474\,c^4 \nonumber \\
 & & -20.7951\,c^5+8.61423\,c^6-2.13978\,c^7+0.291982\,c^8 \nonumber \\
 & & -0.0167994\,c^9\ ,
\label{eq:b2}
\end{eqnarray}
with an absolute error of $\la0.01$ dex (relative error $\la1\%$) for
$c\ga0.5$.

The central intensity, or surface density, in terms of the central volume
density $j_0$ and the King radius $r_0$, is (cf.~\markcite{dja93}Djorgovski
1993a, and the bottom left panel of Fig.~\ref{fig1} above)
\begin{equation}
\log\,{\cal I}_0\equiv \log\,\left(I_0/j_0 r_0\right) =
0.3022 - 7.5726\,\log\,\left[1+0.2180\times 10^{-1.1291\,c}\right]\ ,
\label{eq:b3}
\end{equation}
to within an absolute error of $\la0.005$ dex in the interval $0.5\la c\la
3.6$. [Note that in the limit $c\rightarrow\infty$, this formula gives
${\cal I}_0\rightarrow 2.005$, slightly lower than the correct value for
an isothermal sphere, ${\cal I}_0(\infty)=2.018$.] For these single-mass
models, which have spatially constant mass-to-light ratios,
${\cal I}_0$ is also equal to the dimensionless central {\it mass} surface
density, $\Sigma_0/\rho_0r_0$.

The {\it projected} half-light radius, $R_h$, is related to the King radius,
$r_0\equiv(9\sigma_0^2/4\pi G \rho_0)^{1/2}$, by
\begin{eqnarray}
\log\,{\cal R}\equiv\log\,\left(R_h/r_0\right) & = &
-0.602395+1.36023\,c-1.67086\,c^2+2.65848\,c^3-2.71152\,c^4 \nonumber \\
 & & +1.42555\,c^5  -0.274551\,c^6-0.0381277\,c^7+0.0217849\,c^8 \nonumber \\
 & & -0.00225252\,c^9\ ,
\label{eq:b4}
\end{eqnarray}
which differs by no more than 0.007 dex from the $\log\,{\cal R}$ obtained by
numerically integrating any King model with $0.12\la c\la 3.6$ (bottom right
of Fig.~\ref{fig1} above).

Finally, the dimensionless total luminosity (top panel of Fig.~\ref{fig2})
is given by
\begin{eqnarray}
\log\,{\cal L}\equiv\log\,\left(L/j_0r_0^3\right) & = &
-0.725444+1.90743\,c+4.63720\,c^2-13.0266\,c^3+14.8724\,c^4 \nonumber \\
 & & -9.43699\,c^5+3.39098\,c^6-0.578169\,c^7-0.00512814\,c^8 \nonumber \\
 & & +0.0162534\,c^9-0.00165078\,c^{10}\ .
\label{eq:b5}
\end{eqnarray}
The relative error of this expression is $\la0.5\%$, in $\log\,{\cal L}$,
for any $0.5\la c\la3.6$; it may be compared to the fit given by
\markcite{dja93}Djorgovski (1993a), which applies over a more restricted
range in central concentration. As with equation (\ref{eq:b3}), ${\cal L}$ also
gives the dimensionless {\it mass}, $M/\rho_0r_0^3$, of a single-mass cluster.

\clearpage

\figcaption[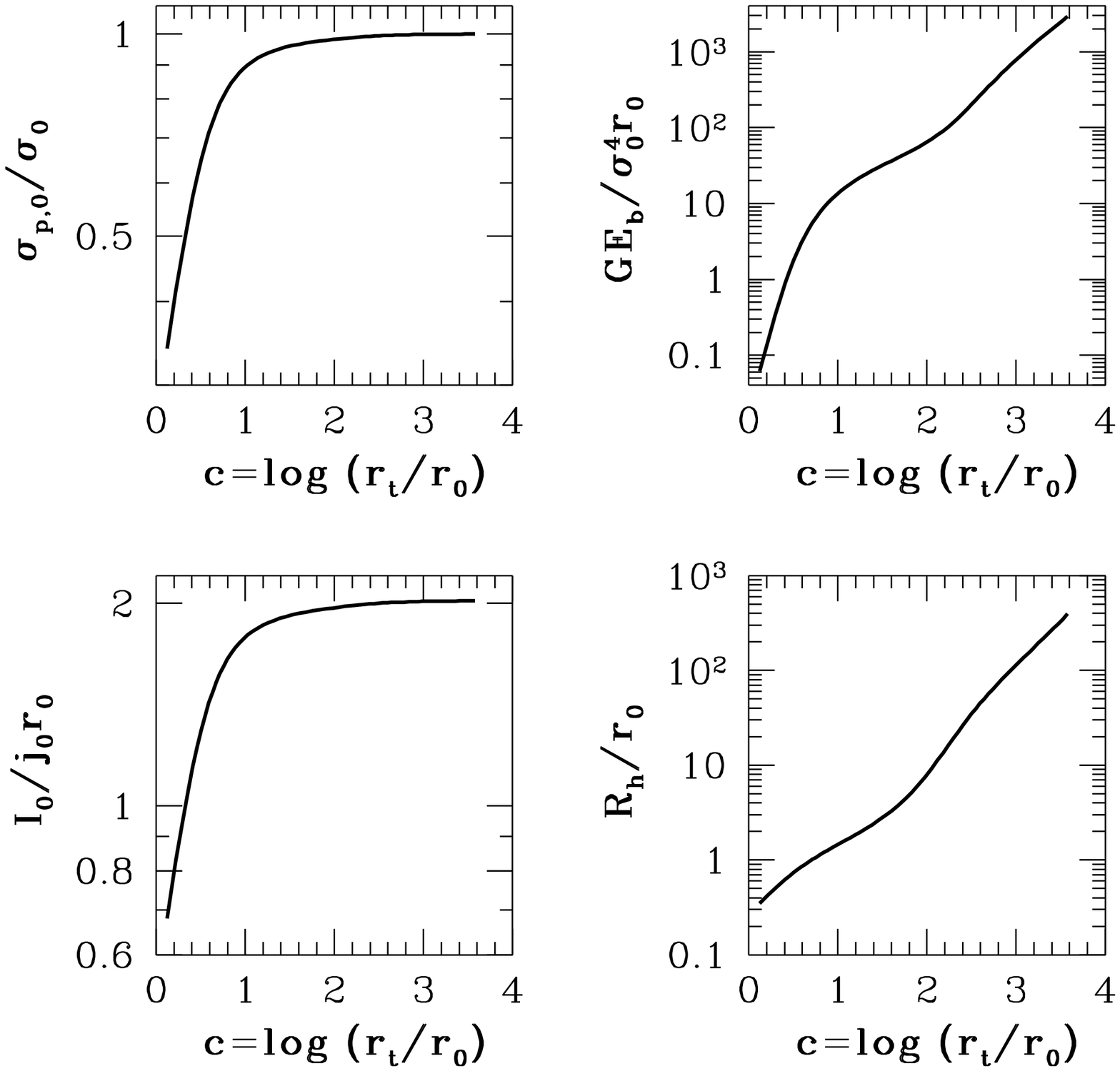]{Dimensionless functions of central concentration for
single-mass, isotropic King (1966) models. Analytic expressions for each are
given in Appendix B. The observed concentrations of Milky Way globulars lie
in the range $0.5\la c\la 2.5$.
\label{fig1}}

\figcaption[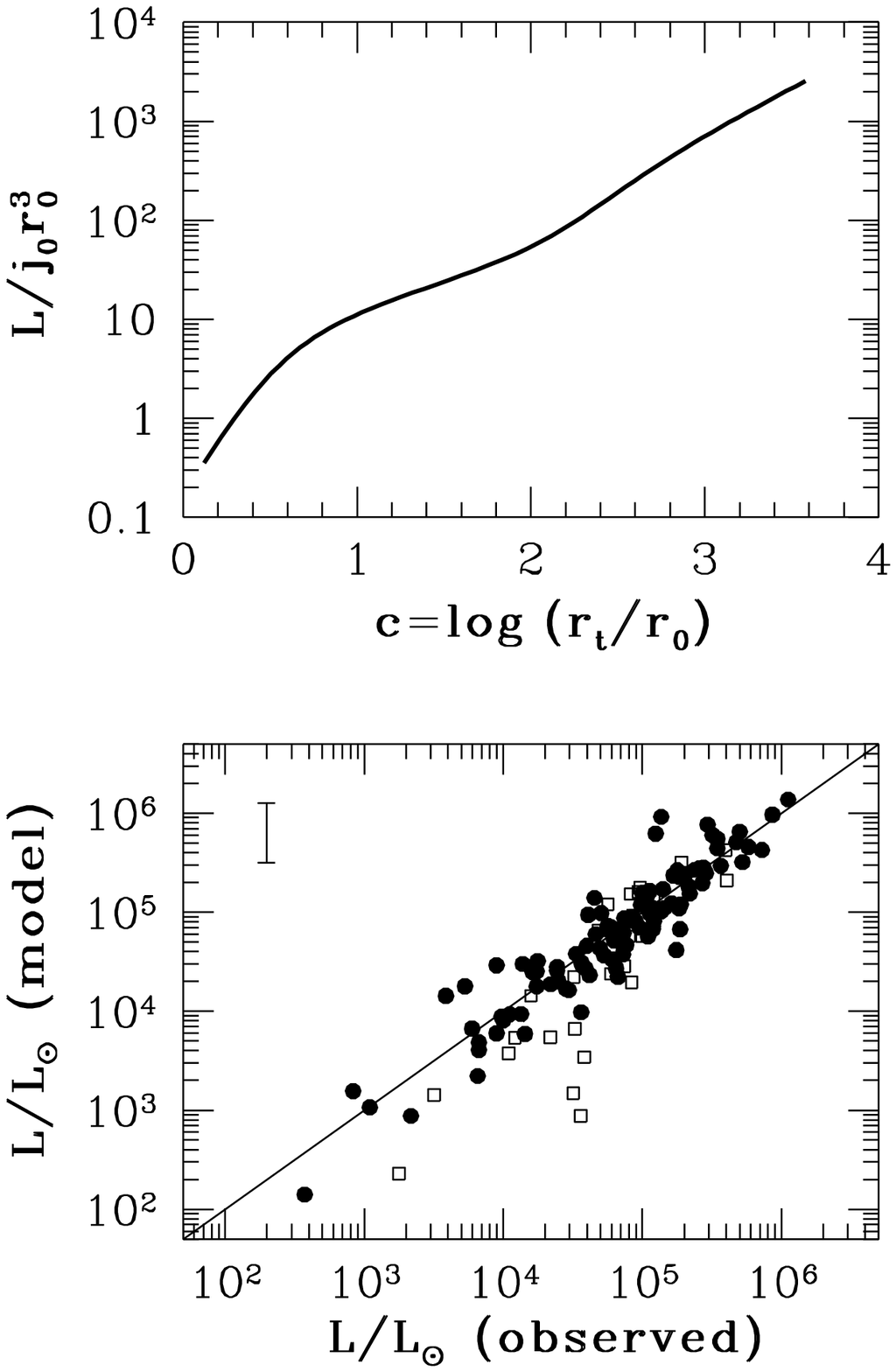]{{\it Top panel}: Dimensionless total luminosity
${\cal L}$ of King models, as a function of central concentration. See
also equation (\ref{eq:b5}). {\it Bottom panel}: Comparison of model
luminosities ${\cal L}(c)j_0r_0^3$ to the directly observed
$L=0.4\left(4.83-M_V\right)$ for globulars from Harris (1996). Filled circles
are regular, King-model clusters; open squares are post--core-collapse
objects. Straight line is the equality $\log\,L_{\rm mod}=\log\,L$, and
the typical observational errorbar on $\log\,L_{\rm mod}$ is shown in the
upper left corner.
\label{fig2}}

\figcaption[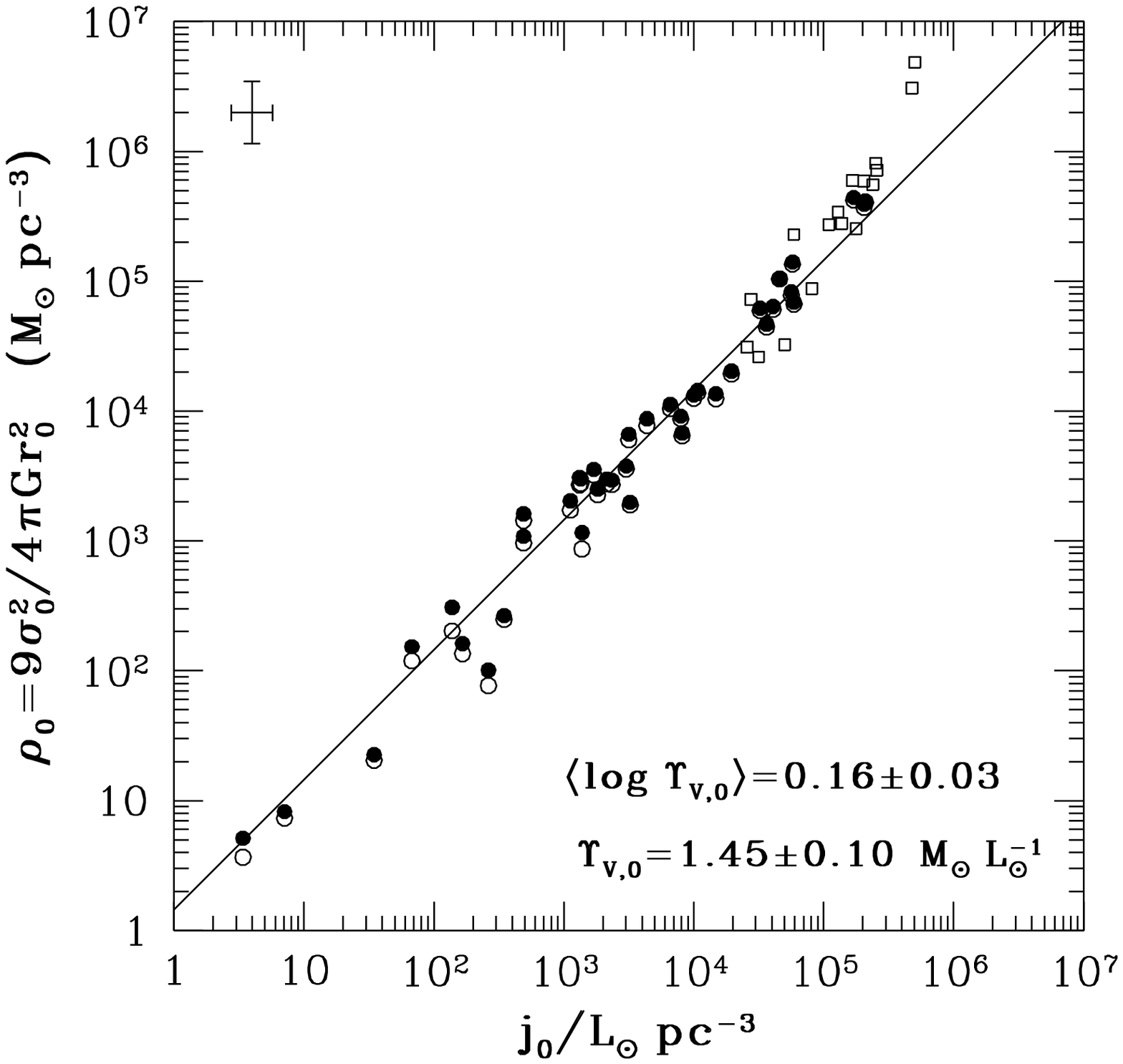]{Determination, according to equation (\ref{eq:23}), of
the mean core mass-to-light ratio of globular clusters in the Milky Way. 
Filled circles correspond to 39 King-model clusters tabulated by Pryor
\& Meylan (1993); open circles have $\rho_0$ computed for these objects using
the directly observed $\sigma_{p,0}$ instead of the model velocity scale
$\sigma_0$. Open squares represent PCC clusters. The solid line has the
equation $\log\,\rho_0=0.16+\log\,j_0$, obtained from a least-squares fit of
the regular-cluster data only. The errorbars at upper left show typical
observational uncertainties in $\rho_0$ and $j_0$.
\label{fig3}}

\figcaption[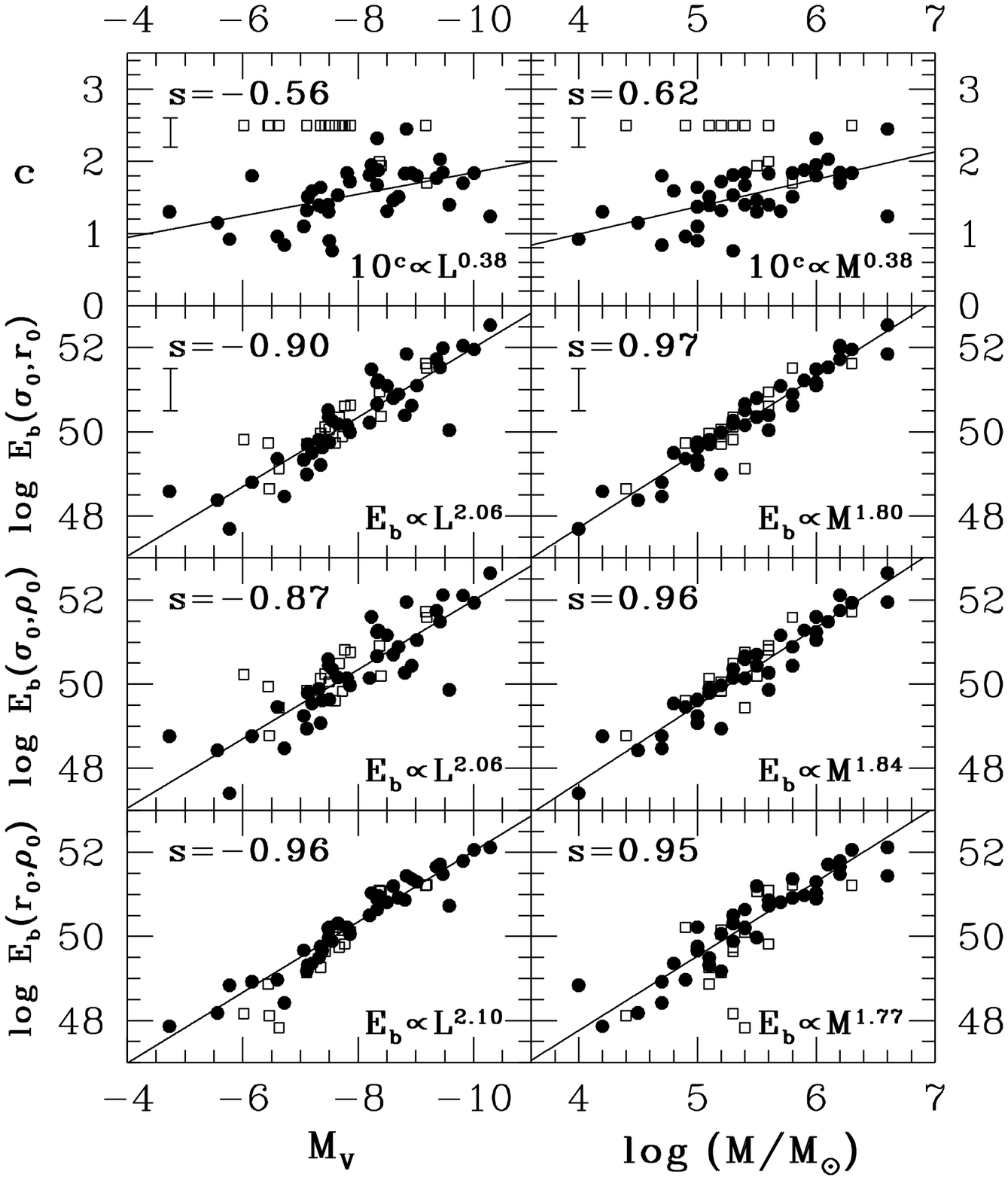]{Concentration parameters and binding energies, as
functions of total luminosity, for globulars with measured central velocity
dispersions (Pryor \& Meylan 1993). Filled circles correspond to
regular, King-model clusters; open squares, to PCC  objects. The three lower
rows present the results of calculating $E_b$ according to each of equations
(\ref{eq:31}). Integrated absolute magnitudes are taken from the catalogue of
Harris (1996), while total masses are obtained by Pryor \& Meylan (1993)
from fits of multi-mass, anisotropic King models.
\label{fig4}}

\figcaption[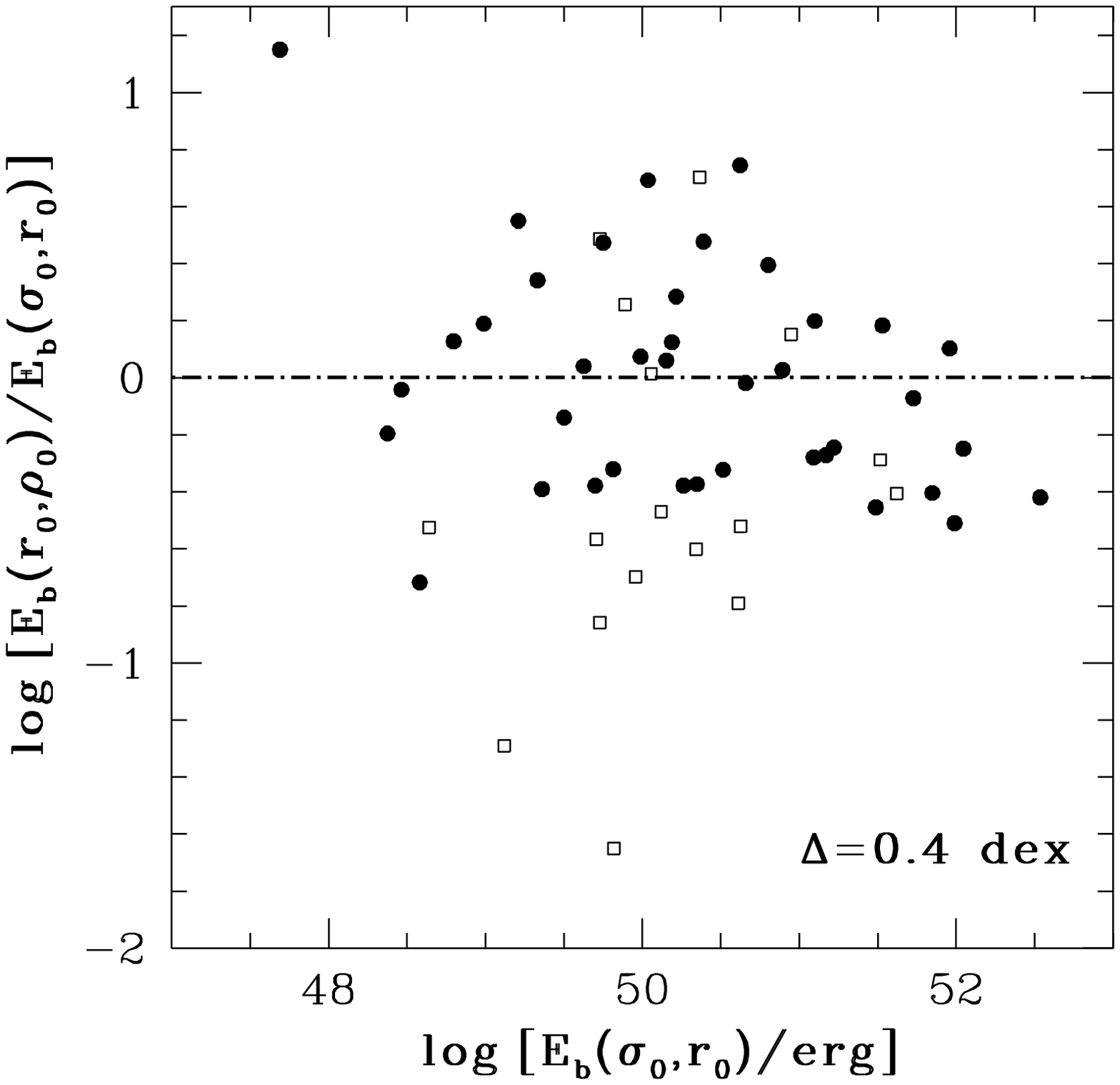]{Confirmation that the binding energy computed using
equation (\ref{eq:31}c) (with $\log\,\Upsilon_{V,0}\equiv0.16$) is equal to
that obtained from (\ref{eq:31}a) for the regular globulars (filled circles)
in the catalogue of Pryor \& Meylan (1993). Open squares again correspond to
the PCC clusters in their sample. This plot is equivalent to Fig.~\ref{fig3}.
\label{fig5}}

\figcaption[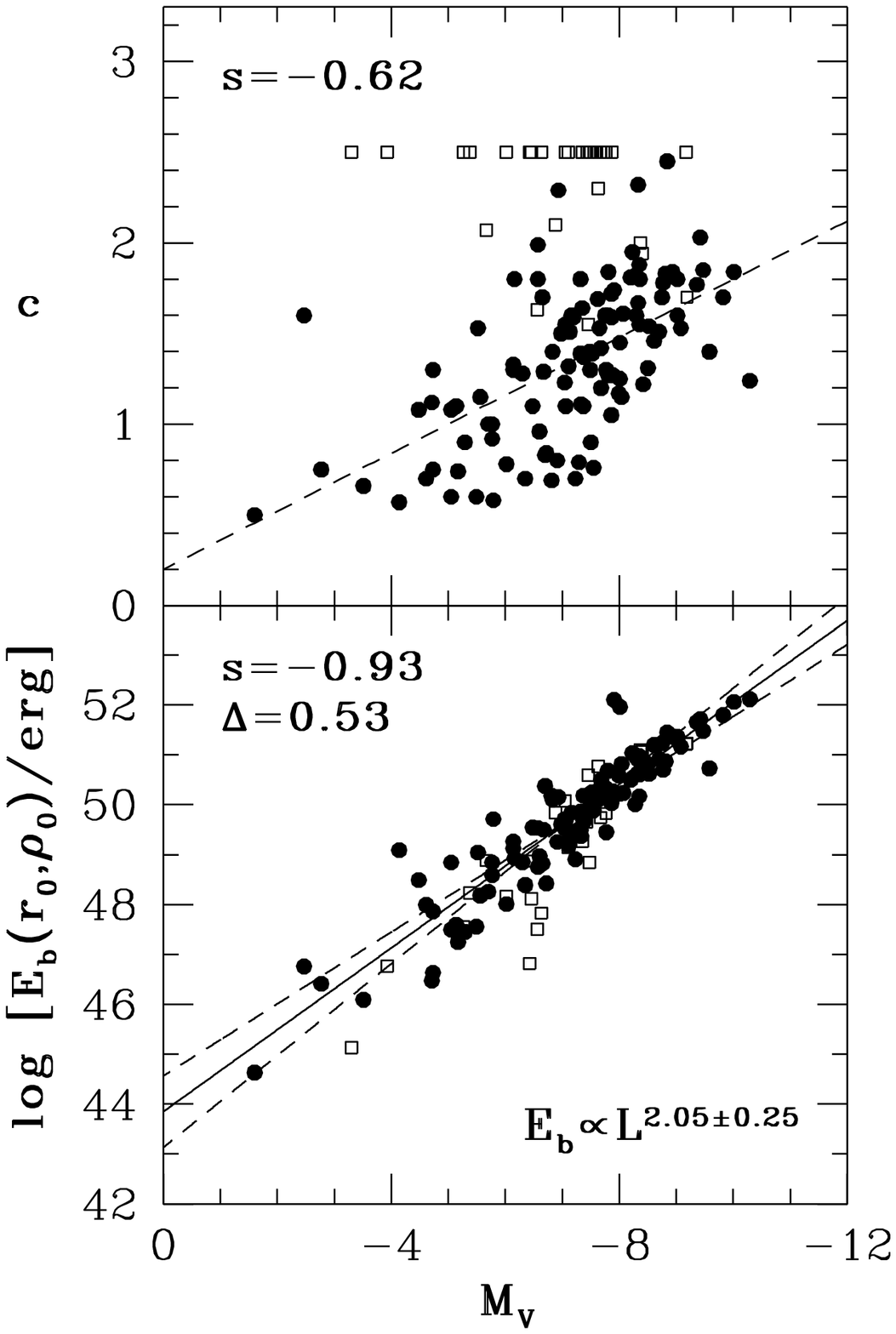]{Central concentrations and binding energies of 109
King-model globulars and 30 PCC clusters (filled circles and open squares)
in the catalogue of Harris (1996). The straight line in the top panel traces
the rough relation $c=-0.57+0.4\,\log\,L$. The solid line in the bottom panel
represents the least-squares fit $\log\,E_b=39.89+2.05\,\log\,L$; dashed lines
show the $3\sigma$ limits (from eq.~[\ref{eq:33}]), $\log\,E_b=41.09+1.8\,
\log\,L$ and $\log\,E_b=38.69+2.3\,\log\,L$.
\label{fig6}}

\figcaption[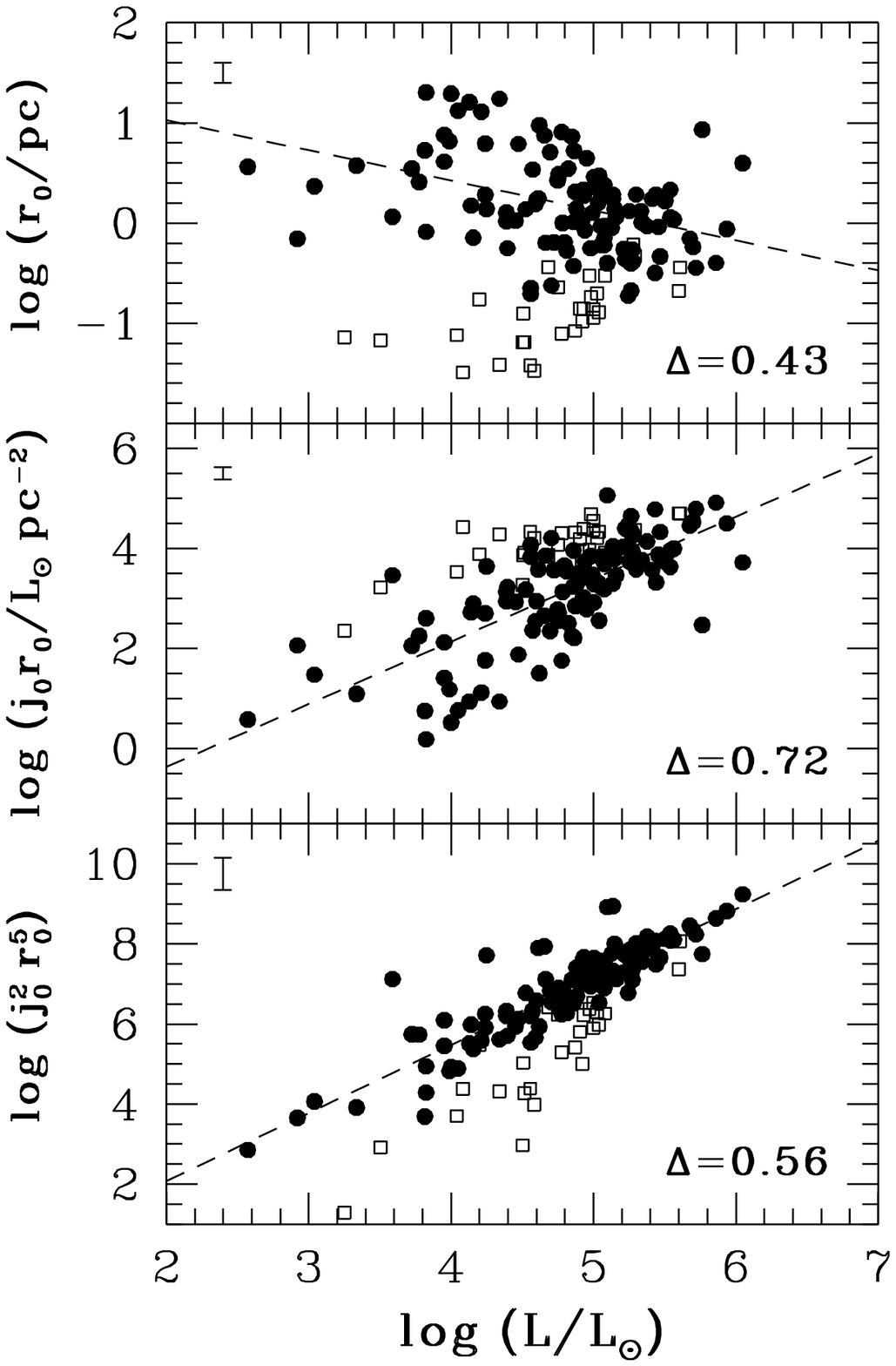]{Monovariate correlations $\log\,r_0$ vs.~$\log\,L$
and $\log\,j_0r_0$ vs.~$\log\,L$ for Galactic globulars in the Harris (1996)
catalogue, as poor reflections of the more fundamental correlation between
$\log\,E_b$ and $\log\,L$. Filled circles are regular clusters; open squares
are PCC objects. The straight line in the top panel has $\log\,r_0=
1.63-0.3\,\log\,L$; in the middle panel, $\log\,j_0r_0=-2.86+1.25\,\log\,L$;
in the bottom panel, $\log\,j_0^2r_0^5=-1.33+1.7\,\log\,L$.
\label{fig7}}

\figcaption[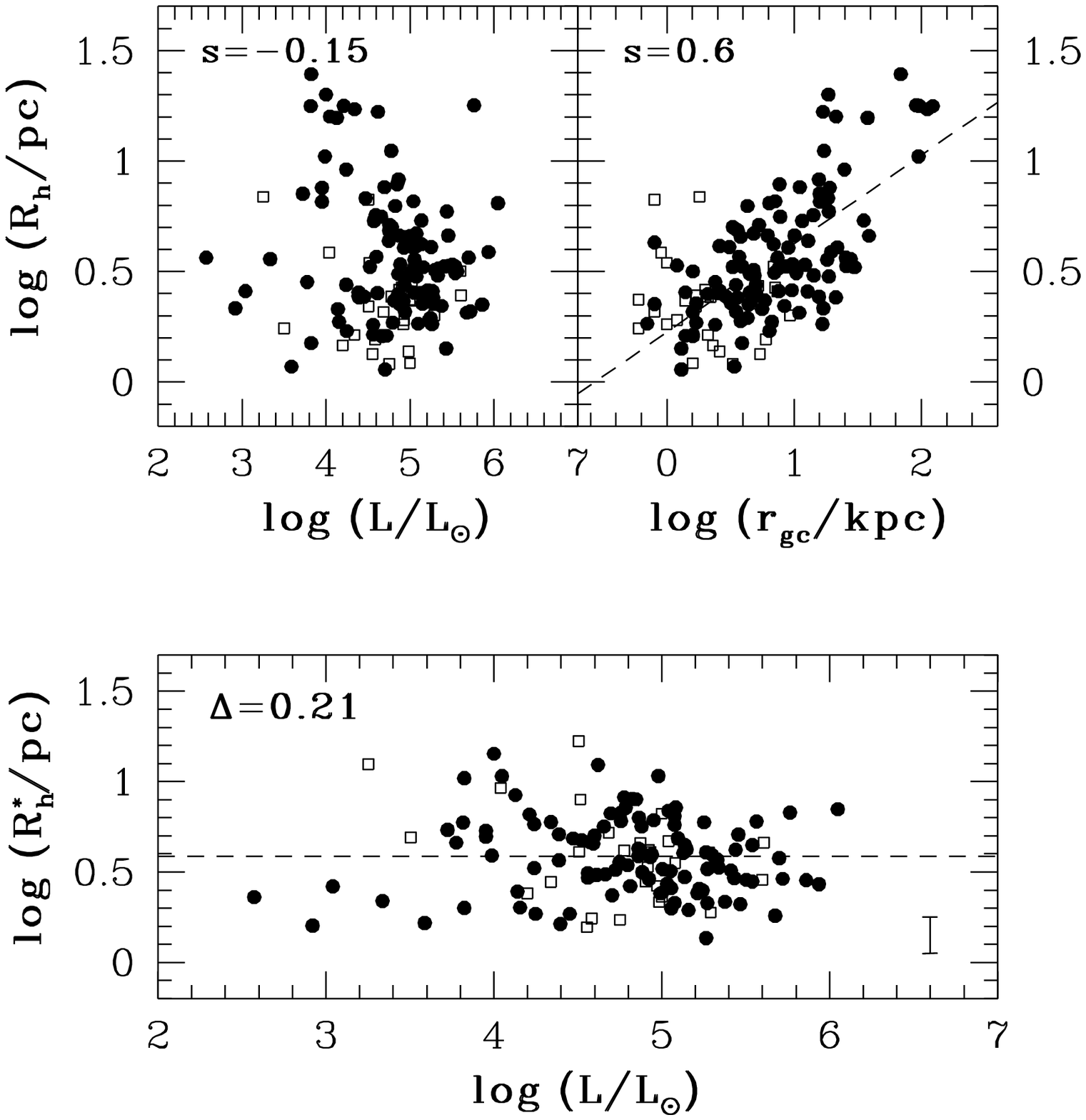]{Dependence of cluster half-light radius on luminosity 
and Galactocentric radius. Point types are the same as in Fig.~\ref{fig7}.
$R_h$ is seen to increase systematically with $r_{\rm gc}$ (the line in
the top right panel is $\log\,R_h=0.226+0.4\,\log\,r_{\rm gc}$), but is
essentially independent of $L$ at a given position in the Milky Way.
$R_h^*$ is the normalized quantity $R_h(r_{\rm gc}/8\,{\rm kpc})^{-0.4}$;
the dashed line in the bottom plot has $\log\,R_h^*\equiv0.59$.
\label{fig8}}

\figcaption[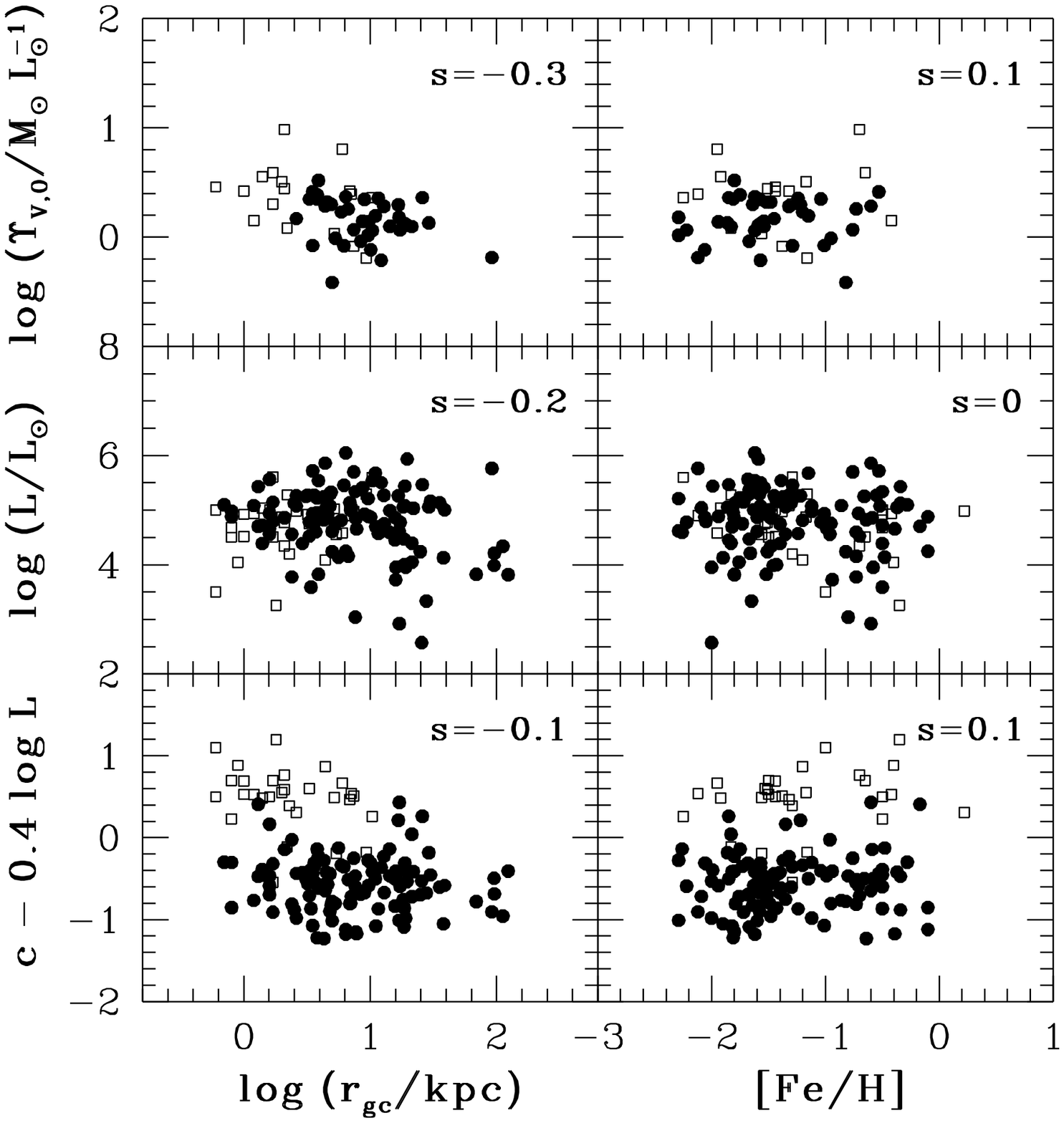]{Insensitivity of three of the basic King-model cluster
parameters to the ``environmental'' factors of Galactocentric position and
cluster metallicity. Open squares again represent PCC clusters; note that $c$
has been {\it arbitrarily} set to 2.5 for many of these.
\label{fig9}}

\figcaption[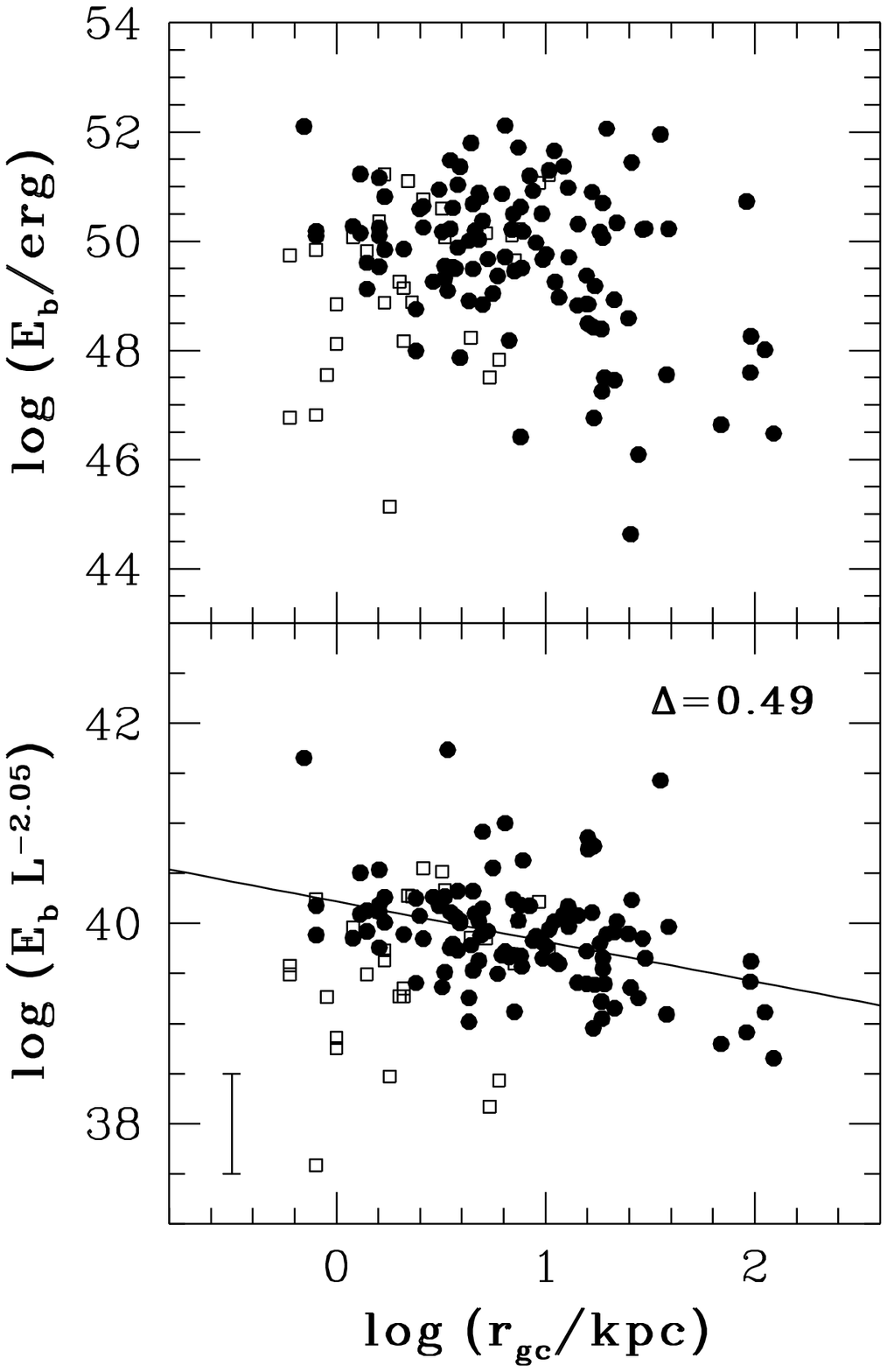]{Dependence of cluster binding energy on Galactocentric
radius. Straight line in the bottom panel represents the fit $E_b\,
L^{-2.05}\propto r_{\rm gc}^{-0.4}$, as in equation (\ref{eq:34}), for
regular clusters from the Harris (1996) catalogue (filled circles). The
observational errorbar shown for $\log\,E_b$ is $\pm0.5$ dex.
\label{fig10}}

\figcaption[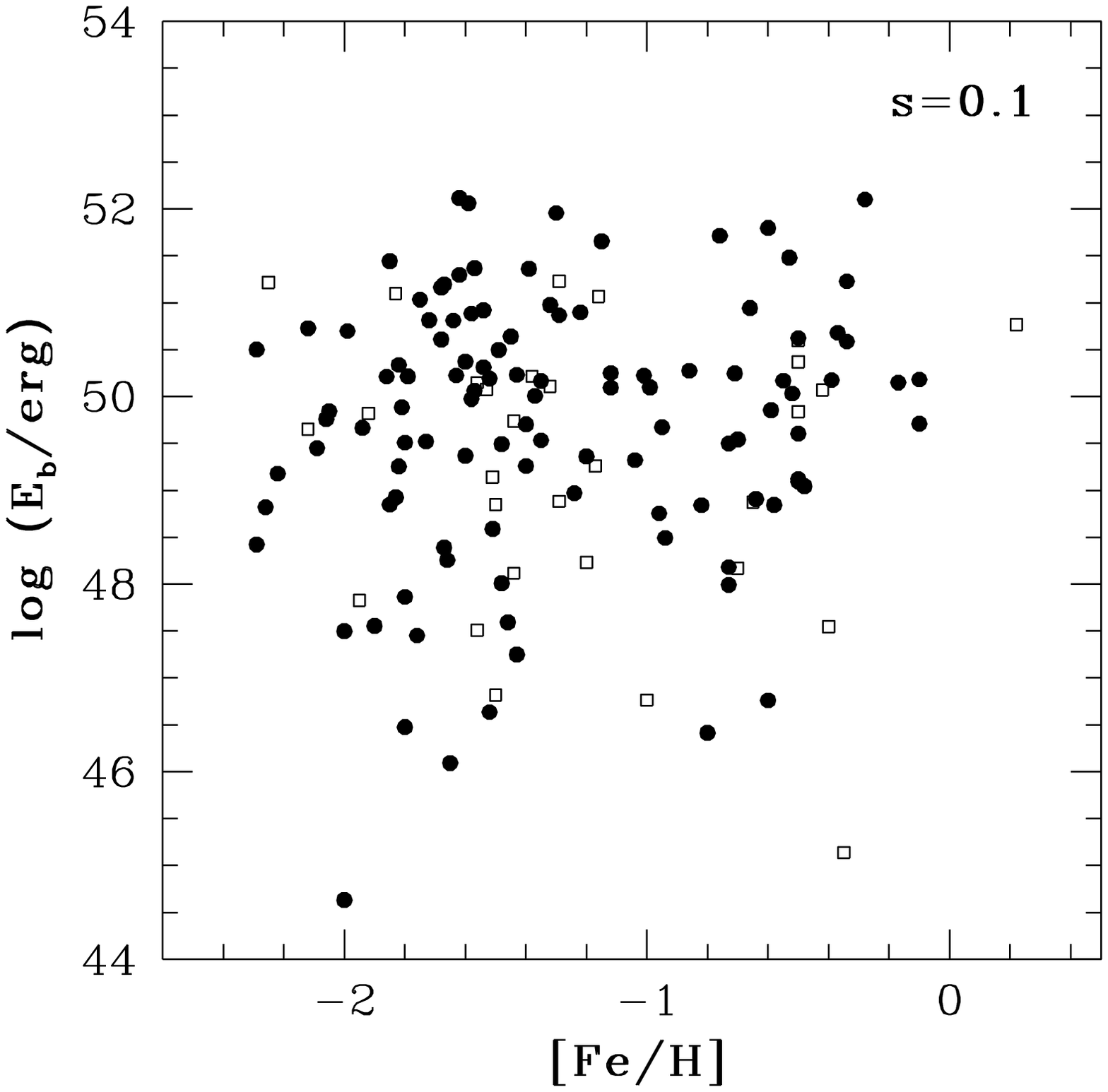]{Independence of binding energy and metallicity for the
139 clusters taken from Harris (1996).
\label{fig11}}

\figcaption[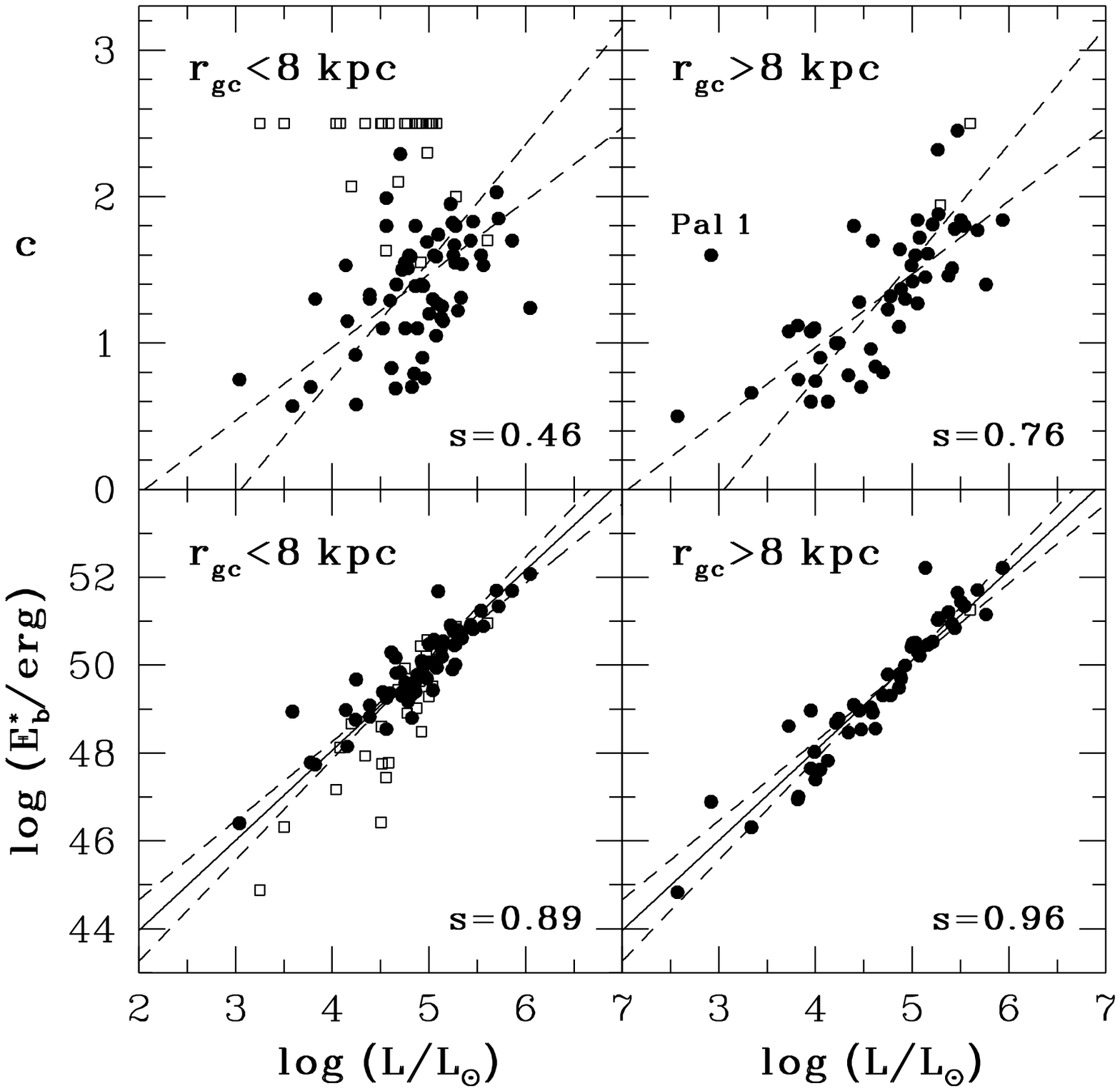]{Concentrations $c$ and normalized binding energies
$E_b^*\equiv E_b(r_{\rm gc}/8\,{\rm kpc})^{-0.4}$ for regular and PCC
Galactic globulars within and beyond $r_{\rm gc}=8$ kpc. Dashed lines
in both top panels are $c=-1.03+0.5\,\log\,L$ and $c=-2.44+0.8\,\log\,L$.
The r.m.s.~scatter in $c$ is $\Delta=0.29$ dex (excluding Palomar 1) at
$r_{\rm gc}>8$ kpc, and $\Delta=0.35$ for $r_{\rm gc}<8$ kpc. Solid line in
the bottom panels is the fit $\log\,E_b^*=39.86+2.05\,\log\,L$, and dashed
lines there are the $3\sigma$ limits (from eq.~[\ref{eq:34}]) $\log\,E_b^*
=41.06+1.8\,\log\,L$ and $\log\,E_b^*=38.66+2.3\,\log\,L$. The r.m.s.~scatter
in $\log\,E_b^*$ is $\pm0.5$ dex.
\label{fig12}}

\figcaption[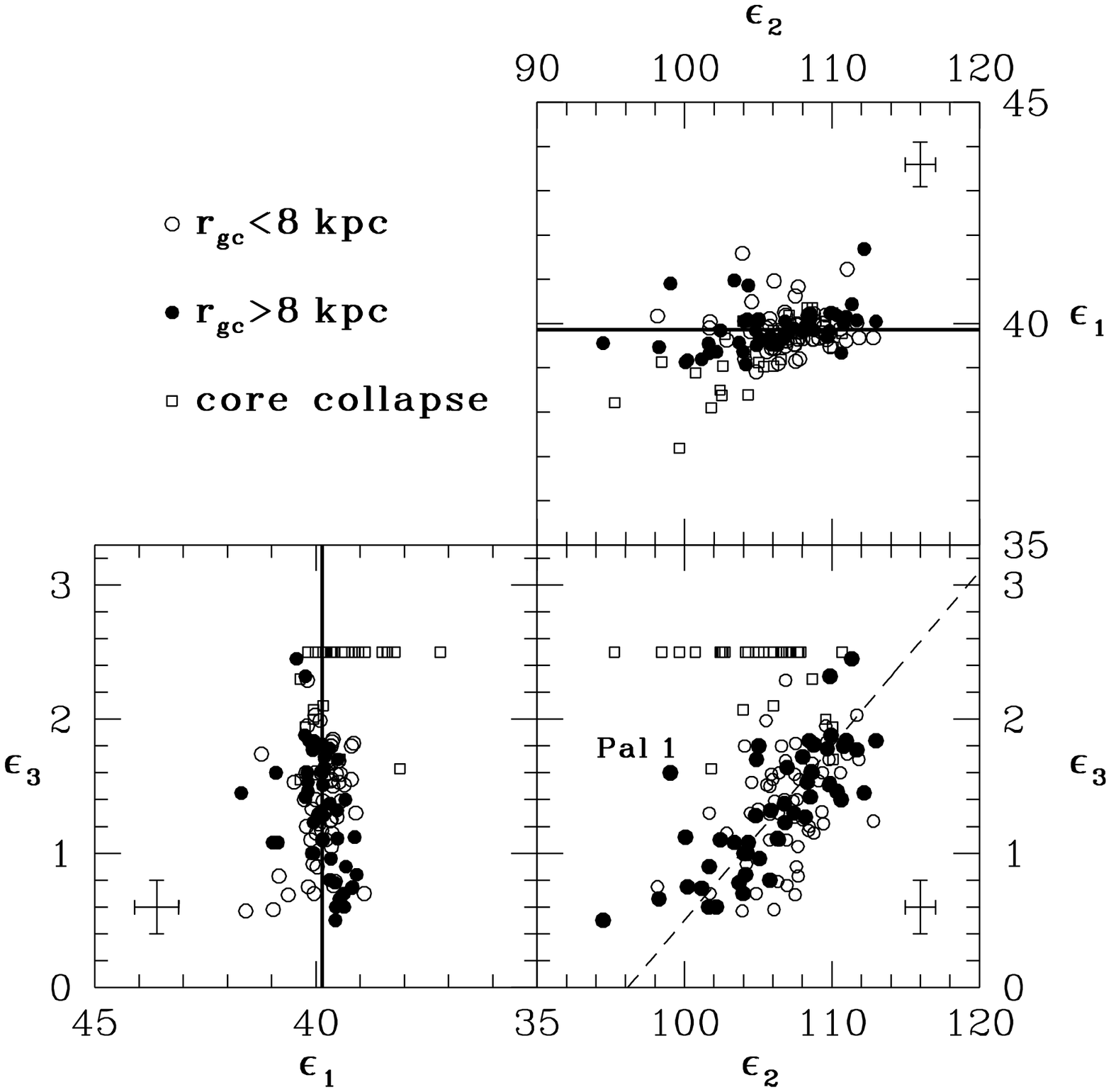]{Fundamental plane of Galactic globular clusters (109
regular and 30 core-collapsed) in the catalogue of Harris (1996). The
orthogonal axes in $\epsilon$-space are defined in equation (\ref{eq:eps}).
The $(\epsilon_1,\epsilon_2)$ and $(\epsilon_1, \epsilon_3)$ cross-sections of
this volume provide edge-on views of the fundamental plane, defined by
$\epsilon_1=39.86$ (eq.~[\ref{eq:34}], drawn as bold solid lines). A face-on
view of the fundamental plane is found in the $(\epsilon_2,\epsilon_3)$ slice
of $\epsilon$-space. Clusters do not fill this plane uniformly but are
confined, particularly at large galactocentric distances, to a fairly narrow
band within it. The dashed line is a fit $\epsilon_3=-12.5+0.13\,\epsilon_2$
to the globulars (excluding Palomar 1) with $r_{\rm gc}>8$ kpc. Most of the
PCC clusters appear to fall away from the fundamental plane here, but these
objects are generally {\it not} well described by King models.
\label{fig13}}

\figcaption[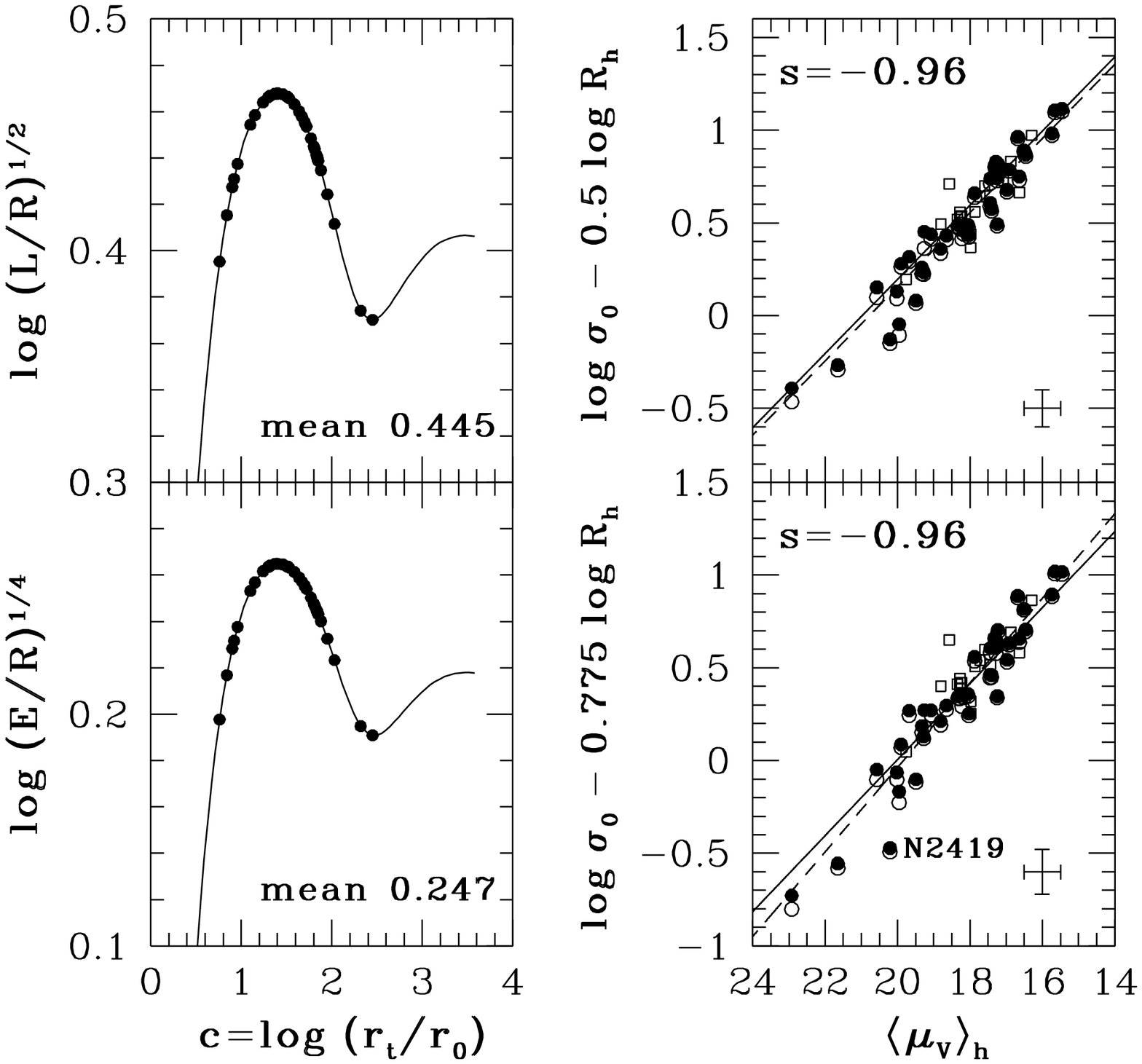]{{\it Left panels}: Effects of cluster non-homology on
two empirical, bivariate correlations involving {\it half-light}
parameters of Galactic globulars. The functions ${\cal L}(c)$, ${\cal R}(c)$,
and ${\cal E}(c)$ are given individually in Figs.~\ref{fig1} and \ref{fig2},
and in Appendix B. Points are placed at the observed $c$-values of the 39
regular clusters in the catalogue of Pryor \& Meylan (1993). {\it Right
panels}: Observed correlations in the Pryor \& Meylan cluster sample. Filled
circles are their regular clusters, and the 17 open squares are their PCC
clusters. Open circles plot the uncorrected $\sigma_{p,0}$ for the regular
objects, rather than the model velocity scales $\sigma_0$. Solid lines in the
top and bottom panels represent equations (\ref{eq:41}b) and (\ref{eq:42}b);
dashed lines are least-squares fits to the data (filled circles). The distant
cluster NGC 2419 is noted as an outlier (see Fig.~\ref{fig18}).
\label{fig14}}

\figcaption[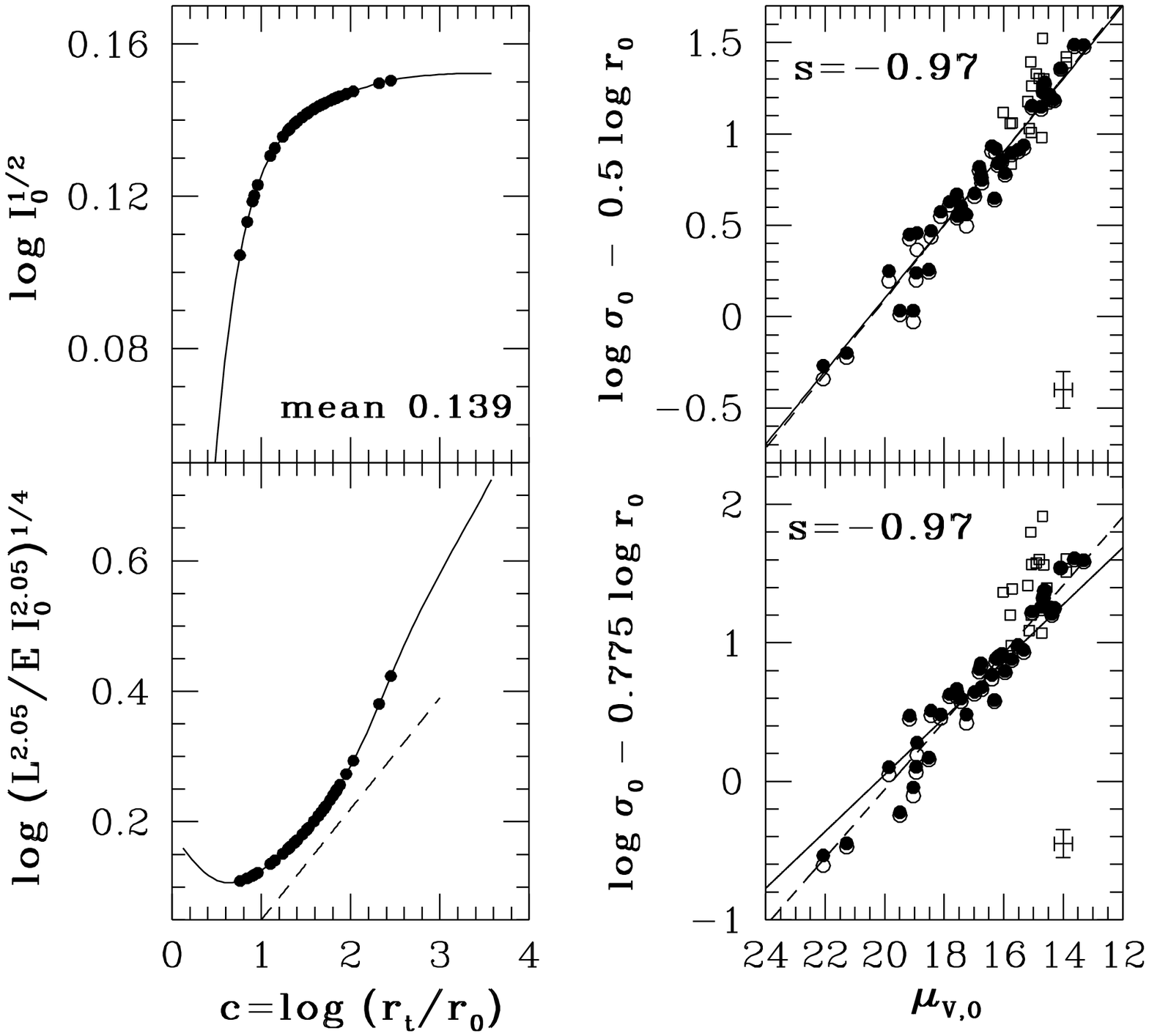]{{\it Left panels}: Effects of cluster non-homology on
two empirical, bivariate correlations involving {\it core}
parameters of Galactic globulars. The functions ${\cal I}_0(c)$, ${\cal L}(c)$,
and ${\cal E}(c)$ are given individually in Figs.~\ref{fig1} and \ref{fig2},
and in Appendix B. The dashed line in the bottom panel shows the rough
scaling $0.25\,\log\,\left({\cal L}^{2.05}/{\cal E}{\cal I}_0^{2.05}\right)
\approx const.+0.17\,c$. {\it Right panels}: Correlations observed for the
Pryor \& Meylan (1993) clusters. Point types have the same meaning as in
Fig.~\ref{fig14}. Solid line in the top panel represents equation
(\ref{eq:41}a) and is indistinguishable from a least-squares fit to the
regular-cluster data. Solid line in the bottom panel is equation (\ref{eq:42}a)
with an intercept of 4.147, estimated by assuming a constant $\langle 0.25\,
\log\,({\cal L}^{2.05}/{\cal E}{\cal I}_0^{2.05})\rangle=0.204$ for all
39 King-model clusters. The least-squares fit (dashed line) is quite
different ($\log\,\sigma_0-0.775\,\log\,r_0=-0.246\,\mu_{V,0}+4.860$)
because of the significant non-homology in this case.
\label{fig15}}

\figcaption[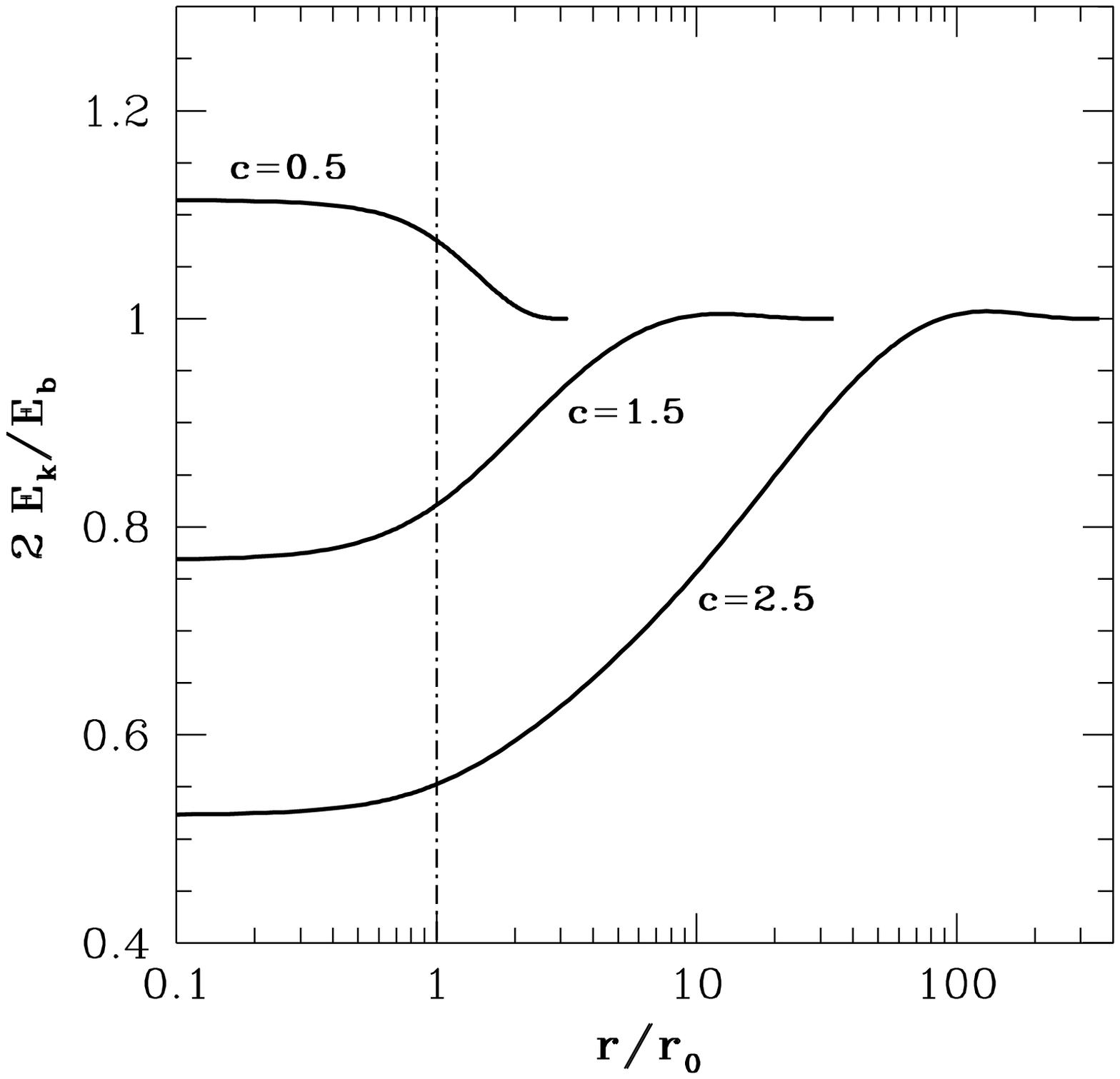]{Virial ratio $2E_k/E_b$ as a function of radius inside
three single-mass, isotropic King (1966) models bracketing the range of central
concentration observed for Galactic globulars. Cluster cores, defined by $r\le
r_0$, in general do {\it not} obey a simplified virial theorem of the form
$2E_k=E_b$.
\label{fig16}}

\figcaption[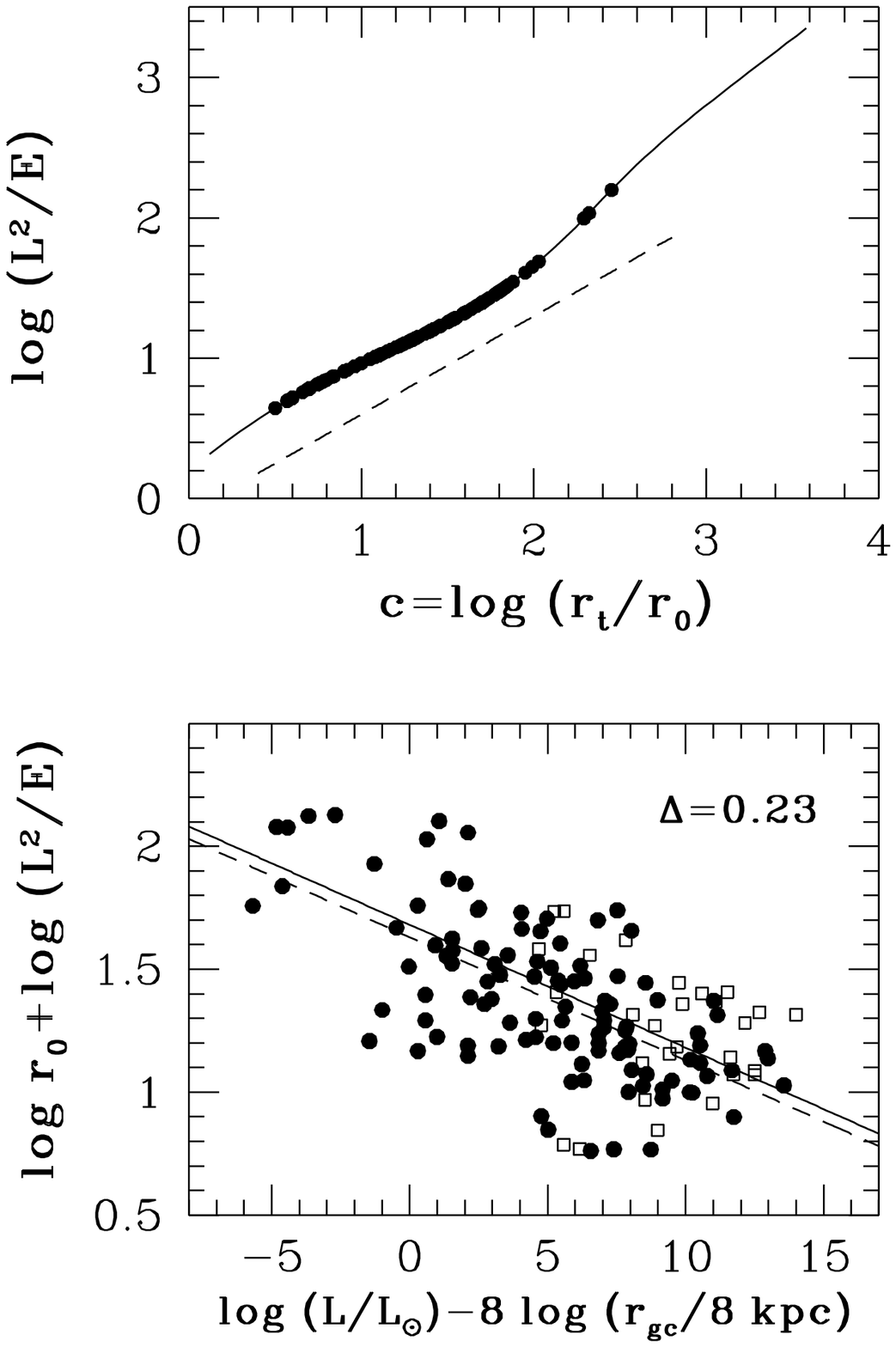]{Source of the monovariate correlation between
$\log\,r_0$ and $\log\,L$ in Fig.~\ref{fig7}. {\it Top panel}: The non-homology
term $\log\,({\cal L}^2/{\cal E})$ as a function of central concentration.
Solid curve is obtained from numerical integrations of King models, points are
placed at the observed $c$-values of the 109 regular clusters in the catalogue
of Harris (1996), and the dashed line shows $\log\,({\cal L}^2/{\cal E})\approx
const.+0.7\,c$. {\it Bottom panel}: The full correlation involving $\log\,r_0$,
$c$, $\log\,L$, and $\log\,r_{\rm gc}$. Data for 30 PCC clusters from Harris
(1996) are also shown (open squares). Solid line is equation (\ref{eq:51}a):
$\log\,r_0+\log\,({\cal L}^2/{\cal E})=1.681-0.05\,\left[\log\,L-8\,
\log(r_{\rm gc}/8\,{\rm kpc})\right]$; the dashed line, with an intercept of
1.63, is the least-squares fit to the regular-cluster data (filled circles).
\label{fig17}}

\figcaption[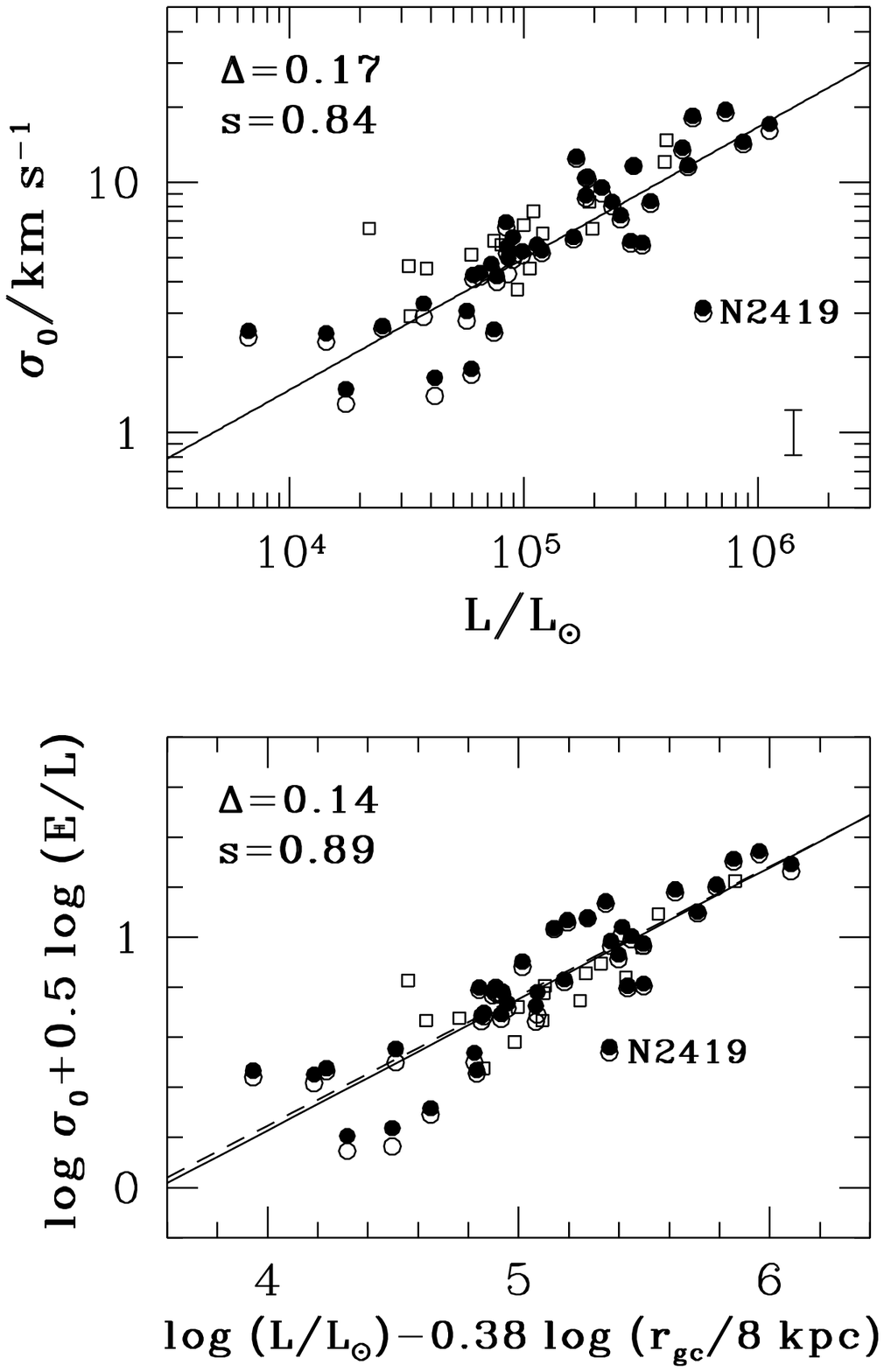]{{\it Top panel}: Monovariate correlation between
$\log\,\sigma_0$ and $\log\,L$ for clusters from Pryor \& Meylan (1993).
Point types have the same meaning as in Fig.~\ref{fig14}. The observational
errorbar on $\log\,\sigma_0$ is $\pm0.09$ dex. The straight line is equation
(\ref{eq:51}f) after substituting appropriate averages for the ratio
${\cal E}(c)/{\cal L}(c)$ and $r_{\rm gc}$ (see text). {\it Bottom panel}: The
full correlation expected on the basis of the fundamental-plane constraints on
$\Upsilon_{V,0}$ and $E_b^*$. Solid line is equation (\ref{eq:51}f); dashed
line is a least-squares fit to the regular-cluster data (filled circles).
\label{fig18}}

\figcaption[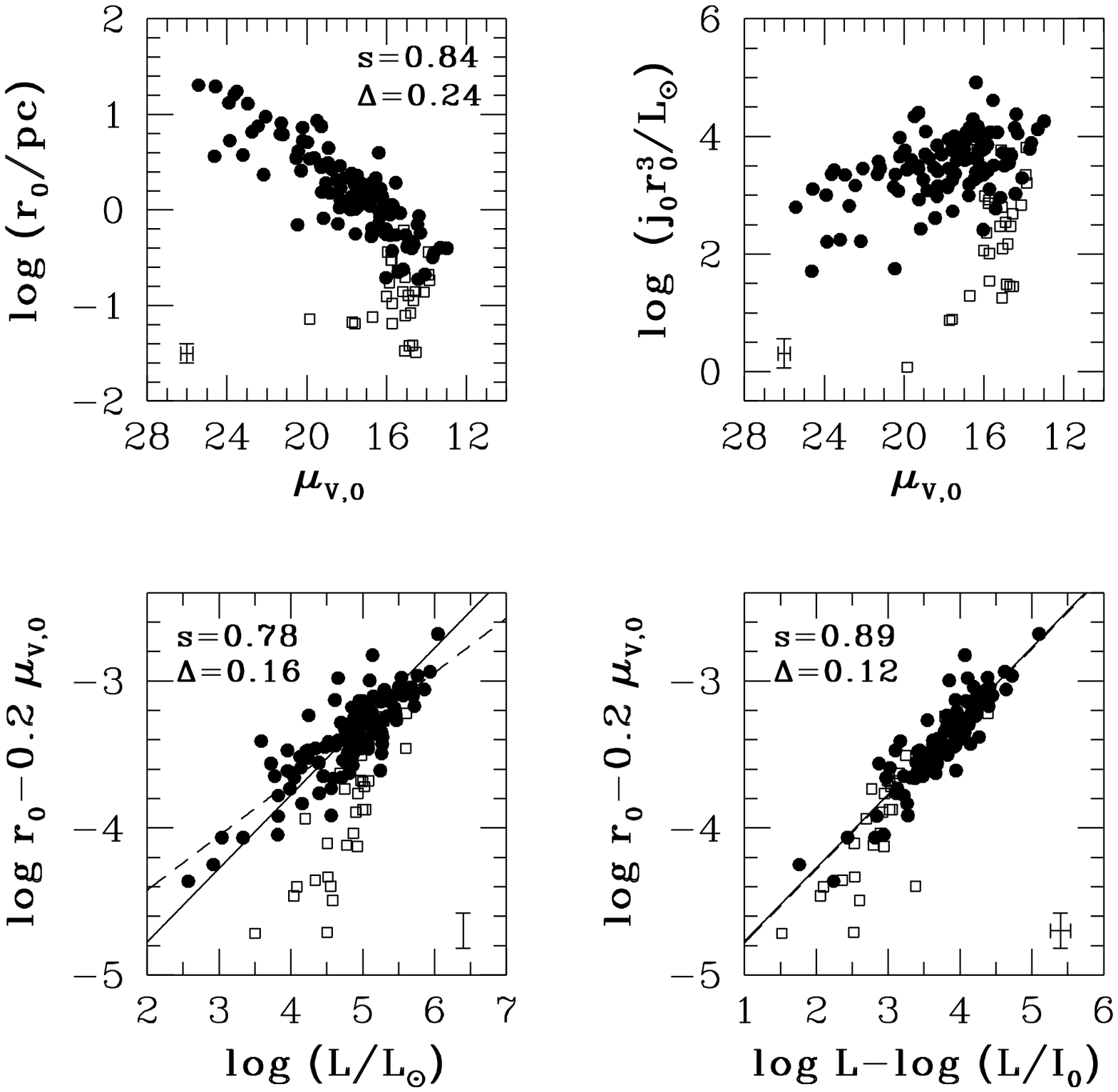]{{\it Top left}: Monovariate correlation between
$\log\,r_0$ and $\mu_{V,0}$ for globulars in the catalogue of Harris (1996).
Solid points refer to the 109 King-model clusters; open squares, to the 30 PCC
objects. The observed r.m.s.~scatter in $\log\,r_0$ exceeds the observational
errorbar of $\pm0.1$ dex. {\it Top right}: Real variations in the ``core
mass,'' $M_0\propto j_0r_0^3$. {\it Bottom left}: Deviations from the line
expected if core mass were a constant are not random, but correlated with
total luminosity. {\it Bottom right}: The first three panels are explained as
results of the behavior of {\it total} luminosity as a function of $c$ in King
models, combined with the definition of core surface brightness. The straight
line is equation (\ref{eq:52}).
\label{fig19}}

\setcounter{figure}{0}

\plotone{fig1.eps}
\figcaption[fig1.eps]{}

\plotone{fig2.eps}
\figcaption[fig2.eps]{}

\plotone{fig3.eps}
\figcaption[fig3.eps]{}

\plotone{fig4.eps}
\figcaption[fig4.eps]{}

\plotone{fig5.eps}
\figcaption[fig5.eps]{}

\plotone{fig6.eps}
\figcaption[fig6.eps]{}

\plotone{fig7.eps}
\figcaption[fig7.eps]{}

\plotone{fig8.eps}
\figcaption[fig8.eps]{}

\plotone{fig9.eps}
\figcaption[fig9.eps]{}

\plotone{fig10.eps}
\figcaption[fig10.eps]{}

\plotone{fig11.eps}
\figcaption[fig11.eps]{}

\plotone{fig12.eps}
\figcaption[fig12.eps]{}

\plotone{fig13.eps}
\figcaption[fig13.eps]{}

\plotone{fig14.eps}
\figcaption[fig14.eps]{}

\plotone{fig15.eps}
\figcaption[fig15.eps]{}

\plotone{fig16.eps}
\figcaption[fig16.eps]{}

\plotone{fig17.eps}
\figcaption[fig17.eps]{}

\plotone{fig18.eps}
\figcaption[fig18.eps]{}

\plotone{fig19.eps}
\figcaption[fig19.eps]{}

\end{document}